\documentstyle[12pt,psfig,amsmath,amssymb]{article}

\newdimen\singlebaseskip		
\newdimen\doublebaseskip		
\font\eightrm=cmr8			
\font\sc=cmcsc10 scaled\magstep0	

\font\bbbrm=cmbx10 scaled\magstep1
\font\bbrm=cmbx10

\font\tenbmit=cmmib10		
\font\sevenbmit=cmmib10 at 7pt	
\font\fivebmit=cmmib10 at 5pt	
\mathchardef\BOLDalpha="710B 	
 \mathchardef\BOLDbeta="710C 	
\mathchardef\BOLDgamma="710D 	
  \mathchardef\BOLDrho="711A 	
\def\={\overline}		
\def\cms{\ifmmode {\rm\,cm\,\,s^{-1}} \else {$\cms$} \fi} 
\def\cmss{\ifmmode {\rm\,cm\,\,s^{-2}} \else {$\cmss$} \fi} 
\def\deg{\ifmmode {^\circ} \else {${}^\circ$} \fi}	
\def\ergg{\ifmmode {\rm\,erg\,\,g^{-1}} \else {$\ergg$} \fi} 
\def\eg{{\it e.g.\/}}		
\def\etal{{\it et al. \/}}	
\def\gcms{\ifmmode {\rm\,g\,\,cm^{-2}} \else {$\gcms$} \fi} 
\def\gcmc{\ifmmode {\rm\,g\,\,cm^{-3}} \else {$\gcmc$} \fi} 
\def\ie{{\it i.e.\/}}		
\def\kms{\ifmmode {\rm\,km\,\,s^{-1}} \else {$\kms$} \fi} 
		% ...
\def\lsun{\ifmmode {\rm\,L_\odot} \else {$\lsun$} \fi}	
\def\msun{\ifmmode {\rm\,M_\odot} \else {$\msun$} \fi}	
\def\rsun{\ifmmode {\rm\,R_\odot} \else {$\rsun$} \fi}	
\def\s{\ifmmode \widetilde \else \~\fi} 
\def\spose#1{\hbox to 0pt{#1\hss}} %
\def\Dt{\spose{\raise 1.5ex\hbox{\hskip3pt$\mathchar"201$}}}    
\def\dt{\spose{\raise 1.0ex\hbox{\hskip2pt$\mathchar"201$}}}    
\def\gta{\mathrel{\spose{\lower 3pt\hbox{$\mathchar"218$}}
     \raise 2.0pt\hbox{$\mathchar"13E$}}}
\def\lta{\mathrel{\spose{\lower 3pt\hbox{$\mathchar"218$}}
     \raise 2.0pt\hbox{$\mathchar"13C$}}}

\newcount\notenumber\notenumber=1
\def\foot#1{\raise3pt\hbox{\eightrm \the\notenumber}
     \hfil\par\vskip3pt\hrule\vskip6pt
     \noindent\raise3pt\hbox{\eightrm \the\notenumber}
     #1\par\vskip6pt\hrule\vskip3pt\noindent\global\advance\notenumber by 1}
\def\note#1{\footnote{$^{\the\notenumber}$}{#1}\global\advance\notenumber by 1}
\def\alph#1{\ifcase#1\or a\or b\or c\or d\or e\or f\or g\or h\or i\or j\or
	k\or l\or m\or n\or o\or p\or q\or s\or t\or u\or v\or w\or x\or
	y\or z\else #1\fi}
\def\Alph#1{\ifcase#1\or A\or B\or C\or D\or E\or F\or G\or H\or I\or J\or
	K\or L\or M\or N\or O\or P\or Q\or S\or T\or U\or V\or W\or X\or
	Y\or Z\else #1\fi}

\def\Roman#1{\expandafter\uppercase\expandafter{\romannumeral #1}}
\newcount\sectno    \sectno=0
\newcount\subno     \subno=0
\newcount\subsubno  \subsubno=0
\newcount\eqnmbrsec \eqnmbrsec=0
\newcount\eqnmbr    \eqnmbr=0
\def\sectionbeg#1{\penalty-200\bigskip\par 
	\advance\sectno by 1
	\subno=0\subsubno=0\eqnmbrsec=0
	\noindent{\bbbrm \Roman\sectno. #1}
	\nobreak\smallskip\par}

\def\subsectbeg#1{\penalty-200\medskip\par
	\advance\subno by 1\subsubno=0
	\noindent\hskip 0.25truein {\bbrm \Alph\subno. #1}
	\nobreak\smallskip\par}
\def\subsubbeg#1{\penalty-200\medskip\par
	\advance\subsubno by 1
	\noindent\hskip 0.50truein {\it \number\subsubno. #1}
	\nobreak\smallskip\par}

\def\eqnam#1{\xdef#1{\the\eqnmbr}}		  
\def\eqnamA#1{\xdef#1{{\it A}\the\eqnmbr}}		  
\def\eqnumrun{\global\advance\eqnmbr by 1 \the\eqnmbr}
\def\eqnumsec{\global\advance\eqnmbrsec by 1 \the\sectno.\the\eqnmbrsec}
\def\refindent{\par\noindent\parskip=4pt\hangindent=3pc\hangafter=1 }
\def\refpaper#1#2#3#4#5{\refindent{\sc#1}. #2. {\it #3\/} {\bf#4}, #5.}
\def\refbook#1#2{\refindent{\sc#1}. #2.}
\def\aanda#1#2#3#4{\refpaper{#1}{#2}{Astron. Astrophys.}{#3}{#4}}
\def\aj#1#2#3#4{\refpaper{#1}{#2}{Astron. J.}{#3}{#4}}

\def\apj#1#2#3#4{\refpaper{#1}{#2}{Astrophys. J.}{#3}{#4}}

\def\icarus#1#2#3#4{\refpaper{#1}{#2}{Icarus}{#3}{#4}}

\def\jgr#1#2#3#4{\refpaper{#1}{#2}{J. Geophys. Res.}{#3}{#4}}
\def\mnras#1#2#3#4{\refpaper{#1}{#2}{Mon. Not. Roy. Astron. Soc.}{#3}{#4}}
\def\nature#1#2#3#4{\refpaper{#1}{#2}{Nature}{#3}{#4}}

\def\pss#1#2#3#4{\refpaper{#1}{#2}{Plan. Space Sci.}{#3}{#4}}
\def\ptps#1#2#3#4{\refpaper{#1}{#2}{Prog. Theor. Phys. Suppl.}{#3}{#4}}
\def\rmp#1#2#3#4{\refpaper{#1}{#2}{Rev. Modern Phys.}{#3}{#4}}
\def\science#1#2#3#4{\refpaper{#1}{#2}{Science}{#3}{#4}}

\def\ssr#1#2#3#4{\refpaper{#1}{#2}{Sol. Sys. Res.}{#3}{#4}}

\def\jgres#1#2#3#4{\refpaper{#1}{#2}{Geophys. Res. Lett}{#3}{#4}}
\def\refpaper#1#2#3#4#5{\refindent{\sc#1}. {\it #3\/} {\bf#4}, #5.}

\def\refpaper#1#2#3#4#5{\refindent{\sc#1}. #2. {\it #3\/} {\bf#4}, #5.}
\def\zamm#1#2#3#4{\refpaper{#1}{#2}{Z. Angew. Math. Mech.}{#3}{#4}}

\begin{document}

\begin{titlepage}
\begin{center}
\begin{Large}
{\bf A Gas-poor Planetesimal Capture Model for the Formation of Giant Planet
Satellite Systems}\\
\end{Large}
\vskip 0.5 in
Paul R. Estrada \\
NASA Ames Research Center, Mail Stop 245-3 \\
Moffett Field, California 94305\\
estrada@cosmic.arc.nasa.gov\\
\vskip 0.1in
and\\
\vskip 0.1in
Ignacio Mosqueira \\
NASA Ames/SETI Institute, Mail Stop 245-3 \\
Moffett Field, California 94305\\
mosqueir@astrosun.tn.cornell.edu\\
\vskip 0.25in
Submitted to Icarus: May, 2005\\
Revised: August, 2005\\
\vskip 0.4in
98 manuscript pages, including... \\
11 Figures and 3 appendices \\
\end{center}
\end{titlepage}

\begin{titlepage}
\begin{center}
Proposed Running Title: ``Planetesimal Capture Model'' \\
\vskip 0.75in
Correspondence Address: \\
\vskip 0.1in
Paul R. Estrada \\
Mail Stop 245-3 \\
NASA Ames Research Center \\
Moffett Field, CA 94305-1000 \\
\vskip 0.75in
{\it E-mail}: estrada@cosmic.arc.nasa.gov \\
{\it Phone}: (650) 604-6001 \\
{\it FAX}:   (650) 604-6779 \\
\end{center}
\end{titlepage}

\begin{abstract}
Assuming that an unknown mechanism (\eg, gas turbulence) removes most of the
subnebula gas disk in a timescale shorter than that for satellite formation,
we develop a model for the formation of regular (and possibly at least some
of the irregular) satellites around giant planets in a gas-poor environment. 
In this model, which follows along the lines of the work of Safronov \etal 
(1986), heliocentric planetesimals collide within
the planet's Hill sphere and generate a circumplanetary
disk of prograde and retrograde satellitesimals
extending as far out as $\sim R_H/2$. At first, the net angular momentum
of this proto-satellite swarm is small, and collisions among
satellitesimals leads to loss of mass from the outer disk, and delivers
mass to the inner disk (where regular satellites form) in
a timescale $\lesssim 10^5$ years. This mass loss may
be offset by continued collisional capture of sufficiently small $< 1$ km
interlopers resulting from the disruption of
planetesimals in the feeding zone of the giant planet. As the planet's feeding
zone is cleared in a timescale $\lesssim 10^5$ years, enough
angular momentum may be delivered to the proto-satellite swarm to account
for the angular momentum of the regular satellites of Jupiter and Saturn.
This feeding timescale is
also roughly consistent with the independent constraint that the Galilean
satellites formed in a timescale of $10^5-10^6$ years, which may be
long enough to accomodate Callisto's partially differentiated state
(Anderson \etal 1998; 2001). In turn, this formation timescale can be used 
to provide plausible constraints on the surface density of solids in the
satellitesimal disk (excluding satellite embryos 
$\sim 1$ g cm$^{-2}$ for satellitesimals of size
$\sim 1$ km), which 
yields a total
disk mass smaller than the mass of the regular satellites, and means
that the satellites must form in several $\sim 10$ collisional cycles.
However, much more work will need to be conducted
concerning the collisional evolution
both of the circumplanetary satellitesimals
and of the heliocentric planetesimals following giant planet
formation before one can assess the significance of this agreement.
Furthermore, for enough mass to be delivered to form the regular
satellites in the required timescale one may need to rely on (unproven) 
mechanisms to replenish the feeding zone of the giant planet. We
compare this model to the solids-enhanced
minimum mass (SEMM) model of Mosqueira and Estrada (2003a,b), and
discuss its main consequences for Cassini observations of the
Saturnian satellite system.

\end{abstract}
\vspace{0.5in}

\newpage
\section{Introduction}
\label{sec:intro}

The similarities between planetary and satellite systems
led to early suggestions that any theory for planetary formation
should incorporate regular satellite formation as a natural byproduct
(Alfven 1971; De \etal 1977). More recently the
propensity has been to argue against a unifying, generally applicable
formation model for planets and satellites (\eg, Stevenson \etal 1986).
Yet, whether or not regular satellites and planets formed in an analogous
manner, it is often remarked that the scaled bulk
characteristics of angular momentum and mass are shared
by the regular satellites of Jupiter, Saturn and Uranus (\eg,
Pollack \etal 1991). The first to suggest that regular satellites formed
out of a collisionally-captured gravitationally-bound swarm of
circumplanetary satellitesimals was O. Yu Schmidt (1957). This idea
was subsequently explored in a number of publications
(Safronov and Ruskol (1977); Ruskol 1981, 1982; Safronov \etal 1986).
However, it is fair to say that
while these works provide valuable insight into the physical mechanisms
possibly involved in the formation of regular satellites of giant
planets out of
an extended circumplanetary satellitesimal swarm, they fall short of
advancing a specific model that can account for the observed properties
of the satellite systems. In this contribution, we revisit this approach
with the aim of providing a testable model as well as a basis for
comparison with our published solids-enhanced minimum mass (SEMM)
satellite formation model (Mosqueira and Estrada
2003a,b; hereafter, MEa,b).

The primary difference between the SEMM model
and the present gas-poor planetesimal
capture (GPPC) model is in the treatment of turbulence. It is commonplace
in the planetary formation 
literature to assume that Keplerian disks are turbulent and
to model the resulting anomalous angular momentum transport in terms
of an $\alpha \sim 10^{-3}$ parameter\footnote{It should be kept in mind,
however, that some workers favor the $\beta$ parameter 
(\ie, Richard and Zahn 1999)}
(Shakura and Sunyaev 1973) regardless of disk conditions or distance from
the central object. While this assumption may or may not (given
the scatter in the data and the existence of alternative explanations)
be justified in the case of the solar nebula by
the observation that disks around young protostars
accrete at low rates $\sim 10^{-8} M_{\odot}$ yr$^{-1}$ (Hartmann \etal
1998), there is no reason to suppose that similar $\alpha$ values need also
apply to circumplanetary nebulae at the time of satellite formation. Indeed,
even those turbulent mechanisms (such as MHD [Balbus and Hawley 1998]
and baroclinic instabilities [Li \etal 2000; Klahr and Bodenheimer 2003])
that may operate at least at some
locations and times in the case of the planetary disk may fail in the
dense (but not massive), cold and mostly isothermal circumplanetary disk
(MEa).\footnote{One would expect subnebula turbulence while
there is significant Roche-lobe gas inflow, but it is difficult
to see how this source of turbulence can be relied upon to remove the 
circumplanetary gas disk.}

It is often suggested that the disk's radial shear may drive turbulence
for sufficiently high Reynolds number (Zahn 1991; Dubrulle 1993; Richard and
Zahn 1999). If true, such a mechanism
would apply quite generally to accretion disks (both planetary and
satellite disks would be affected); however, so far numerical
simulations (Hawley \etal 1999; Mosqueira \etal 2003)
have failed to show the presence of a non-linear shear instability,
apparently due to the stabilizing Coriolis terms in flows with
positive radial gradients in specific angular momentum (cf. Rayleigh
stable flows), even in the presence of both a radial and a vertical
shear (Rudiger \etal 2002).
In fact, it has been claimed that a carefully constructed
Taylor-Couette experiment shows stability in the case of positive
radial gradients in specific angular momentum even for high Reynolds
number (Schultz-Grunow 1959)\footnote{Though it is unclear whether this
would remain
true for even higher Reynolds number. Experimentally, the onset of
turbulence in pipe flows cannot be described simply by
linear stability analysis.}. The difficulty
stems not only from the degree of turbulence, but also whether such
turbulence, if present, may be characterized by an $\alpha$
parameter.

Given the lack of observational or theoretical quantitative
constraints for the
gas inflow rate through the gap or planetary accretion rate
in the final stages of giant planet formation (Lubow \etal 1999),
characterization of the subnebula evolution at the time of satellite 
formation is clearly beyond reach at the present time. One may
tackle this hurdle either by assuming that turbulence decays
so that Keplerian disks must pass through quiescent phases (which enables
planetesimal and satellitesimal formation),
or by assuming that satellites
begin to form once the subnebula has been mostly depleted of gas (which
facilitates their survival) presumably due to sustained
turbulence of unknown origin (possibly hydrodynamic
shear turbulence). In the former case, satellite survival hinges on 
gap-opening (by satellites with mass ratio to the
primary $\mu \sim 10^{-4}$), and this mechanism itself
sets the value for the gas surface density of the satellite disk (MEb).
In the latter case, which is the focus of the present paper,
the gas surface density is low but unspecified, though
the presence of some gas may help to explain the observations.
Here satellite formation is dealt with
in a manner somewhat analogous to the consensus
treatment for the terrestrial planets. 

The second major difference between the two models concerns the mechanism
controlling the timescale for satellite formation. MEa,b includes
satellitesimal migration due to gas drag and tidal torques, and forms
satellites by a combination of Safronov-type binary accretion and
drift augmented accretion in an extended, two-component, largely
quiescent subnebula. This leads to a model in which
satellitesimals ($\sim 100$ km) form quickly ($\sim 1000$ years) but
full grown satellites ($\sim 1000$ km) take significantly longer to
form ($10^4-10^6$ years depending on their location; Mosqueira
\etal 2001), in a
time set by the gas drag time of satellitesimals in a solids-enhanced
minimum mass subnebula\footnote{Gas flow through the giant planet's
gap (Lubow \etal 1999)
does not necessarily imply that the time for disk formation
exceeds that of satellite formation. For instance, one possible
scenario is for Jupiter and Saturn to clear the
disk in a timescale $< 10^5$ years in a weakly
turbulent nebula (Bryden \etal 1999, 2000).}
(enhanced by a factor $\sim 10$, which leads
to a gas surface density $\sim 10^4$ g cm$^{-2}$). Furthermore, this
formation timescale is consistent with the Type I migration timescale
of full grown satellites in such a disk provided 3-D effects are considered
(which leads to a factor $\sim 10$ slower migration; \eg, Bate \etal 2003),
and with the criterion for gap-opening in an inviscid disk (with
aspect ratio $\sim 0.1$; Rafikov 2002). 

In contrast, in
this paper the formation timescale for the regular satellites is
set by the timescale over which planetesimal fragments are
collisionally captured by the circumplanetary satellitesimal
swarm, which is ultimately
tied to the timescale for a giant planet to clear its feeding zone (Mosqueira
\etal 2000).
Given that the gas-surface density in this model is unspecified but is
presumably quite small ($< 100$ g cm$^{-2}$), the migration of satellites
due to either gas drag or tidal torques does not pose a threat to their
survival; yet, one may still use the gas disk to circularize the orbits of
the regular satellites, or to sweep away collisional
debris from regions of the disk that are
observed to be either empty or significantly mass-depleted (such is
the case with {\it most} of the Saturnian satellite system).

The third major difference we draw attention to in this introduction
(see also the comparison table in Appendix C as well
as MEa,b) concerns the interpretation of the compositional gradient
of the Galilean satellites. Two main explanations have been
suggested for this observation. A number of publications ascribe the
high-silicate fraction of Europa (Io might have lost
its volatiles due to the Laplace resonance alone) compared
to Ganymede and Callisto to the
subnebula temperature gradient due to the giant planet's luminosity at
the time of satellite formation following envelope collapse
(\eg, Pollack \etal 1976; Lunine and Stevenson
1982). On the other hand, Shoemaker (1984) argued that this compositional
gradient might be due to the increase of impact velocities 
and impactor flux of Roche-lobe
interlopers deep in the
planetary potential well, leading to preferential volatile depletion
in the case of Europa and Io. That is, in this view all the Galilean
satellites started out ice-rich, but some lost more volatiles than
others.

Several more recent papers attempt to explain the volatile composition of the
Galilean satellites based on turbulent models of the Jovian subnebula.
Makalkin \etal (1999) identify the difficulties involved in explaining
the Galilean satellite composition gradient using turbulent subnebula models,
but do not attempt to produce a specific model of satellite formation.
Mousis \etal (2002) and Mousis and Gautier (2004)
employ an evolutionary turbulent subnebula and deliver solids to
form satellites by ``embedding'' solar nebula planetesimals that
incorporate clathrate hydrates. However, how heliocentric planetesimals become
embedded in the subnebula is never addressed, and the mechanism
for trapping volatiles is uncertain (Lunine \etal 2000).
On the basis of this clathration theory, Hersant \etal (2004) have reproduced
the abundances of C, N and S in Saturn available in 2004, and predicted that
only Xe is substantially enriched in Titan's atmosphere.
However, the CIRS instrument aboard Cassini
(Flasar \etal 2005a) has found a C/H ratio about twice higher than predicted
by Hersant \etal (2004), and no noble gases other than radiogenically
produced $^{40}$Ar (and trace amounts of $^{36}$Ar) were detected in Titan's
atmosphere by the GCMS aboard Huygens (Niemann \etal 2005). 
Hersant \etal (2005) explain these
discrepancies in terms of the uncertainty in the C/H ratio and argue that
Xe may remain trapped in Titan's interior.

Canup and Ward (2002) posit a gas-starved
steady-state subnebula such that a balance is attained between
gas inflow through the giant planet's gap and turbulent removal from
the circumplanetary disk.
In this model a combination of viscous heating and heating from
Jupiter's luminosity prevent condensation of ices close to the planet,
which explains the Galilean satellite compositional gradient. However,
these authors assume that the gas inflow forms a compact
disk around planet, which then viscously spreads. Yet, the gas inflow
through the giant planet's gap would be endowed with enough specific
angular momentum to form an extended circumplanetary gas disk (MEa;
note that this paper considers gas flow through the gap in
connection with Callisto), which
would result in satellites forming in very cold regions of the disk with
their full complement of ices.

Finally, Alibert \etal (2005) forms satellites in a disk essentially like
that of MEa, except that they retain disk evolution due to weak
turbulence instead of allowing for turbulence decay. Like MEa, they employ
an inner disk out to $\sim 30 R_J$ and an outer disk out to $\sim R_H/5$, and
similar inner disk but larger outer disk (note that this 
transition is not as pronounced as in MEa) gas surface densities. Moreover,
their assumed satellite formation timescale 
($10^4-10^6$ yrs depending on location;
Mosqueira \etal 2001; MEa)
is controlled by disk dynamics (as supposed to the gas inflow timescale
as in Canup and Ward 2002), which means that satellite observations
would receive very similar explanations to those outlined in MEa.  
However, unlike in MEa,b
these initial conditions are not justified. Moreover,
too much mass is located in the cold, outer regions of
the disk, so that one would expect satellites to form far from the planet
(where they are not found). On the other hand, if such satellites
migrated large distances it would then become difficult to explain how
rocky satellites close to the planet managed to survive. Also,
these authors base their temperature profiles on the
treatment of Papaloizou and Terquem (1999), which does not
include heating by the central object. Thus, Alibert \etal (2005)
attempt to explain the Galilean satellite 
compositional gradient using viscous heating alone.
As a result, their model yields implausibly
low temperatures both for the nebula and the subnebula as
gas dissipation takes place\footnote{This assumes shielding
by the inner disk. Whether this is feasible
remains to be shown. At any rate, this might again argue against
a compositional gradient based on subnebula temperature.}
(c.f. their
Fig. 6 has a temperature of $T \sim 30$ K at $10 R_J$ and $T \sim 10$ K at
$150 R_J$ for $t = 1$ Myr). 
Furthermore, their model relies on a much different 
Type I migration attenuation factor $f_I$ for the nebula than for 
the subnebula.

In the context of our decaying-turbulence
SEMM model, the
Galilean satellite compositional gradient is used to set a
subnebula temperature at the location of (gap-opening)
Ganymede of $T\sim 250$ K (or
aspect ratio of $\sim 0.1$ as stated above) controlled by the giant planet's
luminosity.\footnote{It is important to stress that because this is a decaying
turbulence model, the disk's temperature is determined by the
planet's luminosity and not by viscous heating.} This may imply a planetary
luminosity between $10^{-4}-10^{-5} L_{\odot}$ for a planetary radius
of $1.5 - 2.0 R_J$ (where $R_J$ is Jupiter's radius) following envelope 
collapse (work in progress).
Here we simply note that deep in the planetary potential
well one might expect a stochastic compositional component
due to high speed $\sim 10$ km s$^{-1}$ impacts with large $\sim 100$ km
Roche-lobe interlopers (but see Mosqueira and Estrada 2005 for the
case of Iapetus, which may be exempt from this argument due to its distance
from the planet, size and isolation).
Such hyperbolic collisions might conceivably remove volatiles from
the mantle of a differentiated satellite and place them on
neighboring satellites (possibly analogous to the impact
that may have stripped Mercury's mantle; Benz \etal 1988). 
This might explain the volatile depletion in Europa
relative to its outer neighbours
(given its density [Jacobson \etal 2005],
a similar argument may apply to Enceladus in the Saturnian system;
alternatively, a different fractionation mechanism may apply to both) 
and obviate the need for 
a temperature of formation constraint. In this
paper we adopt this point of view. In this connection, it is worth
noting that recent work suggests that close-in Amalthea may have incorporated
water-ice (Anderson \etal 2005), which is a definite possibility in the
framework of the present model.

Finally, it is useful to compare
the method of delivery of solids in our two models.
We expect that the regular satellites form at the tail-end of planetary
accretion so that, at least in the core accretion model for the formation
of the giant planet, almost all of the solids mass in the solar nebula starts
out in the form of $\sim 10$ km
planetesimals (\eg, Weidenschilling 1997). However,
following giant planet formation fragmentation may cascade a significant
fraction of the mass in the feeding zone of the giant planets back
to smaller sizes (Stern and Weissman 2001; Charnoz and Morbidelli 2003).
Planetesimals in the meter to kilometer mass range may deliver mass
to the regular satellites of the giant planets either by ablating
in the giant planet envelope prior to envelope collapse, or by ablating
through the circumplanetary gas disk (MEa; Mosqueira and Estrada 2005).
In the context of this paper, planetesimals in this size range are both
easier to capture into the circumplanetary disk, and may also be used to
replenish the feeding zone of the giant planet by gas drag.

In Section \ref{subsec:outline} we summarize the main points of the
GPPC model. The creation of the circumplanetary satellitesimal disk is 
discussed in more detail starting in
Section \ref{sec:planmod}. Readers who are more interested in the 
results may wish to skip to Section \ref{subsec:evolswarm}. In Section
\ref{subsec:angmom}, we present results of the angular momentum calculation.
Finally, in Section \ref{sec:concl} we present discussion of the results
and our conclusions. Several appendices are provided at the end of the
paper for reference.

\subsection{Model Outline}
\label{subsec:outline}

In this section, we summarize the GPPC model.
In the spirit of Safronov \etal (1986), we will label quantities that are
related to the circumsolar disk with the subscript ``1'', and
those that are related to the circumplanetary disk with the subscript ``2''.
The reader is referred to Appendix A for a list of symbols. 

1. The creation of a protosatellite ``swarm''\footnote{Although this 
terminology 
was used by Safronov to describe the circumplanetary disk, we shall adopt
the less ambiguous choice of calling swarm objects satellitesimals.} will
produce a disk of 
prograde and retrograde satellitesimals
extending as far as circumplanetary orbits
are stable $\sim R_H/2$ (Safronov {\it et al}. 1986).
Initially, the capture rate is dominated by inelastic collisions within the 
planet's Hill radius of similarly-sized heliocentric planetesimals (see 
Section \ref{subsec:formswarm}). As the satellitesimal disk becomes more 
massive its 
volume density $\rho_2$ overtakes the heliocentric planetesimal density 
$\rho_1$. Once this happens, the
dominant mass capture mechanism becomes collisions between planetesimals
and larger satellitesimals (see Section \ref{subsec:evolswarm}).

2. The net specific angular momentum $\ell_z$ of this satellitesimal
swarm is likely to
be small; collisions between prograde and retrograde satellitesimals
will result in a close-in, prograde satellite disk.
Close to the giant planet, hypervelocity collisions with incoming
planetesimals may shatter embryo-sized or smaller
objects ($\sim 10^2-10^3$ km), and alter satellite compositions,
which may ultimately result in a diversity of
outcomes (\ie, Jovian-like vs. Saturnian-like satellite systems).

3. Collisions between outer disk satellitesimals can lead
to fragmentation, accretion or removal from the outer disk to the
inner disk, where this material will be accreted by satellite embryos.
Removal of material from the outer disk is balanced by planetesimal fragment
collisional capture (by larger circumplanetary satellitesimals). Here we rely
on fragmentation to decrease the size of heliocentric planetesimals, which
are easier to capture.

4. In the outer disk we expect that numerous $\sim 1$ km 
objects\footnote{However,
later we obtain solutions such that the satellitesimal
population is made up of rubble. In general, satellitesimals in the
meter to kilometer size range are used to 
capture planetesimal collisional fragments in the same size range.}
(of which irregular satellites may be surviving members) contribute a
low optical depth $\tau_{z2} \sim 10^{-5}$, which
yields a surface density $\sigma_2 \sim 1$ g cm$^{-2}$ and volume density
$\rho_2 \sim 1.4\times 10^{-14} (\sigma_1/\sigma_{\rm{MM}})$ 
g cm$^{-3}$, where $\sigma_1 = \sigma_{\rm{MM}} = 3.3$ \gcms corresponds to the 
solids surface density of a minimum mass solar nebula (MMSN; Hayashi 1981).
This means that even for the case of a solids MMSN, 
$\sigma_1 > \sigma_2$ but typically $\rho_1 <
\rho_2$, as noted in 1. To avoid fast satellite formation
the mass in the extended disk of solids
at any one time (excluding satellite embryos) must be a small fraction of
a Galilean satellite ($\sim 10^{26}$ g).
This can lead to a long $10^5-10^6$ years
satellite formation time (presumably consistent with
Callisto's partially differentiated state) controlled by planetesimal feeding.

5. As the giant planet clears its feeding zone
in a $\sim 10^5$ year timescale small planetesimals\footnote{This population
of objects deserve special treatment since they will have smaller inclinations
than eccentricities. Ohtsuki and Ida (1998) discuss the effects of
small planetesimals on planetary spin rates.} 
are fed from the outer regions of
the feeding zone and deliver enough angular momentum to the disk
to form the regular satellites at their observed locations.

\vspace{0.1in}
{\it Satellitesimal disk collapse time:}
Given typical satellitesimals of size $\sim 1$ km, and a
total disk mass a fraction $\sim 0.1$ of a Galilean satellite (which 
corresponds to $\sigma_2 \sim 1$ g cm$^{-2}$), 
the collapse time of the disk is of order
$\tau_{coll} \sim P_2/\tau_{z2} \sim 10^5$ years, where
$P_2$ is a characteristic orbital period of satellitesimals, and
$\tau_{z2} \sim 10^{-5}$ is the optical depth
contributed by the larger particles (see Section \ref{subsec:evolswarm}).
The outer disk provides
materials to the inner disk out of which the regular satellites
form. 
Disk material must then be replenished to the
outer disk by planetesimal capture,
and one might need $N_c = \tau_{acc}/\tau_{coll} \sim 10$ such collisional
cycles to complete the accretion of the Galilean satellites in
$\tau_{acc} \sim 10^5-10^6$ years, which may be long enough to explain 
Callisto's partially differentiated state. 

\vspace{0.1in}
{\it Planetesimal capture:}
In order to show that a GPPC model is viable one needs to address
whether such a disk would be able to capture a sufficient amount of mass
to form the Galilean satellites.
Safronov \etal (1986) considered the following delivery mechanisms:

\begin{itemize}
\item I. Collisions between planetesimals within $R_H$, which may have led
to the formation of a circumplanetary disk.

\item II. Collisions between large satellitesimals and smaller 
planetesimals, which may result in capture of the planetesimal and a mass 
inflow rate, $I_{lb}$, and collisions between planetesimals and many smaller
satellitesimals, which may result in capture of the planetesimal (if it
encounters a mass comparable to itself) and a mass 
inflow rate $I_{sp}$.

\item III. Erosion or delivery of ejecta from a collision of a planetesimal 
with a smaller satellitesimals, which may provide a mass inflow rate, $I_e$.
\end{itemize}

\noindent
The first method of mass delivery
(I) is addressed in the formation of the circumplanetary
disk in Section \ref{subsec:formswarm}. As indicated, this mechanism 
is
dominant until the solids volume density in the satellitesimal disk $\rho_2$ 
exceeds that of the solar nebula, $\rho_1$. In our model mechanism (II)
determines whether enough mass can be delivered to the system in the
planetesimal feeding time. Although mechanism (III) may provide a source
of material, the amount of mass captured by collisional erosion is
uncertain\footnote{Note that capture of material due to erosion differs
from larger planetesimals captured into the circumplanetary disk by
smaller satellitesimals, which we treat in Section \ref{subsec:evolswarm}.}, so
we do not use $I_e$ or address it any further.
We can estimate the mass delivered to the disk on a timescale of 
$\tau_{acc}\sim 10^6$ years using (see Section \ref{subsec:evolswarm})

\begin{equation}
M = 2\pi\tau_{acc} \int_{R_0}^{R_D} I(R)\,p(R)\,R\, dR,
\end{equation}

\noindent
where $I(R)$ is the total mass inflow collision rate
per unit area, $p(R)$ is a capture 
probability, and $R_0$ and $R_D$ define the inner and outer edges of the 
circumplanetary disk, respectively.
Unless otherwise noted, we generally take $R_0$ to be the radius of the 
planet $R_P$, and $R_D = R_H/2$. The inflow rate $I(R)$ can be
expressed in terms of the disk optical depth $\tau_{z2}$, solar nebula
solids density $\rho_1$, impact velocity of incoming planetesimals $v_1$, and 
the planetesimal and satellitesimal cutoff radii $r_{1,max},r_{2,max}$ 
(Safronov {\it et al.} 1986), where (see Section \ref{subsec:evolswarm})

\begin{equation}
I(R) = I_{sp}+I_{lb} = \rho_1 v_1 \tau_{z2} \left(\frac{r_{2,max}}
{r_{1,max}}\right)^{1/2}\left[\tau_{z2} - \ln{\tau_{z2}}\right],
\end{equation}

\noindent
and a mass distribution with power law exponent of
$q_1 = q_2 = 11/6$ has been assumed for both the planetesimal and 
satellitesimal populations. 
Here, $\tau_{z2} = 3\sigma_2/4\rho_s r_{2,max} = \pi r^2_{2,max} \sigma_2/
m_{2,max}$ is the vertical optical depth\footnote{This definition
of the optical depth is only true if {\it all} the particles in the disk
are of size $r_{2,max}$. 
For this reason, $\tau_{z2}$ should be considered more 
a probability of collision with a satellitesimal of size
$r_{2,max}$ than an optical depth.} of the satellitesimal disk 
($\rho_s$ is the
satellitesimal density), which is assumed
to be set by the size of the largest satellitesimal $r_{2,max}$ (Safronov
\etal 1986). Ruskol (1975) gives a functional form for the probability
distribution $p(R)$ which measures the probability that a collision will result
in capture (as supposed to fragmenting and scattering into escape orbits; 
see Section \ref{subsec:formswarm})

\begin{equation}
p(R) = \frac{\theta_1}{l + 2\theta_1}(1 - R/R_H),
\end{equation}

\noindent
where $\theta_1$ is the planetesimals
Safronov parameter, and $l = R/R_P$. We can then
evaluate the mass delivered by planetesimals primarily striking larger or 
comparable-sized satellitesimals and obtain

\begin{equation}
M \sim 10^{26} \left(\frac{\sigma_1}{3.3\, \rm{g}\,\rm{cm}^{-2}}
\frac{\sigma_2}{1\,\rm{g}\,\rm{cm}^{-2}}\right)\left(\frac{10\,\rm{km}}
{r_{1,max}}\frac{1\,\rm{km}}{r_{2,max}}\right)^{1/2} \rm{g}.
\end{equation}

\noindent
This is comparable to a
Galilean satellite mass for reasonable choices of parameters (but see Section
\ref{subsec:evolswarm} for further discussion).

However, an outstanding
issue is whether planetesimals may be fed into the system over a
timescale $\gtrsim 10^5$ years. This is because the giant planet may clear
planetesimals from its feeding zone on a shorter timescale. The
formation of this gap was not taken into account in the above estimate
of the planetesimal capture rate. This can be seen in the following simple
way. A reasonable estimate for the total amount of mass in planetesimals in
the feeding zone of the giant planets is $\sim 10 M_{\oplus}
\sim 10^{29}$ g. Suppose we assume
that every planetesimal and satellitesimal has a size of $1$ km, and that
inelastic collisions between them always lead to planetesimal 
capture into circumplanetary
orbit. Then we can ask how many times $n_p$ will a planetesimal cross the
circumplanetary disk within $R_H/2$ before being scattered by the
giant planet. Ignoring gravitational focusing and taking the feeding zone
of the giant planet to be of width $4\sqrt{3}R_H$, we obtain 

\begin{equation}
n_p = \frac{0.1}
{576 \sqrt{3} \pi} \left( \frac{a}{R_H} \right)^5 \sim 10 - 100.
\end{equation}

\noindent
Given an optical depth of $\tau_{z2} \sim 10^{-5}$ the total amount of material
captured is $\sim 10^{26}$ g. But this material would be fed
in a timescale 

\begin{equation}
\tau_{ej} \sim \frac{0.1}{\Omega_1}
\left( \frac{M_{\odot}}{M_J} \right)^2 \sim 10^5 \,\,\rm{years}, 
\end{equation}

\noindent
which may be too short given our chosen system constraints. 
Also, since it is unrealistic to assume perfect
capture (\ie, $p(R) < 1$) and accretion efficiencies, 
one may have to replenish
the feeding zone of the giant planet with planetesimal fragments brought
in by gas drag and enhance the solid content of the nebula with
respect to the minimum mass solar nebula. Although we discuss this issue 
in Section \ref{sec:concl}, further work
will need to be done to assess whether this is feasible.

\vspace{0.1in}
{\it Angular momentum of the satellitesimal disk:}
A key issue is
whether the delivery of planetesimals provides sufficient angular momentum
to the disk to account for the angular momentum budget of the regular
satellites of the giant planets.
Assuming that a significant fraction
of the mass is in small planetesimals $r_1 < 1$ km with
large Safronov parameter $\theta_1 >> 1$ (so that $p(R) \lesssim 1$), 
we can calculate the amount of
angular momentum delivered to the satellitesimal disk as
the giant planet clears its feeding zone.
A rough estimate of the angular momentum
budget
may be obtained assuming a Rayleigh distribution of planetesimal velocities
and neglecting the gravitational effects of the planet. The specific
angular momentum delivered by planetesimals
is given by (see Section \ref{subsec:angmom})

\begin{equation}
\left<\ell_z\right> = \frac{2\pi}{\dot{M}} \int_{R_0}^{R_D} \epsilon 
{\cal{I}}_L(R)R\,dR
\end{equation}

\noindent
and the mass inflow rate is

\begin{equation}
\dot{M} = 2\pi\int_{R_0}^{R_D} {\cal{I}}(R)p(R)R\,dR
\end{equation}

\noindent
where $\cal{I}_L(R)$ is
the angular momentum delivery rate per unit area given perfect angular
momentum deposition efficiency $\epsilon = 1$, and ${\cal{I}}(R)$ is the 
mass collision inflow
rate per unit area.
This gives a net specific angular momentum that is too small to create a
disk of sufficient size to make the satellites
unless most of the planetesimals originate
from the outer regions of the planet's feeding zone.
If we use a distribution of semi-major axes
depleted inside an annulus of half-width $\sim 1 R_H$ centered on the giant
planet, and a velocity dispersion of planetesimals
$v_1 \sim 6$ km s$^{-1}$ ($\theta_1 \sim 100$), we find
(see Section \ref{subsec:angmom})

\begin{equation}
\left<\ell_z\right> \sim 0.3\Omega R_H^2 \rightarrow R_{coll} =
\frac{\left<\ell_z\right>^2}{G M_J} \sim 20 R_J,
\end{equation}

\noindent
where $R_{coll}$ is the centrifugal collapse radius, and $M_J$ is the mass
of Jupiter. This suggests that it may be possible for the extended disk 
of prograde and retrograde satellitesimals to collapse to
the observed Galilean satellite system size.

\section{Planetesimal Model}
\label{sec:planmod}

Turbulence may act both to remove gas from the subnebula and to replenish it
by feeding gas through the giant planet's gap. However, we do not rely
on this gas inflow as a source of material to make satellites 
for several reasons. First, the inflow rate is nearly unconstrained
beyond the fact that it must be very weak if ice-rich satellites are to
survive (orders of magnitude slower than in Lubow \etal 1999; see, \eg,
Makalkin \etal 1999,
Canup and Ward 2002). Second, the high-specific
angular momentum of this component would lead to the formation of an
extended disk around the planet\footnote{It is important to
note that slower inflow
rates correspond to larger gaps and higher specific angular momentum for
the inflow. Moreover, turbulence would cause the disk to spread out.},
whereas regular satellites are observed to be
close-in. Third, the solid content of this gas component is unknown, and a 
significant fraction of this material might
accrete onto the planet along with the gas. Fourth, most of
the mass of solids in the planetary disk is likely to be in objects larger than
the gas coupling size\footnote{Gas drag may protect sufficiently small
particles, which may grow instead of fragmenting. It could
be argued that the collisional cascade will proceed to even smaller particles
in the presence of turbulence, but one would then need to explain how
such objects formed in the first place.},
which means that
planetesimal dynamics should be treated.

\subsection{Formation of the Circumplanetary Satellitesimal Swarm}
\label{subsec:formswarm}

Planetesimal circumplanetary collisional (or collisionless) capture
can be investigated as a three-body problem with
planetesimals treated as massless particles. 
Inelastic collisions of heliocentric planetesimals within 
the Hill sphere of the 
giant planet can lead to capture of solids because inelastic
collisions decrease relative velocities. In the 3-body
problem one can define the origin to lie at the center of mass of the system, 
and write down the Jacobi constant as

\begin{equation}
\label{equ:jacob}
C_J =  \Omega_1^2(x^2 + y^2) + \frac{2GM_\odot}{a} + \frac{2GM_P}{R} - v^2,
\end{equation}

\noindent
Inelastic collisions can increase this quantity (by decreasing $v^2$)
to values indicative of a circumplanetary orbit, which
allows for the possibility of planetesimal capture by the planet's 
gravitational field. Here the $x$-axis is directed along the line connecting
the finite masses, $a$ is the planetesimals distance from the Sun, $R$ its
distance from the giant planet, and $\Omega_1$ is the uniform angular velocity
of the system.
Planetesimal collisions
are inelastic because there is loss of kinetic energy through
a number of mechanisms such as fragmentation, heating, melting,
evaporation, and the
like. Thus, collisions can occur that lead to a
change in velocity $0 < |\Delta v| < v$
such that the variation in the Jacobi constant (constant absent a
collision) is
$\Delta C_J \sim -2v\Delta v \gtrsim 0$. 

We can obtain an analytical
estimate of this change in energy by assuming that similar-sized particles
collide while travelling at similar speeds 
$v\sim v_{10} = \sqrt{GM_P/\theta_1 R_P}$. Scaling the system to units such 
that the
gravitational constant $G = 1$, distances are in AU, masses are in terms
of the mass ratio of secondary to primary $\mu \equiv M_P/M_\odot$, and
time in terms of the orbital period (1 AU) over $2 \pi$, then
in the case of Jupiter one finds 

\begin{equation}
\label{equ:delCjup}
\Delta C_J \approx v^2 \approx 2\times 10^3\frac{\mu}{\theta_1} \sim 0.7-0.4 
\,\,\,\,\rm{for}\,\, \theta_1 = 3-5.
\end{equation}

\noindent
We can compare this to the difference in the Jacobi constant of circular orbits 
at $30 R_J$ (a representative size of the Jovian satellite system) with
respect to
the L1 Lagrange point $\Delta C_J \sim 0.04$. Thus, inelastic collisions 
likely allow for the capture of planetesimals for Jacobi constants close 
to the value at L1 into orbits like the Galilean satellites with capture
into orbits at further distances even more energetically feasible.

However, mass differences between the colliding 
particles, as well as their initial energies leads
us to introduce a capture probability.
Ruskol (1975) derives a heuristic
capture probability that applies to collisions between similar-sized particles,
which she defines as those particles
whose masses do not differ by more than a factor
of two. Incoming planetesimals are assumed
to have speed $v_{10}$ at infinity and as they
approach the giant planet they reach a speed $v_1 = \sqrt{v_{10}^2 + v_e^2}$,
where $v_e = \sqrt{2GM_P/R}$. The velocity vectors at infinity and at
distance $R$ are randomly
oriented. For a collision taking place at $R$, 
loss of energy from the collision allows for capture
of the colliding bodies or their collisional
fragments into circumplanetary orbits if they retain a
kinetic energy fraction $f_C < (v_e/v_1)^2$, or more simply
if their speed after the collision is less than
the escape speed
at distance $R$. The condition for capture is (Ruskol 1975)

\begin{equation}
\label{equ:capcon}
f_C < \frac{2\theta_1 R_P}{R + 2\theta_1 R_P}.
\end{equation}

\noindent
This indicates that the allowable fraction of post-collisional
kinetic energy 
decreases with distance from the planet.
Assuming that the probability of capture is proportional to this term,
we can write

\begin{equation}
\label{equ:probcap}
p(R) = \beta f_C (1 - R/R_H),
\end{equation}

\noindent
where the parameter $\beta < 1$ accounts for the
average fractional amount of material captured in the collision. For our
model, we assume $\beta = 1/2$; however, this parameter remains uncertain
and depends on physical processes which we do not address here. According
to this heuristic expression,
the probability of capture is highest close to the planet and 
vanishes at the planet's Hill radius\footnote{We note that it may be 
energetically possible for collisions between comparable-sized 
planetesimals just {\it outside} the Hill sphere to be captured.
However, in this paper, we restrict ourselves to
the condition that only collisions within the Hill sphere can lead to capture
as explicitely assumed in the heuristic expression in Eq. 
(\ref{equ:probcap}).}.

We can now estimate the amount of mass that can be 
delivered to the circumplanetary
disk via ``free-free'' collisions (\ie, collisions between
heliocentric planetesimals occuring within the Hill radius of
the planet, as supposed to ``free-bound'' collisions between
heliocentric planetesimals and circumplanetary satellitesimals).
As pointed out by Ruskol (1975), inelastic collisions between
heliocentric planetesimals of disparate masses will not contribute
significantly to the delivery of mass to the
circumplanetary disk. Hence, one need only consider
collisions between planetesimals of comparable mass in the range
$m/\chi < m < \chi m$, where
$\chi \sim 2$. For the planetesimal mass spectrum,
we use a power law of the form $n(m) = C m^{-q}$,
where the constant $C$ can be determined by the condition for the volume 
density of planetesimals:

\begin{equation}
\label{equ:dens}
\rho = \int n(m) m\, dm.
\end{equation}

\noindent
For the case of the circumsolar disk we have

\begin{equation}
\label{equ:C1}
C_1 = \rho_1 \times
\begin{cases}
\frac{2-q_1}{m_{1,max}^{2-q_1}-m_{1,min}^{2-q_1}}	&\,{\rm{for}}\,q_1\neq 2;\\
\left[ \ln{(m_{1,max}/m_{1,min})}\right]^{-1}	&\,{\rm{for}}\,q_1=2.\\
\end{cases},
\end{equation}

\noindent
where $m_{1,min}$ and $m_{1,max}$ are the lower and upper cutoffs of the
planetesimal mass distribution. An analogous expression can be derived for the
satellitesimal distribution. One might 
expect that at the time of 
giant planet formation, most of the mass of solids will be in planetesimals
in the $1-10$ km size range (Weidenschilling 1997). This 
may change significantly as the solids in the solar nebula evolve due
to collisional grinding following giant planet formation, but we
still use this size range as a reference (see Section \ref{subsec:evolswarm} 
for more discussion). At any rate, it is fair to say that
smaller planetesimals make our model
more probable.

As the lower limit, we take $m_{1,min}$ to be the gas decoupling size
at the location of the giant planet. This is the size for which
$\Omega_1 t_s \sim 1$, where $t_s$ is the stopping time. 
The drag force on a particle depends on its size $r$
and the drag regime.
The mean free path of a gas molecule is
$\lambda_G \approx (a/1\,\rm{AU})^{11/4}$ cm for the standard minimum mass 
solar 
nebula model (\eg, Hayashi 1981), which gives $\lambda_G \sim 1$ m at the 
location
of Jupiter.
The stopping time is $t_s = \rho_s r/\rho c$ (Weidenschilling 1977), 
where $\rho$
and $c$ are the local gas density and sound speed.
At the location of Jupiter, the decoupling size is $\sim 1$ m in the Epstein
flow regime.

The probability of a planetesimal collision per unit length is given by

\begin{equation}
\label{equ:expm1}
{\cal{Y}}(m) = \int_{m/\chi}^{\chi m} \xi \pi r^{\prime 2} n_1(m^{\prime})\,
dm^{\prime}
= \left(\frac{3}{4\pi\rho_s}\right)^{2/3}\frac{\xi\pi\rho_1 (2-q_1)}
{q_1 - 5/3} \gamma_1 \frac{m^{5/3-q_1}}{m_{1,max}^{2-q_1}-m_{1,min}^{2-q_1}},
\end{equation}

\noindent
which is valid for $q_1 > 5/3$. In this
expression, the cross-section factor $\xi \approx 2$ for comparable-sized
bodies, and $\gamma_1 = \chi^{q_1-5/3}-\chi^{5/3-q_1}$.
Given Eq. (\ref{equ:C1}), the rate of planetesimal mass undergoing collisions 
per unit volume can be expressed as 
(Ruskol 1975)

\begin{equation}
\label{equ:J1}
J_1 = \int_{m_{1,min}}^{m_{1,max}}{\cal{Y}}(m) v_1 n_1(m)\,m\,dm
= \frac{3}{4}\frac{\rho_1^2v_1}{\rho_s r_{1,eff}}\gamma_1.
\end{equation}

\noindent
The effective radius $r_{1,eff}$ is given by

\begin{equation}
\label{equ:reff}
r_{1,eff} = 
\begin{cases}
\frac{ (q_1-5/3)(11-6q_1) }{ 3\xi(2-q_1)^2 }
\frac{ (r_{1,max}^{6-3q_1}-r_{1,min}^{6-3q_1})^2 }{ r_{1,max}^{11-6q_1}-
r_{1,min}^{11-6q_1} }		&\,q_1 <11/6;\\
\frac{2}{\xi} \frac{ (r_{1,max}^{1/2}-r_{1,min}^{1/2})^2 }{ \ln{(r_{1,max}/r_{1,min})} }	&\,q_1 = 11/6;\\
\frac{ (q_1-5/3)(6q_1-11) }{ 3\xi(2-q_1)^2 } 
\frac{ (r_{1,max}r_{1,min})^{6q-11}(r_{1,max}^{6-3q_1}-r_{1,min}^{6-3q_1})^2 }
{r_{1,max}^{6q_1-11}-r_{1,min}^{6q_1-11}}		&\,q_1 > 11/6.\\
\end{cases},
\end{equation}

\noindent
Depending on the value of the power law exponent $q_1$, $r_{1,eff}$ can be 
determined by
the maximum particle size ($q_1 \leq 11/6$), or the minimum particle size
($q_1 > 11/6$). In the case 
$q_1 = 11/6$, most of the mass (a 
fraction $f_m = 1 - (r/r_{1,max})^{3(2-q_1)} 
\sim 70$\%) lies in the top decade of
the mass distribution (\eg, $\sim 1-10$ km for $r_{1,max} = 10$ km). 

The flux of material 
${\dot{\cal{F}}_1}$ captured into the circumplanetary disk
per unit volume per unit time due to free-free collisions 
is then the product of Eq. (\ref{equ:probcap}) 
and Eq. (\ref{equ:J1}). For $q_1=11/6$, the mass delivery rate per unit
volume can then be written as

\begin{equation}
\label{equ:Sig1_116}
\dot{{\cal{F}}}_1 = p_1(R)J_1 = p_1 \rho_1 v_1 \gamma_1 
\frac{\tau_{z1}}{H_1} \ln{\left(\frac{r_{1,max}}{r_{1,min}}\right)},
\end{equation}

\noindent
which has been written in terms of the local vertical optical depth of 
solids in the nebula
$\tau_{z1} = 3\sigma_1/4\rho_s r_1$ (see footnote 9), where 
$\sigma_1 = \rho_1 H_1$ is the
surface density of solids in the planetary zone, 
$H_1 = \sqrt{2}v_1/\Omega \approx P_1v_1/4$ is the local 
scale-height of solids in the nebula, and $P_1$ is the planet's
orbital period. A rough estimate of the amount of mass 
delivered per year to the circumplanetary disk is

\begin{equation}
\label{equ:flux1}
\begin{split}
\dot{M}_1 = \int {\dot{\cal{F}}}_1 \,dV \approx 
24\pi\gamma_1\frac{(\sigma_1\theta_1R_J)^2}{P_1\rho_s r_{1,max}}
\ln{\left(\frac{r_{1,max}}{r_{1,min}}\right)}\int_{1/2\theta_1}^{x_H}
\frac{x}{1+x}\left(1-x/x_H\right)\,dx \\
\sim 2\times 10^{18} \left(\frac{\sigma_1}{3.3\,\rm{g}\,\rm{cm}^{-2}}\right)^2
\left(\frac{10\,\rm{km}}{r_{1,max}}\right)\,\rm{g}\,\,\rm{yr}^{-1},
\end{split}
\end{equation}

\noindent
where $x_H = R_H/2\theta_1 R_P$, $\rho_s = 1$ g cm$^{-3}$, and we 
assume $\theta_1 = 4$. The timescale
for the giant planet to clear its feeding zone is $\Delta t\sim 10^5$ 
years (\eg, 
Tanaka and Ida 1997; Charnoz and Morbidelli 2003),
so this
is presumably the timescale in which the circumplanetary disk should be
populated. Assuming our fiducial choices for the disk parameters, one
finds that the total mass delivered then is $\Delta t \times \dot{M}_1 
\sim 10^{23}$ g, which is considerably smaller than a Galilean satellite. 
This would seem to pose a problem; however, we show below
that once the circumplanetary disk becomes sufficiently 
massive, capture of planetesimals by inelastic collisions with
circumplanetary disk particles (free-bound collisions) becomes
the dominant mass delivery
mechanism. 

\subsection{Planetesimal Capture Due to the Circumplanetary Disk}
\label{sec:capswarm}

The capture of material due to free-free collisions may
form a circumplanetary disk around the giant planet. As 
the density
of the circumplanetary disk $\rho_2$ increases, it may eventually become 
larger than $\rho_1$. 
Once this happens, the primary delivery mechanism may become
free-bound collisions (Ruskol 1975, Safronov \etal 1986).

Along the lines of the prior section, except that satellitesimals are confined
to a circumplanetary disk with scale-height $H_2$,
we first estimate the mass rate
per unit {\it area} undergoing
collisions between planetesimals and
satellitesimals of {\it comparable} size. In analogy to Eq. (\ref{equ:expm1}),
the probability of a collision of a planetesimal crossing the circumplanetary
disk in the normal direction is given by

\begin{equation}
\label{equ:expm2}
{\cal{E}}(m) = \int_{m/\chi}^{\chi m} \xi \pi r^{\prime 2} H_2
n_2(m^{\prime})\,dm^{\prime}
= \left(\frac{3}{4\pi\rho_s}\right)^{2/3}\frac{\xi\pi\sigma_2 (2-q_2)}
{q_2 - 5/3} \gamma_2 \frac{m^{5/3-q_2}}{m_{2,max}^{2-q_2}-m_{2,min}^{2-q_2}},
\end{equation}

\noindent
which is valid for $q_2 > 5/3$. Here,
$n_2 = C_2m^{-q_2}$ is the mass distribution of satellitesimals with the
coefficient $C_2$ defined as in Eq. (\ref{equ:C1}), 
$\gamma_2 = \chi^{q_2-5/3}-\chi^{5/3-q_2}$, and $H_2$ and $\sigma_2$ are the 
scale-height and surface density of solids in the protosatellite disk, 
respectively. The total number of planetesimals impacting the 
circumplanetary disk per unit
area over the mass interval $m$ to $m+dm$ is $v_1 n_1(m) dm$. 
Integration
over $m$ gives the mass rate per unit area of collisions between
planetesimals and satellitesimals of comparable size

\begin{equation}
\label{equ:I2}
I_2 = \int_{m_{2,min}}^{m_{2,max}} {\cal{E}}(m)v_1 n_1(m)\,m\,dm 
= \frac{3}{4}\frac{\rho_1\sigma_2v_1}{\rho_s r_{2,eff}}\gamma_2 \approx
H_2J_2,
\end{equation}

\noindent
The effective
radius $r_{2,eff}$ can be expressed in terms of the appropriate limits

\begin{equation}
\label{equ:reff2}
r_{2,eff} = {\cal{R}}
\begin{cases}
\frac{(11/3-q_1-q_2)(q_2-5/3)}{\xi(2-q_1)(2-q_2)}
\frac{1}{r^{(11-3q_1-3q_2)}_{2,max}-r^{(11-3q_1-3q_2)}_{2,min}}	
&\,q_1+q_2 < 11/3; \\
\frac{q_2-5/3}{3\xi(2-q_1)(2-q_2)}
\frac{1}{\ln{(r_{2,max}/r_{2,min})}}	&\,q_1+q_2 = 11/3; \\
\frac{(q_1+q_2-11/3)(q_2-5/3)}{\xi(2-q_1)(2-q_2)}
\frac{(r_{2,max}r_{2,min})^{11-3q_1-3q_2}}{r^{(11-3q_1-3q_2)}_{2,max}-
r^{(11-3q_1-3q_2)}_{2,min}}	&\,q_1+q_2 > 11/3; \\
\end{cases},
\end{equation}

\begin{equation}
\label{equ:reff2_2}
{\cal{R}} \equiv (r_{1,max}^{6-3q_1}-r_{1,min}^{6-3q_1})
(r_{2,max}^{6-3q_2}-r_{2,min}^{6-3q_2}).
\end{equation}

\noindent
The choice of $r_{2,min}$ is problematic given that the gas surface density
in the circumplanetary disk is left unspecified. 
We shall, unless otherwise noted, arbitrarily choose this size to be
$\sim 1$ cm, though our results are not dependent on this parameter given
our choices for $q_2 = 11/6$ and $r_{2,max} \sim 1$ km.

We now consider the conditions under which capture by the 
circumplanetary disk begins to dominate the capture rate.
We can express the capture rate of material
due to collisions of comparable-sized bodies as

\begin{equation}
\label{equ:totrate}
\dot{{\cal{F}}_{cs}} = \dot{{\cal{F}}_1} + \dot{{\cal{F}}}_2
= p_1J_1 + \frac{p_1}{2}J_2 = \frac{3}{4}\frac{\rho_1v_1}{\rho_s}
\left(\frac{\rho_1}{r_{1,eff}}\gamma_1 + \frac{\rho_2}{2r_{2,eff}}\gamma_2
\right),
\end{equation}

\noindent
where we have assumed that the typical amount of mass 
involved in a free-bound collision 
is roughly half that of the free-free 
case (Ruskol 1975). This is because we are only
considering comparable-sized collisions for both capture rates. 
From Eq. (\ref{equ:totrate}), we see that
in the case of collisions between particles of
comparable size the condition that
the dominant capture mechanism involves the circumplanetary disk is

\begin{equation}
\label{equ:rho2rho1}
\rho_2 \gtrsim 2 \rho_1 \frac{\gamma_1}{\gamma_2}\frac{r_{2,eff}}{r_{1,eff}}
= 2\rho_1 \left(\frac{r_{2,max}}{r_{1,max}}\right)^{1/2}\frac{\ln{(r_{1,max}/
r_{1,min})}}{\ln{(r_{2,max}/r_{2,min})}} \sim \rho_1,
\end{equation}

\noindent
where the next to last term on the right side applies for $q_1 = q_2 = 11/6$.
Independent of the uncertainty in the lower bound of the size distribution
for satellitesimals, we see that the circumplanetary disk dominates the 
capture of
planetesimals when the volume density of solids in the disk becomes comparable
to that of the solar nebula (Safronov \etal 1986).

To determine the implied volume density of solids in the circumplanetary disk 
that satisfies this condition, we estimate
the mean surface density of solids in the protosatellite disk to be

\begin{equation}
\label{equ:sigma2}
\sigma_2 = \frac{4M_d}{\pi(R^2_H-4R^2_P)} \sim \frac{4\dot{M}_1\Delta t}
{\pi(R^2_H-4R^2_P)} \sim 0.01\,\rm{g}\,\rm{cm}^{-2},
\end{equation}

\noindent
where we have taken the mass delivered in $\Delta t = 10^5$ years due to
both mechanisms to be approximately $\dot{M}_1 \Delta t$, 
we have made use of Eq. (\ref{equ:flux1}) with $\theta_1 = 4$,
and we have ignored collisional removal. This means that
$\sigma_2 << \sigma_1$. 
Since the critical density such that capture by the circumplanetary
disk becomes dominant is
$\rho_{2,crit} = \sigma_2/H_{2,crit} = \rho_1$, for our 
disk parameters
the scale-height of solids in the circumplanetary disk 
must be less than

\begin{equation}
\label{equ:H1}
H_{2,crit} \lesssim 100 \left(\frac{\sigma_1}{3.3\,\rm{g}\,\rm{cm}^{-2}}\right)
\left(\frac{10\,{\rm{km}}}{r_{1,max}}\right)\,R_J,
\end{equation}

\noindent
and the critical density must be greater than 

\begin{equation}
\rho_{2,crit} \gtrsim 1.4\times 10^{-14}\left(\frac{\sigma_1}{3.3\,\rm{g}\,
\rm{cm}^{-2}}\right)\,\rm{g}\,\rm{cm}^{-3}.
\end{equation}

\noindent
We stress that this is the condition for disk capture to become dominant.
Once this happens the subnebula surface and volume densities may
increase significantly. As we show later,
in general $\sigma_2 \lesssim \sigma_1$ to
be consistent with chosen system constraints.

\subsection{Total Mass Delivered to the Circumplanetary Disk}
\label{subsec:evolswarm}

We now remove the requirement that colliding particles
be of comparable mass, calculate
the mass inflow rate due to the entire mass
spectrum, and then use this quantity to estimate the mass delivered to the
circumplanetary disk over the planetesimal feeding timescale. 

Here there are two cases to consider. For a given planetesimal of mass $m$, we
estimate the mass inflow collision rates per unit area (henceforth inflow 
rates) $I_{sp}$ and $I_{lb}$ due to interactions with
satellitesimals with $m^{\prime} < m$, and $m^{\prime} > m$,
respectively.
These rates include the contribution due to
similar-sized bodies\footnote{In prior sections, we estimated the mass
inflow rates for two
mechanisms for comparable sized bodies.
In the case
of collisions between planetesimals in the Hill radius of the giant planet,
collisions between comparable sized objects 
are considered because
these incur the largest velocity changes
resulting in an increased likelihood of capture. In the case of collisions
between planetesimals and comparable-sized satellitesimals,
the mass inflow rate for $q_1 = q_2 = 11/6$ using Eq.
(\ref{equ:I2})-(\ref{equ:reff2_2}) is

$$
\label{equ:Ics}
I_{cs} \simeq 0.2 \frac{\rho_1 v_1 \sigma_2}{\rho_s (r_{1,max}r_{2,max})^{1/2}}
\ln{\left(\frac{r_{2,max}}{r_{2,min}}\right)} \simeq 0.25\rho_1v_1\tau_{z2}
\left(\frac{r_{2,max}}{r_{1,max}}\right)^{1/2}
\ln{\left(\frac{r_{2,max}}{r_{2,min}}\right)},
$$

\noindent
which is in agreement with Eq. (6) in Safronov \etal (1986).}.
In the regime we are in, we will
see that the inflow rate $I_{lb} >> I_{sp}$ since the optical depth
$\tau_{z2} << 1$ given the constraints of the problem.

For an incoming planetesimal of mass $m$
to stand a chance $p(R)$ of being
captured by satellitesimals smaller than $m$, the
condition is that $\pi r^2 H_c \rho_c \geq m$ (Safronov \etal 1986), which
simply says that in order for the mass $m$ to have a
chance of being captured into the
circumplanetary disk 
by particles smaller than itself, it must encounter a mass comparable or larger
than itself when crossing the disk.
The characteristic size for which all particles with $r<r_c$ stand a
chance of being captured
into the satellitesimal disk is

\begin{equation}
\label{equ:rc}
r_c/r_{2,max} = \left[\tau_{z2}\left\{1-(r_{2,min}/r_c)^{6-3q_2}\right\}
\right]^{\frac{1}{3q_2-5}},
\end{equation}

\noindent
where we have taken (cf. Eq. [\ref{equ:C1}])
$\rho_c = C_2(m^{2-q_2}-m_{2,min}^{2-q_2})/(2-q_2)$.
If we assume for
simplicity that $r_c >> r_{2,min}$, then for our fiducial case of $q_2 = 11/6$
one has $r_c \approx \tau^2_{z2}r_{2,max}$, which suggests that unless the disk
solids surface density $\sigma_2$ is very large, only the smallest particles
are assured of capture. 
For a given $q_2$ in which most of 
the mass
is in the largest particles, the inflow collision rate is small (\ie, for
$\tau_{z2} << 1$). Under these circumstances, there is a
probability that {\it all} bodies below this size are captured into the
disk (so that ${\cal{E}}(m) = 1$). The inflow rate on
the circumplanetary disk is then just given by an
integral over the number of impacting bodies on the disk's surface

\begin{equation}
\label{equ:Isp}
I_{sp} = \int_{0}^{m_c} v_1 n_1(m) m\, dm \simeq
\rho_1 v_1 \tau_{z2}^{\frac{6-3q_1}{3q_2-5}}\left(\frac{r_{2,max}}
{r_{1,max}}\right)^{6-3q_1} =
\rho_1 v_1 \tau_{z2} \left(\frac{r_{2,max}}{r_{1,max}}\right)^{1/2},
\end{equation}

\noindent
where $q_1 = q_2 = 11/6$, and we have made use of Eq. (\ref{equ:rc}) in
integrating from $0$ to $m_c$. For a sufficiently broad
size distribution for the satellitesimals, $I_{sp}\sim I_{cs}$
(see footnote [12]).
Given that in order to satisfy the long timescale of
formation of Callisto $\tau_{z2} \sim 10^{-5}$, $I_{sp}$ is likely to be
small as indicated above. We can see this if we examine a case such that
$\tau_{z2} = 1$, which means $r_c=r_{2,max}$.
Then all planetesimals with $r \leq r_{2,max}$ could be captured into the
disk. However, this condition is
quite restrictive. For instance, a disk with the mass of the Galilean
satellites out to $\sim 30 R_J$ has $\sigma_2 \sim 10^3$ g cm$^{-2}$, which
implies $r_{2,max} \sim 1-10$ m. Even if such a disk could be
produced, the formation time for the satellites would be short,
which would be inconsistent with Callisto's partially
differentiated state. 

Now we consider the mass delivery to the circumplanetary disk due to the
interaction of planetesimal fragments
with larger satellitesimals. 
The expected number of collisions of a
disk crossing planetesimal of mass
$m$ with satellitesimals of mass $m^{\prime} > m$
is

\begin{equation}
\label{equ:explb}
\begin{split}
{\cal{E}}(m; m^{\prime}>m) = \int_{m}^{m_{2,max}} \xi \pi r^{\prime 2} H_2
n_2(m^{\prime})\,dm^{\prime} \simeq \,\,\,\,\,\,\,\,\,\,\,\,\,\,\,\,\,\,\,\,\,\,\,\,\,\,\, \\
\left(\frac{3}{4\pi\rho_s}\right)^{2/3} \pi\sigma_2 \frac{2-q_2}{q_2-5/3}
\frac{m^{5/3-q_2}}{m_{2,max}^{2-q_2}} \left[1 -\left(\frac{m}{m_{2,max}}\right)^{q_2-5/3}\right],
\end{split}
\end{equation}

\noindent
which is Eq. (11) of Safronov \etal (1986), where we have corrected an
error in one of the exponents in that expression. Thus, the frequency of
interaction of bodies with $m < m_{2,max}$ is
proportional to $m^{5/3-q_2}$ (Ruskol 1975). Integrating from $m_{2,min}$
to $m_{2,max}$ we find the three cases of interest for the mass inflow rate of
planetesimals colliding with satellitesimals of larger mass

\begin{equation}
\label{equ:Ilb}
I_{lb} = \rho_1 v_1 \tau_{z2}
\begin{cases}
\frac{2-q_2}{11/3-q_1-q_2}\left(\frac{r_{2,max}}{r_{1,max}}\right)^{6-3q_1}
A_{lb}  &\,q_1 + q_2 < 11/3; \\
\frac{3(2-q_1)(2-q_2)}{q_2-5/3}\frac{\ln{\left(\frac{r_{2,max}}{r_c}\right)}-
\frac{1}{(3q_2-5)}\left[1-\left(\frac{r_c}{r_{2,max}}\right)^{3q_2-5}\right]}
{r_{1,max}^{6-3q_1}r_{2,max}^{5-3q_2}}  &\,q_1 + q_2 = 11/3; \\
\frac{2-q_2}{q_1+q_2-11/3}\left(\frac{r_{2,max}}{r_{1,max}}\right)^{6-3q_1}
B_{lb}  &\,q_1 + q_2 > 11/3; \\
\end{cases},
\end{equation}

\begin{equation}
\label{equ:AlbBlb}
\begin{split}
A_{lb} = 1 - \frac{2-q_1}{q_2-5/3}\left(\frac{r_c}{r_{2,max}}
\right)^{11-3(q_1+q_2)} + \frac{11/3-q_1-q_2}{q_2-5/3}\left(
\frac{r_c}{r_{2,max}}\right)^{6-3q_1} \\
B_{lb} = \frac{2-q_1}{q_2-5/3}\left(\frac{r_{2,max}}{r_c}
\right)^{3(q_1+q_2)-11} + \frac{q_1+q_2-11/3}{q_2-5/3}\left(
\frac{r_c}{r_{2,max}}\right)^{6-3q_1} - 1
\end{split},
\end{equation}

\noindent
where the lower limit of integration has been taken to be $r_c$.
For our chosen values of
$q_1 = q_2 = 11/6$, the inflow
rate due to larger bodies is

\begin{equation}
\label{equ:Ilb2}
I_{lb} \simeq \rho_1 v_1 \tau_{z2}\left(\frac{r_{2,max}}{r_{1,max}}
\right)^{1/2}\left[\tau_{z2} + \ln{\tau_{z2}^{-1}} - 1\right],
\end{equation}

\noindent
so that the total inflow rate into the circumplanetary disk 
for $q_1 = q_2 = 11/6$ is

\begin{equation}
\label{equ:totflux}
I = I_{sp} + I_{lb} \approx \rho_1 v_1 \tau_{z2}\left(\frac{r_{2,max}}
{r_{1,max}}\right)^{1/2}\left[\tau_{z2} - \ln{\tau_{z2}}\right].
\end{equation}

\noindent
We see from this expression that for a low optical depth disk the
mass delivery of incoming planetesimals by larger satellitesimals is more 
efficient
than delivery by smaller or comparable sized bodies, as already mentioned.
Note that in Eq. (\ref{equ:totflux}), the
second term becomes zero for values of $\tau_{z2} \geq 1$ since this
corresponds to the situation in which $r_c \sim r_{2,max}$ (\ie, all bodies
smaller than $r_{2,max}$ could be captured into the disk).

The mass accretion rate obtained by integrating Eq.
(\ref{equ:totflux}) over the disk area is

\begin{equation}
\label{equ:massrate}
\begin{split}
\dot{M} = 2\pi\int_{R_0}^{R_D} I(R)\,p(R)\,R\,dR \approx
\,\,\,\,\,\,\,\,\,\,\,\,\,\,\,\,\,\,\,\,\,\,\,\,\,\,\,
\,\,\,\,\,\,\,\,\,\,\,\,\,\,\,\,\,\,\,\,\,\,\,\,\,\,\,\\
\frac{8\pi\sigma_1}{P_1}\left(\frac{r_2}{r_1}\right)^{1/2}
\int_{R_0}^{R_D} 
\tau_{z2}\left[\tau_{z2}-\ln{\tau_{z2}}\right]\,p(R)\,R\,dR
\end{split},
\end{equation}

\noindent
where $p(R)$ is given by Eq. (\ref{equ:probcap}), $l \equiv R/R_P$, and
$r_2$ and $r_1$ are the satellitesimal and planetesimal
size distribution cutoffs.
We have expressed the quantity $\rho_1v_1$ in
terms of the surface density of solids in the nebula $\sigma_1$, and the 
planetary period $P_1$.
%
%

We can now obtain
solutions for the mass captured by the circumplanetary disk assuming
that the mass delivery rate to the disk is balanced by the mass removal
rate to the inner disk (and used to form close-in, prograde regular
satellites).
The mass removal rate is given by $\dot{M} \sim M_{sw}/\tau_{coll}$, where
$M_{sw}$ is the steady state mass of the satellitesimal swarm (which
excludes satellite embryos), 
and the collisional time is $\tau_{coll} \sim P_2/\tau_{z2}$. We
choose the characteristic orbital period in the disk to be
$P_2 \sim 1$ year, which is the orbital period at $R \sim R_H/4$ (where
the maximum of $p(R) \times R$ is located for $\theta_1 \sim 75$). 
Given that the total mass to be delivered in the satellite's 
accretion time $\tau_{acc}$ is $M \sim 10^{26}$ g,
we must have $M = M_{sw} + \dot{M}\times\tau_{acc}$. This simply
means that the disk mass must satisfy $M_{sw} = M/(1 + N_c)$, where
$N_c = \tau_{acc}/\tau_{coll}$ is the number of collisional cycles.
We can also write this equation in the form

\begin{equation}
\label{equ:miter}
\dot{M}\tau_{acc} - \frac{N_c}{1+N_c}M = 0,
\end{equation}

\noindent
which allows us to express $r_2$, $\tau_{z2}$,
and $\sigma_2$ in terms of $N_c$, so that equations
(\ref{equ:massrate}-\ref{equ:miter}) can be solved by
iteration as functions of $N_c$,
for specific choices
of the solar nebula parameters $r_1$, $\theta_1$ and $\sigma_1$. Here
we assume that the
optical depth and solids surface density of the circumplanetary disk
are constant not only in time but with position as well. 

Figures 1 and 2 correspond
to a case such that the satellite accretion time (which is
set by the planetesimal feeding time) is
$\tau_{acc} = 10^6$ years, the solids surface density in the solar
nebula is ten times MMSN, and the
Safronov parameter $\theta_1 = 4$ ($v_{10} = 21$ \kms).
In Fig. 1 we plot the satellitesimal cutoff size $r_2$ (solid
\begin{small}
\marginpar{{\bf Figure 1}}
\end{small}
line) and surface density $\sigma_2$ (dotted line) as a function of $r_1$
that satisfy the system constraints.
In this case solutions for which $N_c \sim 1-100$ correspond to the range of 
$r_1$ shown.
Larger values of $r_2$ correspond to larger $\tau_{coll}$ (smaller $\tau_{z2}$)
and hence fewer
collisional cycles, as indicated in Fig. 1 and by the dotted curve in 
Fig. 2. 
\begin{small}
\marginpar{{\bf Figure 2}}
\end{small}
The solids surface density $\sigma_2$ changes only by an order of
magnitude over the same range in which the satellitesimal maximum cutoff size
varies by a factor of $\sim 1000$.
The greater variation of $r_2$ compared to $\sigma_2$ is a result of the
of the stronger dependence of $r_2 \propto N_c^{-1}(1+N_c)^{-1}$
compared to $\sigma_2 \propto (1+N_c)^{-1}$. It should be
pointed out that there
are solutions for $N_c \sim 1$. This corresponds to a case
in which the steady state disk mass is $\sim 0.5$ the mass
the Galilean satellites. It is unclear, however, whether the
required size of $r_2 \sim 10$ km (for $\theta_1 = 4$) is too large to
be realistic. We do not attempt to model their size distribution,
but a population of such large satellitesimals may be
difficult to maintain given fragmentation from hypervelocity
collisions from Roche-lobe
interlopers. We discuss this point in more detail at the end of this section.
Here we note that the range of possible values of $r_2$ is quite broad, so that
an independent constraint for this quantity is essential. It is interesting
to note that there are solutions such that the circumplanetary
population is composed of rubble $r_2 < 10$ m. This can work
because for these cases the optical depth of the disk is relatively high, so
that enough mass may be captured from the small mass fraction at the
low end of the
incoming planetesimal mass spectrum (given that $r_1 >> r_2$).

Next we do a case in which $\sigma_1$ is two times
the solids MMSN, $\tau_{acc} = 10^6$ years,
and we assume $\theta_1 = 75$ ($v_{10} = 5 \kms$),
implying a fairly ``cold'' population of heliocentric
objects (Figs. 3 and 4). We find solutions
\begin{small}
\marginpar{{\bf Figure 3}}
\end{small}
for $r_1 \sim 5-10$ km, which correspond to $N_c$ in the
range $5-100$. In particular,
\begin{small}
\marginpar{{\bf Figure 4}}
\end{small}
a solution exists where the satellitesimal cutoff size is $r_1 = 1$ km for
a planetesimal cutoff size of $r_1 = 10$ km. In this case, $\tau_{z2} \sim
5\times 10^{-6}$ implying that timescale for the collisional cycle is
$\tau_{coll} \sim 2\times 10^5$ years, so that $N_c \sim 5$ collisional
cycles are required to
complete the formation of the satellites. 
We note that the range of 
variation of both $r_2$ and $\sigma_2$ is significantly smaller for this case,
which suggests
that large $\theta_1$
cases may be easier to constrain.

To understand better the effect
of the velocity dispersion of the heliocentric
planetesimal population on the mass delivery
rate, we plot in Fig. 5
\begin{small}
\marginpar{{\bf Figure 5}}
\end{small}
the total mass delivered to the circumplanetary disk over
and accretional time of $\tau_{acc} = 10^6$ years as a function of the
Safronov parameter $\theta_1$. In this case
we fix $r_1$ and $r_2$ at 10 and 1 km respectively. All of the cases presented
assume that $\tau_{z2} = 5\times 10^{-6}$ so that $N_c = 5$.
The solid curve corresponds to a solids MMSN ($\sigma_{\rm{MM}}$),
the long dashed line
corresponds to a case of $\sigma_1 = 2\sigma_{\rm{MM}}$,
the dotted line to $\sigma_1=10\sigma_{\rm{MM}}$,
and the short-dashed line to $\sigma_1=0.1\sigma_{\rm{MM}}$.
From this plot we can see that for a choice of $\theta_1 = 4$
to have a solution for $r_1 = 10$ km one would require a case with
greater nebula solid content than ten times MMSN
(cf. Fig. 1). Conversely, one could have a slightly colder population
($\theta_1 \sim 8$) that satisfy the constraints for the case of ten times
MMSN.

Finally, in Fig. 6 we plot the mass captured by the circumplanetary disk
\begin{small}
\marginpar{{\bf Figure 6}}
\end{small}
in one collision time for $\tau_{acc} = 10^6$ years 
as a function of collisional cycles $N_c$. This mass simply corresponds to the
steady-state disk mass $M_{sw}$ as it should. 
All prior solutions for a range of $r_1$ lie on this curve. For
instance, the case for twice MMSN and $r_1 = 10$ km for which
$N_c \sim 5$ collisional cycles implies that $M \sim 0.2$ Galilean
masses are captured per collision time $\tau_{coll} \sim 2\times 10^5$ years 
(corresponding to $\dot{M}\sim 5 \times 10^{20}$ g yr$^{-1}$). This means 
that at any given time the total satellitesimal disk mass $M_{sw}$
(excluding embryos) is $\sim 20\%$ of the total mass of the satellites.

An important caveat is that all these cases use
accretion times that are significantly longer that feeding zone
clearing timescale, which means that we are assuming some
mechanism to replenish it. It is also important to point out that at
these late times modeling of the planetesimal mass spectrum
would involve different regimes with different power law exponents
(see, \eg, Charnoz and Morbidelli 2003, Kenyon and Bromley 2004,
Pan and Sari 2004). In the next section we consider small values for 
$r_1 \sim 1$ km to indicate that following giant planet formation a
significant fraction of the planetesimal mass in the feeding zone
of the giant planet may cascade back down to small sizes. 
Although our solutions so far have involved somewhat larger values of $r_1$,
in general only planetesimal fragments $\lesssim 1$ km may be captured and
can be used for satellite formation. 
One might expect even more mass in smaller particles, which as we
point out below would benefit our model.

Finally, one would still need to solve self-consistently for the size of
$r_2$. This size might be set by a
competition between the growth time of satellitesimals
versus the time it takes for catastrophic collisions with incoming
planetesimals to disrupt them.
The growth time of $r_2$ is $\sim \tau_{coll}$ which
may be, as for the cases in Figs. 3 and 4, quite long $\sim 10^5$ years. 
We can estimate the lifetime of a satellitesimal of size $r_2$
as the time it takes for a bigger planetesimal to collide with it.
Given a planetesimal size
distribution with upper cutoff of $r_1 = 10$ km and power law exponent
$q_1 = 11/6$, the
time for a large planetesimal to shatter a $\sim 1$ km
satellitesimal can be estimated to
be $\tau_{frag} \sim 10^{-3}P_1\rho_s(r_1r_2)^{1/2}/
\sigma_1 \sim 10^3$ years for a solids MMSN. This is $\sim 100$
times faster than its growth time, suggesting that the cutoff in $r_2$ should
be smaller than 1 km, and invalidating the assumptions used to obtain
model solutions. This result is a direct consequence of the model constraints
coupled with a small capture probability $p(R) << 1$.
It may be possible to circumvent this problem by speeding
up the accretion of satellitesimals using circumplanetary debris,
and increasing the capture probability by using a large value of
the Safronov parameter $\theta_1$. It is also possible that
a different process, such as the capture of planetesimal fragments
or a gravitational instability, is
responsible for setting the satellitesimal cut-off size, or that
that the satellitesimals are composed of collisional rubble, as
indicated by some of our solutions. 
However, we leave investigation of these possibilities for further work.

So far we have ignored the issue of the angular momentum of the satellitesimal
swarm, and found solutions subject to the constraint that enough mass
be delivered to form the regular satellites absent a gap in the
planetesimal disk. We obtained
solutions such that $r_1 \gtrsim r_2 \sim 1$ km as well
as solutions where $r_1 >> r_2 \sim 1$ m.
This is because lower values of $r_1$ would deliver too
much mass. However, given the uncertainty in the calculation this
can not be taken to be a binding conclusion. Below we attempt to
match both the mass and angular momentum of the satellites. In this
case, we need to deliver enough mass to make the satellites
as the giant planet opens a gap in the planetesimal distribution. 
For this reason, next we
consider cases such that $r_2 \gtrsim r_1$ and $\theta_1 >> 1$, both of
which facilitate capture. Allowing for larger values of $r_1$ would
require us to use more planetesimal 
disk mass by an amount that depends on the (unknown) value of $q$.
For simplicity, below we neglect these complications.

\subsection{Angular Momentum Delivery}
\label{subsec:angmom}

In this section we show that it may be possible to deliver enough
angular momentum to the circumplanetary disk to explain the observed
angular momentum budget of the regular satellite systems.
At first this might appear to be a problem given
that the circumplanetary disk is composed of both prograde
and retrograde satellitesimals
with small net angular momentum.
However, as the giant planet clears planetesimals from its feeding
zone on a timescale comparable to the satellite formation timescale,
inhomogeneities in the planetesimal disk can pump up the angular
momentum delivered to the circumplanetary disk
(a point first made by Lissauer and Kary 1991 in the context of
planetary spins), which may account for the observed radial
extent of the satellite systems provided that most of the solids
that make up the satellites originate from the outer regions of
the giant planet's feeding zone.

Here we assume that the bulk of the angular momentum
that goes into the regular satellites arrives late as the planet
clears its feeding zone, and that the planetesimal population at this
time is composed
of small particles 
$r_1 < r_2 \lesssim 1$ km with a velocity dispersion
characteristic of a fairly cold population ($\theta_1 \sim 100$,
but below we still use the ``high'' dispersion limit)
with lower inclinations than eccentricities ($v_z = 0.1 v_1$)\footnote{
Strictly speaking this implies that $v_1 \lesssim \Omega_1 R_H$.}.
This implies that the capture probability $p(R) \lesssim 1$. 
Furthermore, we assume that on average 
a free-bound collision deposits a significant fraction
$\epsilon \sim 0.5$
of the angular momentum of the interloper.
This is because colliding planetesimals
may deposit angular momentum in the circumplanetary disk
even if they (or their fragments) 
are not captured\footnote{One might consider the
case $\epsilon = p(R)$ as a lower bound on the angular momentum
deposition efficiency. This would require that a planetesimal (or its
fragments)
be captured in order to deliver its angular momentum to the disk.
For our choice of parameters $p(R) \lesssim \epsilon$.}.

We provide an estimate for the amount of angular momentum delivered
to the circumplanetary disk using a method which is related to 
planetary
spin calculations (\eg, Lissauer and Kary 1991; Dones and Tremaine 1993,
hereafter DT93). We
calculate the rate of accretion of mass and angular momentum by
a (2-D) disk of satellitesimals. Our
simplifying assumptions are that the planet is on a circular orbit,
that incoming planetesimals do not interact with each other, and that
$\sigma_1$ is uniform over the planet's (undepleted portion of the) 
feeding zone.
Incoming particle velocities increase as they approach the planet.
This prevents us from considering significantly larger values
than $\theta_1 \sim 100$,
and leads us to adjust the velocity dispersion using
$v_1 = \sqrt{v_{10}^2 + v_{e}^2}$, with $v_{10}$ and $v_e$
as previously defined.\footnote{It should
be noted that the satellitesimals Keplerian velocity
is not much smaller than $v_1$. This could affect the
calculation for the net angular momentum deposition. 
We leave investigation of this complication
for further work.}
In addition, the giant planet's mass and the properties
of the circumplanetary disk are assumed to be constant. With these
assumptions, the problem can be cast in terms of two dimensionless
parameters

\begin{equation}
\label{equ:rands}
b \equiv \frac{R}{R_H},\,\,\,\,\,\,\,\,s\equiv \frac{v_1}{\Omega_1 R_H}
\end{equation}

\noindent
where $b$ is the ratio of radial distance in the 
circumplanetary disk to
the Hill radius, and $s$ measures the importance of the 
differential rotation
of the circumsolar disk relative to the velocities of incoming
planetesimals (DT93). The parameter $b$ is a measure 
of the importance of the planet's gravity relative to that of the Sun at
$R$ in the 
satellitesimal disk. For our problem, $\theta_1 \gtrsim 100$ implies
$v_1 \lesssim 6$ \kms and thus $s \lesssim 6$. At $R_H/4$ (where $p(R) \times
R$ peaks for $\theta_1 \sim 100$),  
$b/s \lesssim 1$ and $bs^2 \gtrsim 1$, implying we are in a regime in
which dispersion is still high and gravity is somewhat important.

The accretion of mass and angular momentum by the 
circumplanetary disk can be 
described by a 3-D distribution $\Phi(R,\ell_z)$, where
$\Phi(R,\ell_z)d\ell_z$ is the collisional inflow
rate per unit area of mass with specific angular
momentum $\ell_z$ in the range $\ell_z$ and $\ell_z + d\ell_z$. For our 
problem, we have

\begin{equation}
\label{equ:Phirect}
\begin{split}
\Phi(R,\ell_z) = 2\int_{0}^{2\pi}d\phi 
\int_{-\infty}^{\infty} d\dot{x}\int_{-\infty}^{\infty} d\dot{y}
\int_{-\infty}^{\infty} d\dot{z} \,\tau(\vartheta) \times\\
f_0(x,y,z\rightarrow 0^+,\dot{x},\dot{y},\dot{z})
\delta(\ell_z - x\dot{y}+y\dot{x}-\Omega_1 R^2)|\dot{z}|H(-\dot{z}),
\end{split}
\end{equation}

\noindent
where the azimuthal angle $\phi$ is measured from the origin of the system
centered on the planet such that $(x,y) = R(\cos{\phi},\sin{\phi})$,
the $x-y$ plane is the planetary orbit plane (which coincides with
the circumplanetary disk plane), the $x$-axis points radially
outward, the $y$-axis points in the direction of orbital motion,
$\delta$ is the Dirac delta function, and $H(-\dot{z})$
is the Heaviside step function, which selects only planetesimals with velocities
directed towards the disk plane. The notation $z\rightarrow 0^+$ is
meant to indicate that we are approaching the disk plane from above ($z > 0$). 
The probability of a free-bound collision in crossing the
disk $\tau(\vartheta)$
is given by

\begin{equation}
\label{equ:tau}
\tau(\vartheta) = \frac{\tau_{z2}}{\left| \sin{\vartheta} \right|} =
\tau_{z2}\left(\frac{v_r^2 + v_\phi^2 + 
\dot{z}^2}{\dot{z}^2}\right)^{1/2}\sim \,\tau_{z2}\frac{\left| v_r\right|}
{\left|\dot{z}\right|},
\end{equation}

\noindent
where $v_r = \dot{x}\cos{\phi} + \dot{y}\sin{\phi}$ is the radial velocity, 
$v_{\phi}$ is the azimuthal velocity,
and we have assumed that the planetesimal radial velocities dominate their
motion (which implies a shallow angle of entry). 
In Eq. (\ref{equ:Phirect}), $f_0$ is a Schwarzchild distribution
function, describing the mass of a uniform disk of particles in a 
three-dimensional phase-space volume element $dxdydzd\dot{x}d\dot{y}d\dot{z}$.
 
\begin{equation}
\label{equ:f0dist}
\begin{split}
f_0(x,y,z,\dot{x},\dot{y},\dot{z}) = 
\,\,\,\,\,\,\,\,\,\,\,\,\,\,\,\,\,\,\,\,\,\,\,\,\,\,\,\,\,\,\,\,\,
\,\,\,\,\,\,\,\,\,\,\,\,\,\,\,\,\,\,\,\,\,\,\,\,\,\,\,\,\,\,\,\,\,
\,\,\,\,\,\,\,\,\,\,\,\,\,\,\,\,\,\,\,\,\,\,\,\,\,\,\,\,\,\,\,\,\, \\
\frac{\rho_0}{(2\pi)^{3/2} v_1^2(1-\frac{1}{2}\Gamma)^{1/2}v_z}
\exp{\left[-\frac{\Omega_1^2 z^2}{2 v_z^2}
-\frac{\dot{x}^2}{2v_1^2} - \frac{(\dot{y}+\Gamma\Omega_1 x)^2}
{2v_1^2(1-\frac{1}{2}\Gamma)} -\frac{\dot{z}^2}{2v_z^2}\right]},
\end{split}
\end{equation}

\noindent
which may apply sufficiently far from the planet. In this expression, 
$\rho_0 = \sqrt{2\pi}\sigma_1/P_1v_z$
is the midplane planetesimal volume density, and we have assumed
that the $x$-component of the velocity dispersion $v_x = v_1$,
and $v_y = v_1(1-\frac{1}{2}\Gamma)^{1/2}$, where 
$\Gamma = -(d\ln{\Omega}/d\ln{R}) = 3/2$ for a Keplerian disk.

Integrating Eq. (\ref{equ:Phirect})
over $\dot{z}$,
defining the velocities
$(\dot{x},\dot{y}) \equiv (v\cos{\psi},v\sin{\psi})$, and replacing
the azimuthal angle $\phi$ with $\chi = \pi + \phi - \psi$ so that
$v_r = -v\cos{\chi}$, we can write in polar coordinates (DT93)

\begin{equation}
\label{equ:Phipolar}
\begin{split}
\Phi(R,\ell_z) = \frac{2\sigma_1\tau_{z2}}{\left[2\pi(1-\frac{1}{2}
\Gamma)\right]^{1/2}v_1^2v_zP_1}
\int_{0}^{2\pi}d\psi \int_{-\pi/2}^{\pi/2} \cos{\chi}d\chi \\
\int_{|\ell_z-\Omega_1 R^2|/R}^{\infty}
v^2e^{-\frac{1}{2}\Psi}\delta(\ell_z - Rv\sin{\chi}-\Omega_1 R^2)
dv
\end{split},
\end{equation}

\noindent
and

\begin{equation}
\label{equ:Psi}
\Psi \equiv \frac{v^2\cos{^2\psi}}{v_1^2} + \frac{\left[v\sin{\psi} -
\Gamma\Omega_1 R\cos{(\chi + \psi)}\right]^2}{v_1^2(1-\frac{1}{2}\Gamma)}.
\end{equation}

Generally, if the velocity dispersion is large enough or for small $b$,
the quantity $b/s << 1$, and Eq. (\ref{equ:Phipolar}) can be
simplified by expanding Eq. (\ref{equ:Psi}) to first order in $b/s$. However,
for our problem $b/s \gtrsim 0.1$ so that we expand Eq. (\ref{equ:Psi})
to fourth order to obtain a better estimate. Carrying out
the expansion and integrating over $\chi$ and $v$ (see Appendix B)

\begin{equation}
\label{equ:Phialp}
\begin{split}
\Phi(R, \alpha) = \frac{2\sigma_1\tau_{z2}}
{R\left[2\pi(1-\frac{1}{2}\Gamma)
\right]^{1/2}v_zP_1}
\int_{0}^{2\pi} \frac{d\psi}{g(\psi)}[(K_0(g) + K_1(g)\alpha +
K_2(g)\alpha^2 + \\
K_3(g)\alpha^3 + 
K_4(g)\alpha^4) e^{-g\alpha^2} +
(K_5(g)\alpha^2 + K_6(g)\alpha^3 + \\
K_7(g)\alpha^4)E_1(g\alpha^2)],
\end{split}
\end{equation}

\noindent
where the $K_i$ are functions of the angle $\psi$, $\Gamma$, and $b/s$.
Here
$\alpha = (\ell_z - \Omega_1 R^2)/\sqrt{2}v_1R$, $E_1$ is the exponential
integral of order 1, and

\begin{equation}
\label{equ:gpsi}
g(\psi) = \cos{^2\psi} + \frac{\sin{^2\psi}}{1-\frac{1}{2}\Gamma}.
\end{equation}

To obtain
the mass {\it capture} rate of planetesimals with
specific angular momentum $|\ell_z| \ge \ell_0$
we define
the mass inflow collision rate per unit area ${\cal{I}}(R,\ell_0)$

\begin{equation} 
\begin{split}
{\cal{I}}(R,\ell_0) 
= \frac{1}{2\pi}\int_{\ell_0}^{\infty}\Phi(R,\ell_z)\,d\ell_z + 
\frac{1}{2\pi}\int_{-\infty}^{-\ell_0}\Phi(R,\ell_z)\,d\ell_z =
\,\,\,\,\,\,\,\,\,\,\,\,\,\,\,\,\,\,\,\,\,\,\,\,\,\,\,\,\,\,\,\,\, \\
\frac{\sqrt{2}v_1R}{2\pi}\left\{\int_{\alpha_0}^{\infty}\Phi(R,\alpha)\,d\alpha 
+ 
\int_{\alpha_0^{\prime}}^{\infty}\Phi(R,\alpha^{\prime})\,d\alpha^{\prime}
\right\}
\end{split},
\end{equation}

\noindent
where $\Phi(R,\alpha^{\prime})$ is identical to Eq. (\ref{equ:Phialp}), except
with the replacements $K_1 \rightarrow -K_1$, $K_3\rightarrow -K_3$,
$K_6\rightarrow -K_6$, and $\alpha^{\prime} = (\ell_z + \Omega_1 R^2)/
\sqrt{2}v_1R$. Integrating over the disk 
area, we get the mass capture rate

\begin{equation}
\label{equ:Mdot}
\begin{split}
\dot{M}(\ell_0) = 2\pi\int_{R_0}^{R_D} {\cal{I}}(R,\ell_0)p(R)R\,dR = 
\,\,\,\,\,\,\,\,\,\,\,\,\,\,\,\,\,\,\,\,\,\,\,\,\,\,\,\,\,
\,\,\,\,\,\,\,\,\,\,\,\,\,\,\,\,\,\,\,\,\,\,\,\,\,\,\,\,\,
\,\,\,\,\,\,\,\,\,\,\,\,\,\,\,\,\,\,\,\,\,\,\,\,\,\,\,\,\, \\
\frac{2\sigma_1\tau_{z2}}{\left(1-\frac{1}{2}\Gamma \right)^{1/2}P_1}
\int_{R_0}^{R_D} \frac{v_1}{v_z}\,p(R)R\,dR\int_{0}^{2\pi}
\frac{d\psi}{[g(\psi)]^{3/2}} \sum_{i=0}^{7}
{\cal{M}}_i(g(\psi),R,\ell_0))
\end{split},
\end{equation}

\noindent
where $p(R)$ is given by Eq. (\ref{equ:probcap}), and 
the ${\cal{M}}_i$ are given in Appendix B.
The total mass delivered to the circumplanetary disk is then given by
$M = M_{sw} + \dot{M}\tau_{acc}$, where as before
$M_{sw}$ is the steady-state
circumplanetary disk mass (excluding satellite embryos).

The angular momentum rate delivered to the circumplanetary disk
$\dot{L}_z$ can be
calculated from the first moment of the distribution $\Phi(R,\ell_z)$.
The angular momentum delivery rate per unit area ${\cal{I}}_L(R,\ell_0)$
assuming perfect deposition efficiency is

\begin{equation}
\begin{split}
{\cal{I}}_L(R,\ell_0) = \frac{1}{2\pi} \int_{\ell_0}^{\infty}\ell_z 
\Phi(R,\ell_z)\,d\ell_z + \frac{1}{2\pi} \int_{-\infty}^{-\ell_0}
\ell_z\Phi(R,\ell_z)\,d\ell_z = 
\,\,\,\,\,\,\,\,\,\,\,\,\,\,\,\,\,\,\,\,\,\,
\,\,\,\,\,\,\,\,\,\,\,\,\,\,\,\,\,\,\,\,\,\,\,\, \\
\Omega_1 R^2{\cal{I}}(R) +
\frac{v_1^2R^2}{\pi}\left\{\int_{\alpha_0}^{\infty}\alpha \Phi(R, \alpha)
\,d\alpha - \int_{\alpha_0^{\prime}}^{\infty}\alpha^{\prime}
\Phi(R, \alpha^{\prime})\,d\alpha^{\prime}\right\}
\end{split},
\end{equation}

\noindent
which when integrated over the circumplanetary disk yields

\begin{equation}
\label{equ:meanLzdot}
\begin{split}
\dot{L}_z(\ell_0) = 2\pi \int_{R_0}^{R_D}\epsilon{\cal{I}}_L(R,\ell_0)R\,dR  =
\frac{2\epsilon\sigma_1\tau_{z2}\Omega_1}{\left(1-
\frac{1}{2}\Gamma\right)^{1/2}P_1} \int_{R_0}^{R_D} \frac{v_1}{v_z}
R\,dR \,\times
\,\,\,\,\,\,\,\,\,\,\,\,\,\,\,\,\,\,\,\,\,\,
\,\,\,\,\,\,\,\,\,\,\,\,\,\,\,\,\,\,\,\,\,\,\,\, \\
\left\{R^2\int_{0}^{2\pi}\frac{d\psi}{[g(\psi)]^{3/2}}
\sum_{i=0}^{7}{\cal{M}}_i(g(\psi),R,\ell_0)) + \frac{\sqrt{2}v_1R}{\Omega_1}
\int_{0}^{2\pi}\frac{d\psi}{[g(\psi)]^{3/2}} \sum_{i=0}^{7}
{\cal{L}}_i(g(\psi),R,\ell_0))\right\}
\end{split},
\end{equation}

\noindent
where the ${\cal{L}}_i$ are given in Appendix B.
The total angular momentum delivered to the circumplanetary
disk is then $L_z = L_i + \dot{L}_z\tau_{acc}$, where $L_i$ is
the initial angular momentum of the disk, which is taken to be small. 
The specific angular momentum is
$\left<\ell_z\right> = L_z/M$.


We integrate Eq. (\ref{equ:Mdot}) and (\ref{equ:meanLzdot})
over angle $\psi$ and radius
using standard numerical techniques (note that, although $v_1$ varies with $R$,
we held the ratio $v_1/v_z$ constant in these calculations). 
In Fig. 7
\begin{small}
\marginpar{{\bf Figure 7}}
\end{small}
we plot both the mass $M$ and $z$-component of
angular momentum $L_z$ delivered to the circumplanetary disk over the
planetesimal feeding time $\tau_{acc} = 10^6$ years 
as a function of the feeding
zone gap-size $R_{gap}=\sqrt{\ell_0/\Omega_1}$ with $R_0 = R_P$ and
$R_D = R_H/2$. 
It is important to note that this is a practical
definition based on the
observation that for $R_{gap} > 2.5 R_H$ the mass inflow rate drops
to nearly zero. Because in the high dispersion weak gravity
limit large magnitude (planetesimals may deposit positive or
negative) angular momentum contributions
are not limited to the outer regions of the feeding zone of the giant planet,
this calculation likely overestimates the increase in specific
angular momentum as a gap is
opened\footnote{Ohtsuki and Ida (1998)
argue that such a boundary should be implemented in terms of
the Jacobi constant.}.
Here we are mainly
interested in the relative behavior of the angular momentum
curves to the mass curves.
The mass and angular momentum are plotted so that
unity corresponds to the total mass and angular momentum 
contained in the Galilean satellites,
$M_{sats}\sim 4\times 10^{26}$ g, $L_{z,sats}\sim 4\times 10^{43}$ 
g cm$^2$ s$^{-1}$, which yields
$\left<\ell_z\right>_{sats}\sim 1\times 10^{17}$ cm$^2$
s$^{-1}$. 
In this figure we plot curves for combinations of
surface density $\sigma_1$ and angular momentum
deposition efficiency $\epsilon$ such that solutions
matching the mass and angular momentum in the 
Galilean satellites were found for gap sizes
$R_{gap} \sim 1 R_H$. 
This implies that for
our chosen parameters the
bulk of the solids that form the Galilean satellites
can be delivered 
as the giant planet
clears a gap in its feeding zone.
This can work provided that a significant fraction
of the mass of the solids is in planetesimal
fragments with $r_1 \lesssim r_2$
with low inclination orbits $v_z = 0.1 v_1$, that collisions
are efficient at depositing angular
momentum $\epsilon \sim 0.5$, and that the nebula
is enhanced in solids with respect to MMSN. 
Given a value for the
satellitesimal size cutoff of $r_2 = 1$ km, 
we then have $\tau_{z2} \sim 10^{-5}$,
$\tau_{coll} \sim 10^5$ years (since the dynamical
time in the satellitesimal disk is $\sim 1$ year),
$N_c \sim 10$, $M_{sw} \sim 0.1 M_{sats}$ as
before (which is the value
the mass curves [dotted lines] approach
as the gap size $R_{gap}$
encompasses the giant planet's feeding zone),
and $L_i < 0.1 L_{sats}$.
We can also define the quantity

\begin{equation}
\label{equ:Lambda}
\Lambda = \sigma_1\tau_{z2}\left(\frac{v_1}{v_z}\right)\tau_{acc} =
\sigma_1P_2\left(\frac{v_1}{v_z}\right)N_c,
\end{equation}

\noindent
where $\sigma_1$ is taken to be in units of the solids MMSN
($\sigma_{\rm{MM}} = 3.3 \gcms$), and $\tau_{acc}$ and $P_2$ are in years.
The pairs of curves in Fig. 7 correspond to
$\Lambda = 280$ ($\epsilon = 0.8$, $\sigma_1 = 2.8 \sigma_{\rm{MM}}$),
$\Lambda = 450$ ($\epsilon = 0.5$, $\sigma_1 = 4.5 \sigma_{\rm{MM}}$),
and $\Lambda = 1200$ ($\epsilon = 0.2$, $\sigma_1 = 12 \sigma_{\rm{MM}}$).
If we change the values of parameters
we can obtain approximately the same curves as
before by keeping $\Lambda$ constant.
One can also
consider longer feeding timescales, but such solutions might be unrealistic.

In Fig. 8 we plot the specific angular momentum as a
\begin{small}
\marginpar{{\bf Figure 8}}
\end{small}
function of the gap size for parameters such that the
mass and angular momentum of the Galilean satellites
are obtained for $\left<\ell_z\right>/\left<\ell_z\right>_{sats} = 1$.
As one might expect, the specific angular momentum curves
first increase with gap size, reach a maximum, and begin
to decrease as the gap size chokes off the mass inflow. It is
interesting to note that for the case of nearly perfect angular
momentum deposition efficiency $\epsilon \sim 1$ it
would seem possible to match the constraints
even without a gap. 
However, it remains the case that to satisfy
the system constraints more generally most of the angular momentum must be 
delivered
as the giant planet opens a gap in the planetesimal distribution.

In Fig. 9 we plot the specific angular momentum delivered
\begin{small}
\marginpar{{\bf Figure 9}}
\end{small}
as a function of the satellitesimal disk size $R_D$ for several values
of the gap size. This figure
shows that one may need an extended disk $\sim R_H/2$ in order to be
able to deliver sufficient mass and angular
momentum to form the regular satellites of Jupiter 
(and Saturn\footnote{Saturn's satellite system might require
one to adapt the present formalism to the case of a
giant planet with significant obliquity.}).

Finally, in Figs. 10 and 11 we increase the size
\begin{small}
\marginpar{{\bf Figure 10}}
\end{small}
of the inner disk boundary $R_0$ and compare the relative
behavior of the angular momentum and mass curves as we did
\begin{small}
\marginpar{{\bf Figure 11}}
\end{small}
before. Strictly speaking, this procedure would be tantamount
to opening a gap in the heliocentric planetesimals in the
limit of small velocity dispersion and weak gravity, which is
inconsistent with the assumptions employed to obtain these
curves. Nevertheless, because
opening a gap would preferentially decrease the inflow rate close
to the planet, it is still instructive to consider this case.
Here we simply re-scale the size of the gap by $R_{gap} = 5 R_0$. The
results are qualitatively similar to before, except
that the effect on $\left<\ell_z\right>$ 
is less pronounced and a solution could
not be found in the case of $\epsilon = 0.2$.

It is important to note that these results employ $\tau_{acc} = 10^6$ yrs.
An outstanding
issue is whether planetesimals may be replenished by
gas drag in the feeding
zones of the giant planets on a
timescale $\gtrsim 10^5$ years.
By this time the
giant planet would have opened a more or less clean gap in the
gas as well as the planetesimal disk
(depending on the unknown turbulence parameter),
which means that even meter-sized particles
may not be well-coupled to the gas
in the feeding zone of the giant
planet. In this limit the gas density may not be responsible
for selecting the sizes of planetesimals, and the collisional cascade
may proceed to possibly even smaller sizes. In regions outside the
gas gap,
resonances may capture inwardly drifting planetesimals, which
might prevent them from replenishing the feeding zone. However, collisions
or stirring by larger particles may free them and allow them to continue
their inward migration. As a result, this problem is unique in that it
spans both gas-free and gaseous regimes.



Ignoring the gap in the gas,
the criterion for feeding zone cleanup is (Tanaka and Ida 1997
and references therein)

\begin{equation}
\sqrt{3}\left(\sqrt{1 + 0.46(T_0h/T_K)^{2/3}} - 1\right)/
\left(1 + 0.46(T_0h/T_K)^{2/3}\right)^{1/4} > \eta/h, 
\end{equation}

\noindent
where
$T_0 = 2m/C_D\pi r^2\rho_g (hv_K)$ is the characteristic
gas drag time, $m$ is the particle mass, $\rho_g$ is the gas density,
$T_K$ is the orbital period of the planet, $h=(M_P/3M_\odot)^{1/3}$,
$C_D = 0.44$, $v_K$ is the Keplerian velocity of the planet, $c$ is the
gas sound speed,
and $\eta = c^2/v^2_K$. In the above equation, it is assumed that $i<e$.
If the criterion for cleanup is not satisfied,
planetesimals outside the planet's orbit can enter the feeding zone of
the planet. For Jupiter, the size that can
replenish the feeding zone of the planet is
$\sim 1$ m, where we have used a minimum mass solar nebula (Hayashi 1981).
But simulations indicate that bigger sizes may be able
to replenish the feeding zone of the giant planet (Tanaka and Ida 1997).
It is also possible that planetary migration may help in this regard, or
that one planet may stir up the planetesimal disk and replenish the feeding 
zone of its neighbor.

\section{Discussion and Conclusions}
\label{sec:concl}

We have focused on
outstanding issues that a planetesimal collisional capture model
faces before it can
provide a viable alternative to the model developed in MEa,b.
Its strength is that even though it assumes that turbulence
can drive the evolution of the subnebula disk in a shorter timescale than
that of satellite formation, a turbulence parameter appears nowhere
in the analysis. 
While several of the
parameters employed herein are at best poorly constrained at this time, none
of them are incalculable, and all are based on sound physical processes.
We have investigated whether the mass and angular momentum
budget in the Galilean satellites and Titan may be
accounted for by a model that delivers material to the
circumplanetary disk in the same timescale as the feeding zone
clearing timescale ($\lesssim 10^5$ years, see, \eg,
Charnoz and Morbidelli 2003). In particular, the angular momentum
calculation is new to this work\footnote{In the present planetesimal
capture framework. MEa provides a related argument in the case
that the angular momentum budget was dominated by the gas component.
Other satellite formation models have yet to provide an account of
the angular momentum budget of the satellites of Jupiter and
Saturn.}, as is the discussion of specific
model parameters that may apply during satellite formation
late ($10^5-10^6$ yrs) after giant planet formation. 
It is possible that inelastic
collisions between planetesimals
within the Hill radius of the giant planet form an extended ($\sim R_H/2$)
circumplanetary
satellitesimal swarm, which then sustains itself by collisionally
capturing sufficiently
small ($< 1$ km) interlopers produced by the fragmentation
of planetesimals following giant planet formation. This process might
result in a comparable surface density of solids in the
subnebula as in the nebula ($\sigma_2 \sim 1$ \gcms),
and enough mass to form the Galilean satellites over several
collisional cycles ($N_c \sim 10$).

This may work only if a significant fraction
of the mass in the satellites
originates from the outer regions 
of the planet's feeding zone, which means that
the model may have difficulties delivering enough material to simultaneously
account for the total masses of the regular satellites of Jupiter and
Saturn.
Yet, because collisions from
Roche-lobe interlopers can pump-up the angular momentum of the circumplanetary
disk whether or not the interaction results in significant mass capture, we
show that
a planetesimal collisional capture model
may deliver sufficient angular momentum to account
for the angular momentum budget of the regular satellites provided enough
mass is also delivered (which may require that a significant
fraction of the planetesimal mass be in small $< 1$ km fragments with
low inclination orbits, as well
as a solids enhanced solar nebula). 
Still, the feeding zone clearing
timescale, which in this model controls the satellite formation timescale,
may be too short to accomodate the result that Callisto
is partially differentiated (Anderson \etal 1998; 2001). This is because
only if the satellite grows sufficiently slowly ($\tau_{acc} \sim
10^5-10^6$ years) can it
radiate away its accretion energy in time to keep it from
melting and differentiating fully
(Kuramoto and Matsui 1994). We have suggested that it may
be possible to replenish the feeding zone
of the giant planet with sufficiently small
($1 - 100$ m) planetesimal
fragments brought in by gas drag, but we have not tested
whether this is really possible.
By this time the giant planet would presumably
have opened a more or less clean gap in the
gas disk, depending on the unknown turbulence parameter $\alpha$, which
makes this a difficult problem to tackle.
Also, resonances may capture inwardly drifting planetesimals,
which might prevent them from replenishing the feeding zone. However,
collisions or stirring by larger particles may free them and allow them
to continue their inward migration. It is important to note that
the model we advance here relies on fragmentation to decrease
the size of planetesimals and make them easier to capture, but we do not
employ collisional erosion directly as a capture mechanism.

Even if enough mass and angular momentum
can be delivered, a planetesimal model still faces the
further obstacle of how to
constrain the model parameters. In particular, we have postponed
modeling the planetesimal and satellitesimal mass spectrum, and
have instead treated
the size cutoffs for the planetesimals and satellitesimals ($r_1$ and
$r_2$) as adjustable parameters, meant only to indicate
characteristic sizes containing a significant fraction of
the solids disk mass. In the case of no gap in the
planetesimal distribution, we found solutions such that 
$r_1 \gtrsim r_2 \sim 1$ km as well as $r_1 >> r_2 \sim 1$ m. 
However, it is premature to conclude that this
is a model requirement. 

Although we have found regions of parameter space that might
plausibly lead to a self-consistent model for regular satellite formation,
until reliable ways are found to constrain the relevant parameters,
it remains uncertain how meaningful this is. In particular, it is difficult
to know what to choose for the size of the circumplanetary disk particles
$r_2$, where most of the satellitesimal mass resides given our assumed
value for $q_2 = 11/6$. Also, our choices both for $q_1$ and $q_2$
are meant to be in line with earlier choices
in Safronov \etal 1986 and other works. However, it is quite likely
that further research will show that other choices are more appropriate
in the present context.
In particular, there may be more mass in small particles at the late
times of interest than implied by our chosen values of $q$. It is
unlikely, however, that this will invalidate the general outline of the
present model given that small particles are easier to capture.
One might argue that irregular satellites may provide
an indication, but it is presently unknown whether the irregular
satellites were collisionally captured, or bear any connection to the
process that led to the formation of the regular satellites. On the other
hand, we do not believe that observations of irregular
satellite properties can be used to rule out the possibility either. Even
if a connection can be made, subsequent collisional and
dynamical evolution of satellitesimals would make it very difficult to extract
useful constraints from the observed population of irregular satellites.

In theory, it might be possible to model the coupled problem of
the fragmentation and growth of circumplanetary satellitesimals and
heliocentric planetesimals, but at present we are far from realizing this
goal. This raises the issue of what
physical process sets the size for $r_2$. If this size is determined by a
competition between collisional disruption by heliocentric
interlopers and Safronov-type growth in the circumplanetary disk,
then a difficulty arises for the model we advance here (at least for
$r_1 \gtrsim r_2$). Namely,
our system constraints lead us to a model in which collisions
between satellitesimals and similarly-sized planetesimals
are more frequent than collisions between similarly-sized satellitesimals.
This is because we seek a mass balance (\ie, satellitesimal
collisions remove a similar amount of mass as gets delivered by captured
planetesimals), yet hypervelocity
collisions with similarly-sized interlopers do not
always result in capture (\ie, $p(R) << 1$). As a result,
satellitesimals may not have sufficient time to grow to the
typical sizes $r_2 \sim 1$ km we use to satisfy our system constraints
(which require long satellitesimal collisional times $\sim 10^5$ years,
such that $N_c = \tau_{acc}/\tau_{coll} \sim 10$).
However, it may be
possible to circumvent this problem if we decrease the
satellitesimal Safronov growth time by accreting
``cold'' circumplanetary debris, and increase $p(R)$ by capturing
sufficiently small heliocentric planetesimals in low inclination orbits
and large Safronov parameter $\theta_1 >> 1$. It is also possible that
the satellitesimal cutoff size is instead set by the sizes of 
the largest captured
planetesimal fragments, or by a gravitational
instability (which would set an upper-size
cut-off of $r_2 \sim 10-100$ m, since the escape
speed from particles of that size would be sufficient for
stability), though we have not yet investigated
the implications of these possibilities. It seems likely that a low velocity
dispersion population of small planetesimal fragments, as might be expected
at the time of satellite formation, might be needed in any case. 
Perhaps the most promising region of parameter space is one in
which both the satellitesimal and planetesimal populations of interest
can be best described as collisional rubble, so that $r_2 < 100$ m
(as hinted at by some of our solutions),
but we leave investigation of that regime for future work. It
should be noted that such a regime would likely also involve different
$q$ values than were employed in this paper.

Much more work would need to be done to constrain this problem
to the point where it can be stated whether a planetesimal collisional
capture satellite formation model is viable. The arguments presented
here constitute only the outline of a promising model. For instance,
it is clearly an oversimplification to assume as we have
done here that collisions between
retrograde and prograde satellitesimals control the evolution of the
swarm in a timescale $\tau_{coll}$. Also, the capture
probability $p(R)$ does not explicitly treat the physics of
collisional fragmentation and may be unrealistic (we are currently
considering other possible choices based on pi-scaling theory
for the fractional mass captured
following a disruptive collision).
Furthermore, we
have avoided the issue of the contribution to the angular momentum
from regions of the disk (between $\sim 0.5 - 0.7 R_H$; \eg, Nesvorny \etal
2003) where
only retrograde orbits are stable\footnote{One would need to argue
that either the capture probability decreases outside $\sim R_H/2$ or
that most of this retrograde material is not delivered to the inner
disk where regular satellites form but rather escapes outward.} and
prograde objects are rapidly lost.
At this time, it is most useful to discuss how Cassini observations
can affect this model's prospects compared to that of
MEa,b (see also Appendix C). The main observational
objection that might be levelled against this model is that it
fails to account for the composition of Saturn's icy satellites.
In particular, the medium-sized Saturnian satellites have densities that
would seem to preclude a solar composition ice/rock ratio (this
is only marginally true for Titan itself), which
one might not expect from a model that derives the bulk of the
solids that make-up the regular satellites directly from heliocentric
orbit. This issue has been noted before in a number of
publications (\eg, Johnson \etal 1987; Lunine and Tittemore 1993; 
Podolak \etal 1993)
largely on the basis of a comparison between the composition of
regular satellites and objects thought to have formed in the outer
solar system such as Pluto-Charon (see, \eg, McKinnon \etal 1997) as
well as Triton (McKinnon and Mueller 1989). The recent Cassini flyby of
Phoebe has yielded a density of $1.6$ \gcmc for this retrograde, irregular
satellite (Porco \etal 2005), which indicates a rock to ice ratio of at least
$50 \%$ (Johnson and Lunine 2005), and reinforces the demarcation
between objects that formed in the outer solar nebula and the giant planet
subnebula.

Mosqueira and Estrada (2005) sketch a possible resolution of this
issue, involving a hypervelocity collision between Titan and a Triton-sized
differentiated
interloper, and the accretion of secondary icy moons out of a volatile-rich
disk formed in the aftermath of the collision. In this scenario Iapetus
may have been scattered by Titan to its present position, and Hyperion
may represent a remnant from the collision which became captured into
resonance and was thus prevented from crossing Titan's orbit and getting
either scattered or accreted by it (which was presumably the fate of
other objects forming in the region). If so, one would
expect an icy composition for both Iapetus and Hyperion.\footnote{In the
model of MEa,b one also expects an icy composition for Iapetus and Hyperion.
In this model satellitesimals forming in the outer disk
are icy (Mosqueira and Estrada 2005) and drift in a long timescale
due to gas drag. As they approach a
satellite they might get captured into resonances, and
collisions among them might grind them down and knock them out of
resonance. Thus, in this model too Hyperion might represent a
collisional remnant. Also, one might expect that Hyperion-like
objects might be typical impactors in the formation of Titan and Callisto.}
Regarding Titan itself, such an energetic impact would very likely
have led to a fully differentiated state (Tonks and Melosh 1992), but
there are a number of reasons to expect that Titan will turn out to
be fully differentiated.
Conversely, in the unlikely event that Cassini data indicated
that Titan like Callisto is only partially differentiated that would
argue {\it against} such an event, but in {\it favor} of
a planetesimal collisional capture model, since it would provide
evidence that the bulk of the material forming Titan was delivered over
a long timescale as we assume here.

It is also important to consider the consequences of this model to the
delivery of volatiles trapped in planetesimals to the atmosphere of
Titan. This delivery of trapped volatiles
is presumably also linked to the Galileo probe
observation that Jupiter's heavy volatile element global
abundances, excluding neon and oxygen, are about $3$ times solar
(Atreya \etal 1999; Mahaffy \etal 2000).
Given that to form the clathrate of Ar,
the model of
Gautier \etal (2001a,b) requires a very low solar nebula temperature
$T \sim 36$ K, which may be too cold for the location of
Jupiter even for passive disk models, 
and the temperature constraint to trap it in
amorphous ice is even lower $T \sim 25 K$ (Notesco and Bar-Nun 2005),
the observed enhacement
appears to argue that planetesimals
trapped argon {\it either} as clathrate or in amorphous ice in
cold regions of the disk before migrating to warmer
regions. If so, our planetesimal collisional capture formation model would
seem to imply Ar/N$_2$ $\sim 0.06$ Owen (2000) and that the molecular
nitrogen in Titan's atmosphere was delivered as trapped $N_2$. However,
before such a conclusion can be reached one must first gain an understanding of
the consequences of planetesimal collisional grinding
on trapped volatiles. Thus,
a planetesimal fragment capture model may be consistent with recent Cassini
observations of Titan (Flasar \etal 2005b; Niemann \etal 2005).

\vspace{0.2in}
\centerline{\bf Acknowledgements}
We thank Dr. Jeffrey N. Cuzzi for a thoughtful review of this manuscript and 
useful conversations during the course
of this work. We thank an anonymous reviewer for comments that
lead to improvements in the presentation of this paper.
This work is supported by a grant from PG\&G
and the National Research Council.

\newpage
\vspace{-4in}
\centerline{\psfig{figure=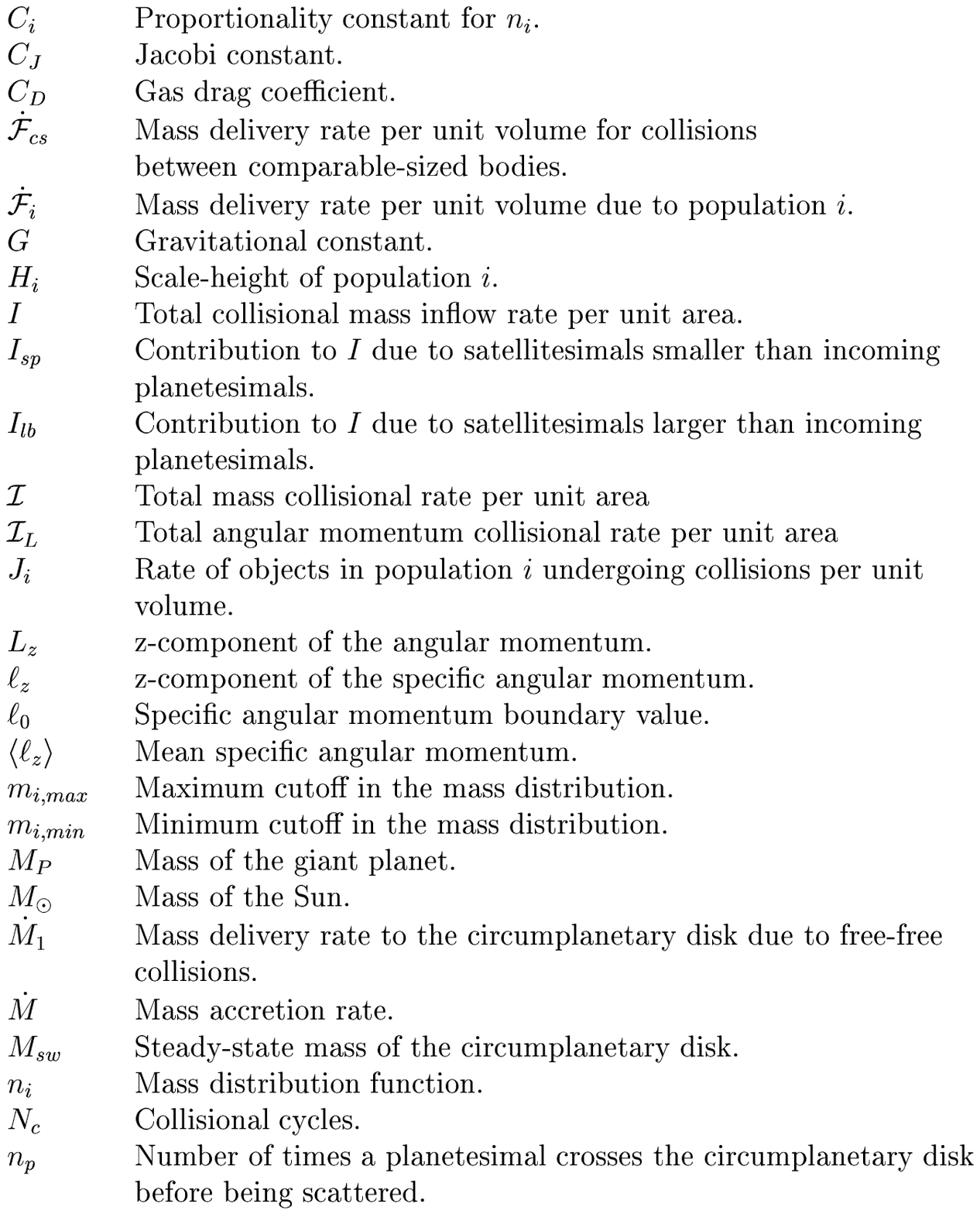}}

\newpage
\vspace{-4in}
\centerline{\psfig{figure=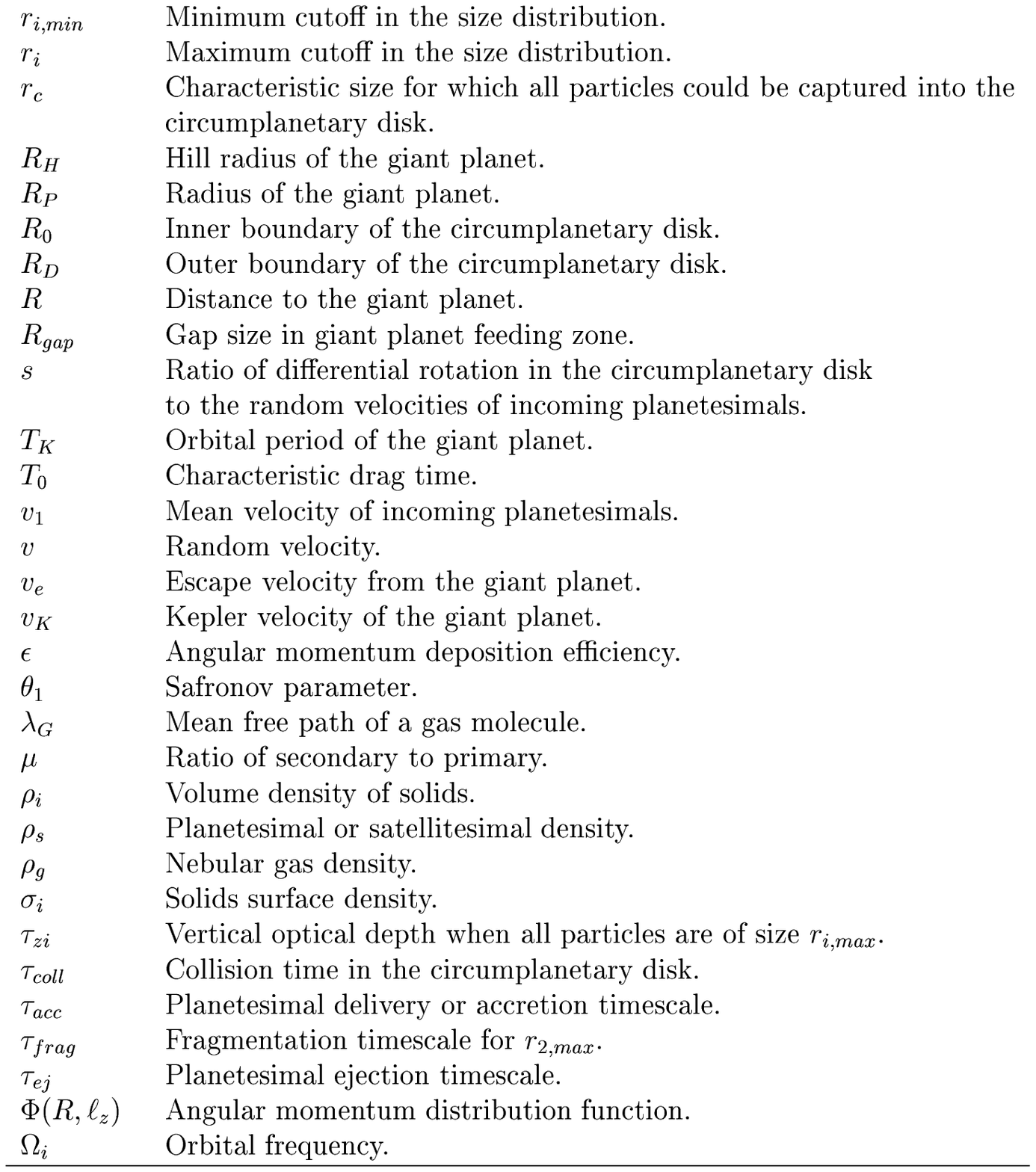}}

\newpage
\vspace{0.5in}
\centerline{\bf Appendix B: 4$^{\rm{th}}$ Order Expansion of $\Phi(R,\ell_z)$} 

In this appendix, we derive the 4th order expansions for the mass
and angular momentum collision rates per unit area which are described by the 
distribution $\Phi(R,\ell_z)$, where $\Phi(R,\ell_z)\,d\ell_z$ is the 
mass collision rate per unit area of planetesimals with $z$-component of 
specific angular momentum $\ell_z$
in the range $\ell_z$ and $\ell_z + d\ell_z$. Our treatment follows that of
DT93 for the case of weak gravity and high velocity dispersion. The mass 
collision rate per unit area ${\cal{I}}(R)$
and the angular momentum collision rate per unit area ${\cal{I}}_L(R)$ 
correspond to the zeroth and first moments

$$
{\cal{I}}(R) = \frac{1}{2\pi}\int \Phi(R,\ell_z)\,d\ell_z,\,\,\,{\cal{I}}_L(R) 
= \frac{1}{2\pi}\int \ell_z\Phi(R,\ell_z)\,d\ell_z,
\eqno{(\rm{B}1)}
$$

\noindent
where integration is over all angular momenta such that $|\ell_z| \geq \ell_0$.

In polar coordinates, the distribution funcion $\Phi(R,\ell_z)$ is given by
Eq. (\ref{equ:Phipolar}) where the quantity $\Psi$ is 

$$\frac{1}{2}\Psi \equiv \frac{v^2\cos{^2\psi}}{2v_1^2} + 
\frac{\left[v\sin{\psi} -
\Gamma\Omega_1 R\cos{(\chi + \psi)}\right]^2}{2v_1^2(1-\frac{1}{2}\Gamma)}
= \frac{1}{2}\Psi_0 - B_1 + B_2,
\eqno{(\rm{B}2)}$$

\noindent
where $\Psi_0 = (v/v_1)^2g(\psi)$ and $g(\psi)$ is defined in Eq. 
(\ref{equ:gpsi}), and $B_1$ and $B_2$ are given by

$$B_1 = \Gamma\frac{v}{v_1}\left(\frac{b}{s}\right)\frac{\sin{\psi}
\cos{(\chi+\psi)}}{1-\frac{1}{2}\Gamma},\,\,\,B_2 = \frac{1}{2}\Gamma^2\left(
\frac{b}{s}\right)^2\frac{\cos{^2(\chi+\psi)}}{1-\frac{1}{2}\Gamma}.
\eqno{(\rm{B}3)}$$

\noindent
We derive an expression for the distribution function $\Phi(R,\ell_z)
\propto e^{-\frac{1}{2}\Psi} = e^{-\frac{1}{2}\Psi_0}e^{B_1-B_2}$ 
by expanding the exponential in terms of $B_1$ and $B_2$ to fourth order 
in the quantity $b/s$.
Using the trigonometric identity $\cos{(\chi+\psi)} = \cos{\chi}\cos{\psi}-
\sin{\chi}\sin{\phi}$ and surpressing terms that are odd functions of $\sin$ 
and $\cos$ which will later vanish upon integration over $\psi$, the expansion
is

$$
\begin{array}{r}
e^{B_1-B_2} \approx 1 - 2f_1\frac{v}{v_1}\left(\frac{b}{s}\right)\sin{\chi} 
+ \left(\frac{b}{s}\right)^2\left[2f_1\left(\frac{v}{v_1}\right)^2 - 
\Gamma\right](f_{21}-2f_{22}\sin{^2\chi}) \\
- 2\left(\frac{b}{s}\right)^3\left[2f_1^2\left(\frac{v}{v_1}\right)^3 -
3\Gamma f_1\left(\frac{v}{v_1}\right)\right]
(f_{21}-f_{32}\sin{^2\chi})\sin{\chi} \\
+ \left(\frac{b}{s}\right)^4\left[\frac{2}{3}f_1^2\left(\frac{v}{v_1}\right)^4
-2\Gamma f_1\left(\frac{v}{v_1}\right)^2 + \frac{1}{2}\Gamma^2\right]
(f_{41}-2f_{42}\sin{^2\chi}+f_{43}\sin{^4\chi})
\end{array}, 
\eqno{(\rm{B}4)}
$$

\noindent
where the functions $f$ are given by

$$
\begin{array}{l}
f_1(g) = g(\psi) - 1,\,\,\,\,\,\,f_{21}(g) = \frac{1-\left(1-\frac{1}{2}\Gamma
\right)g}{1-\frac{1}{2}\Gamma}, \,\,\,\,\,\, 
f_{22}(g) = \frac{1 -\Gamma/4 - \left(1-\frac{1}{2}\Gamma\right)g}
{1-\frac{1}{2}\Gamma} \\
f_{32}(g) = \frac{1 + \frac{1}{3}\left(1-\frac{1}{2}\Gamma\right) - \frac{4}{3}
\left(1-\frac{1}{2}\Gamma\right)g}{1-\frac{1}{2}\Gamma},\,\,\,\,\,\,\,
f_{42}(g) = f_{41}(g) + \frac{3\left[1-
\left(2-\frac{1}{2}\Gamma\right)g + \left(1-\frac{1}{2}\right)g^2\right]}
{1-\frac{1}{2}\Gamma}, \\
f_{41}(g) = \frac{\left[1-\left(1-\frac{1}{2}\Gamma\right)g\right]^2}
{\left(1-\frac{1}{2}\Gamma\right)^2},\,\,\,\,\,\,
f_{43}(g) = f_{41}(g) +
\frac{6\left[1+\frac{1}{6}\left(1-\frac{1}{2}
\Gamma\right) - \frac{1}{3}(7-2\Gamma)g + \frac{7}{6}\left(1-\frac{1}{2}
\Gamma\right)g^2\right]}{1-\frac{1}{2}\Gamma}
\end{array}.
\eqno{(\rm{B}5)}
$$

\noindent
Carrying out the integration over $\chi$ where 
$\sin{\chi}=(\ell_z-\Omega_1 R^2)/vR$ leads to

$$
\Phi(R,\ell_z) = \frac{2\sigma_1\tau_{z2}}{R[2\pi\left(1-\frac{1}{2}\Gamma
\right)]^{1/2}P_1v_1^2v_z}
\int_{0}^{2\pi}d\psi \int_{|\ell_z-\Omega_1 R^2|/R}^{\infty} ve^{-\frac{1}{2}
\Psi_0} \times \,\,\,\,\,\,\,\,\,\,\,\,\,\,\,\,\,\,\,\,\,\,\,\,\,
$$
$$
\left[A_0 + A_2\left(\frac{v}{v_1}\right)^2 + A_4\left(\frac{v}{v_1}
\right)^4 + A_{-2}\left(\frac{v_1}{v}\right)^2 + A_{-4}\left(\frac{v_1}
{v}\right)^4\right]\,dv
\eqno{(\rm{B}6)}
$$

\noindent
where the $A_i$ are functions of the functions $f$, $\Gamma$, and 
$\alpha = (\ell_z-\Omega_1 R^2)/\sqrt{2}v_1R$ arrived at by collecting
terms with common velocity powers in Eq. (B4). This expression can then be 
integrated over the velocity to yield Eq. (\ref{equ:Phialp}), where the
coefficients $K_i$ are given by

$$
\begin{array}{l}
K_0 = 1 - \left[\Gamma f_{21}-\frac{4}{g}f_1f_{21}\right]\left(\frac{b}{s}
\right)^2 + \left[\frac{1}{2}\Gamma^2f_{41} - \frac{4\Gamma}{g}f_1f_{21}^2 +
\frac{16}{3g^2}f_1^2f_{41}\right]\left(\frac{b}{s}\right)^4, \\
K_1 = -2\sqrt{2}f_1\left(\frac{b}{s}\right) + \left[6\sqrt{2}\Gamma f_1f_{21} -
\frac{8\sqrt{2}}{g}f_1^2f_{21}\right]\left(\frac{b}{s}\right)^3,
\,\,\,\,K_2 = -4f_1(2f_{22}-f_{21})\left(\frac{b}{s}\right)^2 \\ 
+4\left[\Gamma f_1(2f_{42}-f_{41}) +
\frac{g}{8}\Gamma^2f_{43} - \frac{4}{3g}f_1^2(f_{42}-f_{41})\right]
\left(\frac{b}{s}\right)^4, \\
K_3 = 8\sqrt{2}f_1^2(f_{32}-f_{21})\left(\frac{b}{s}\right)^3,\,\,\,\,\,\,\,
K_4 = \frac{8}{3}f_1^2(f_{43}-2f_{42}+f_{41})\left(\frac{b}{s}\right)^4, \\
K_5 = 2g\Gamma f_{22}\left(\frac{b}{s}\right)^2 - g\Gamma^2f_{42}
\left(\frac{b}{s}\right)^4,\,\,\,\,\,\,K_6 = -6g\sqrt{2}\Gamma f_1f_{32}
\left(\frac{b}{s}\right)^3, \\
K_7 = -\left[4g\Gamma f_1f_{43} + \frac{g^2}{2}\Gamma^2 f_{43}\right]
\left(\frac{b}{s}\right)^4
\end{array}.
\eqno{(\rm{B}7)}
$$

The mass collision rate per unit area is then given by integration of 
$\Phi(R,\ell_z)$ over all $|\ell_z| \ge \ell_0$ and yields

$$
{\cal{I}}(R,\ell_0) = 
\frac{\sigma_1\tau_{z2}}{\pi\left(1-\frac{1}{2}\Gamma\right)^{1/2}P_1}
\frac{v_1}{v_z}
\int_{0}^{2\pi}\frac{d\psi}{[g(\psi)]^{3/2}} \sum_{i=0}^{7}
{\cal{M}}_i(g(\psi),R,\ell_0),
\eqno{(\rm{B}8)}
$$

\noindent 
with the ${\cal{M}}_i$ given by

$$
\begin{array}{l}
{\cal{M}}_0 = K_0h_0(\psi),\,\,\,{\cal{M}}_1 = \frac{K_1}{2(\pi g)^{1/2}}
\left(e^{-g\alpha^2_0} - e^{-g\alpha^{\prime 2}_0}\right) = K_1h_1(\psi), \\
{\cal{M}}_2 = K_2\left\{\frac{1}{2g}h_0(\psi) + \frac{1}{2(\pi g)^{1/2}}
\left(\alpha_0e^{-g\alpha^2_0} + \alpha^{\prime}_0e^{-g\alpha^{\prime 2}_0}
\right)\right\} = K_2\left\{\frac{1}{2g}h_0(\psi) + h_2(\psi)\right\}, \\
{\cal{M}}_3 = \frac{K_3}{2(\pi g)^{1/2}}\left\{\left(\alpha^2_0+\frac{1}{g}
\right)e^{-g\alpha^2_0} - \left(\alpha^{\prime 2}_0 + \frac{1}{g}\right)
e^{-g\alpha^{\prime 2}_0}\right\} = K_3h_3(\psi), \\
{\cal{M}}_4 = K_4\left\{\frac{3}{4g^2}h_0(\psi) + \frac{1}{2(\pi g)^{1/2}}
\left[\alpha_0\left(\alpha^2_0+\frac{3}{2g}\right)e^{-g\alpha^2_0} +
\alpha^{\prime}_0\left(\alpha^{\prime 2}_0+\frac{3}{2g}\right)e^{-g
\alpha^{\prime 2}_0}\right]\right\} \\
\,\,\,\,\,\, = K_4\left\{\frac{3}{4g^2}h_0(\psi) + h_4(\psi)\right\},\,\,\,
{\cal{M}}_5 = \frac{K_5}{3}\left\{\frac{1}{g}h_0(\psi) + 2h_2(\psi) + 
j_3(\psi)\right\}, \\
{\cal{M}}_6 = \frac{K_6}{4}\left\{2h_3(\psi) + j_4(\psi)\right\},\,\,\,
{\cal{M}}_7 = \frac{K_7}{5}\left\{\frac{3}{2g^2}h_0(\psi) + 2h_4(\psi) + 
j_5(\psi)\right\}
\end{array},
\eqno{(\rm{B}9)}
$$

\noindent
with $j_n(\psi) = (g/\pi)^{1/2}[\alpha^n_0{\rm{Ei}}(-g\alpha^2_0) +
(-1)^{n+1}\alpha^{\prime n}_0{\rm{Ei}}(-g\alpha^{\prime 2}_0)]$, where
the exponential integral ${\rm{Ei}}(-g\alpha^2) = -E_1(g\alpha^2)$, and

$$
h_0(\psi) = 1-\frac{{\rm{erf}}(g^{1/2}\alpha_0) + {\rm{erf}}(g^{1/2}
\alpha_0^{\prime})}{2}.
$$

\noindent
Equation (B8) can now be integrated numerically over the angle $\psi$. 
Likewise, 
the $z$-component angular momentum collision rate per unit area is

$$
\begin{array}{l}
{\cal{I}}_L(R,\ell_0) = 
\frac{\sigma_1\tau_{z2}\Omega_1R^2}{\pi\left(1-\frac{1}{2}\Gamma\right)^{1/2}
P_1}\frac{v_1}{v_z}\times \\
\left\{\int_{0}^{2\pi}\frac{d\psi}{[g(\psi)]^{3/2}} 
\sum_{i=0}^{7}{\cal{M}}_i(g(\psi),R,\ell_0) + \frac{\sqrt{2}v_1}{\Omega_1 R}
\int_{0}^{2\pi}\frac{d\psi}{[g(\psi)]^{3/2}} \sum_{i=0}^{7}
{\cal{L}}_i(g(\psi),R,\ell_0)\right\}
\end{array},
\eqno{(\rm{B}10)}
$$

\noindent
with the ${\cal{L}}_i$ given by

$$
\begin{array}{c}
{\cal{L}}_0 = \frac{K_0}{K_1}{\cal{M}}_1,\,\,\,{\cal{L}}_1 = \frac{K_1}
{K_2}{\cal{M}}_2,\,\,\,{\cal{L}}_2 = \frac{K_2}{K_3}{\cal{M}}_3,\,\,\,
{\cal{L}}_3 = \frac{K_3}{K_4}{\cal{M}}_4, \\
{\cal{L}}_4 = K_4h_5(\psi),\,\,\, 
{\cal{L}}_5 = \frac{K_5}{K_6}{\cal{M}}_6,\,\,\,{\cal{L}}_6 = \frac{K_6}{K_7}
{\cal{M}}_7,\\
{\cal{L}}_7 = \frac{K_7}{6}\left\{4h_5(\psi) + j_6(\psi)\right\} \\
\end{array},
\eqno{(\rm{B}11)}
$$
$$
\begin{array}{c}
h_5(\psi) = \frac{1}{2(\pi g)^{1/2}}\left\{\left[\left(\alpha^2_0 +
\frac{1}{g}\right)^2+\frac{1}{g^2}\right]e^{-g\alpha^2_0} - 
\left[\left(\alpha^{\prime 2}_0 + \frac{1}{g}\right)^2 + \frac{1}{g^2}\right]
e^{-g\alpha^{\prime 2}_0}\right\}
\end{array}.
$$

\noindent
Note that in the limit of $b/s << 1$, and $\alpha_0 = \alpha^{\prime}_0 = 0$,
the integrals over $\psi$ in  Eq. (B8) and (B10) are identical to those in
Eq. (44) and the expression for $\lambda = \left<\ell_z\right>/\Omega
R_P^2$ in Eq. (45) of DT93 (omitting the multiple-hits correction) 
with $R=R_P$ for the case 
of high dispersion and weak gravity.

\newpage
\vspace{-4in}
\centerline{\psfig{figure=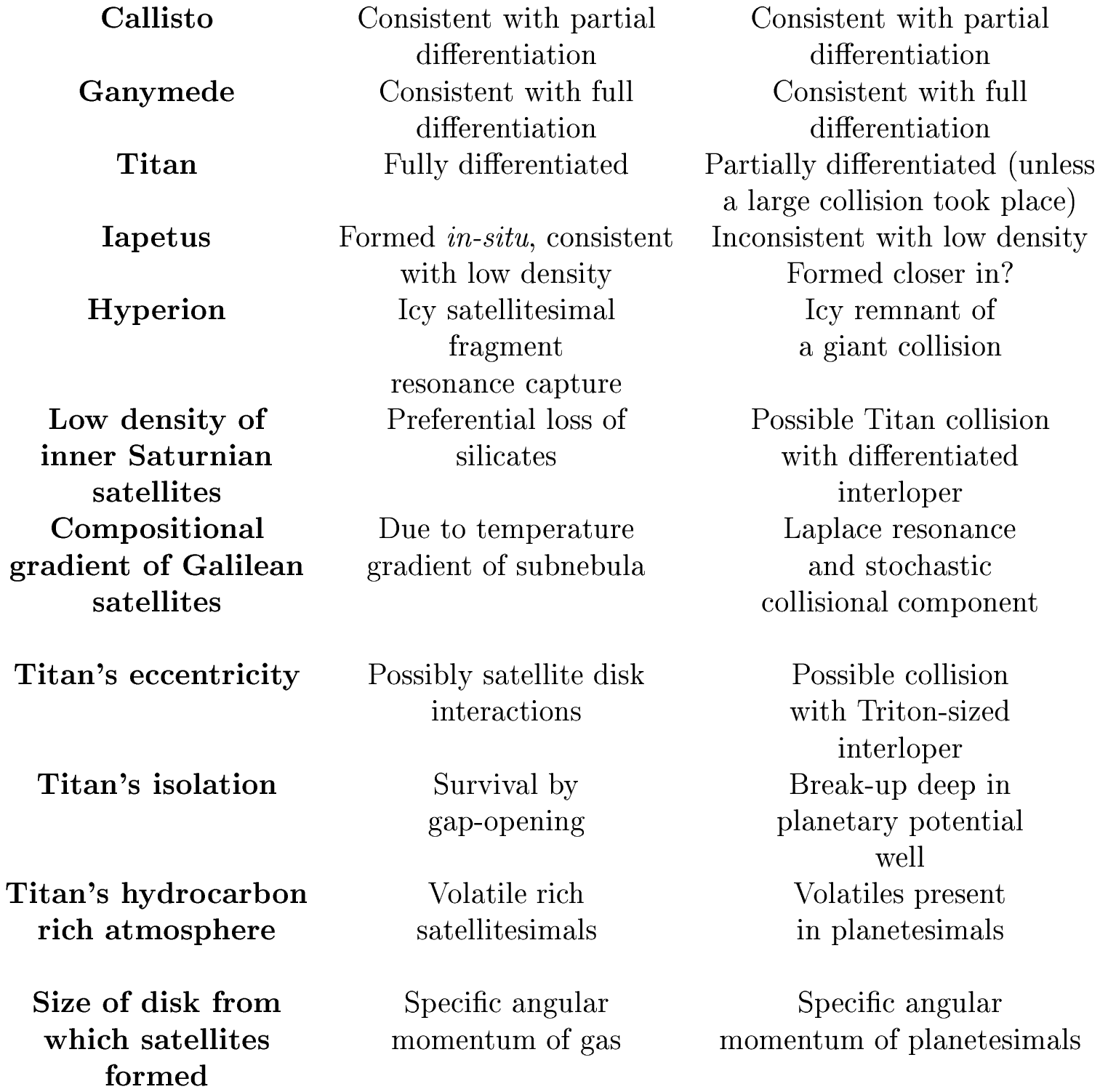}}

\newpage
\vspace{-4in}
\centerline{\psfig{figure=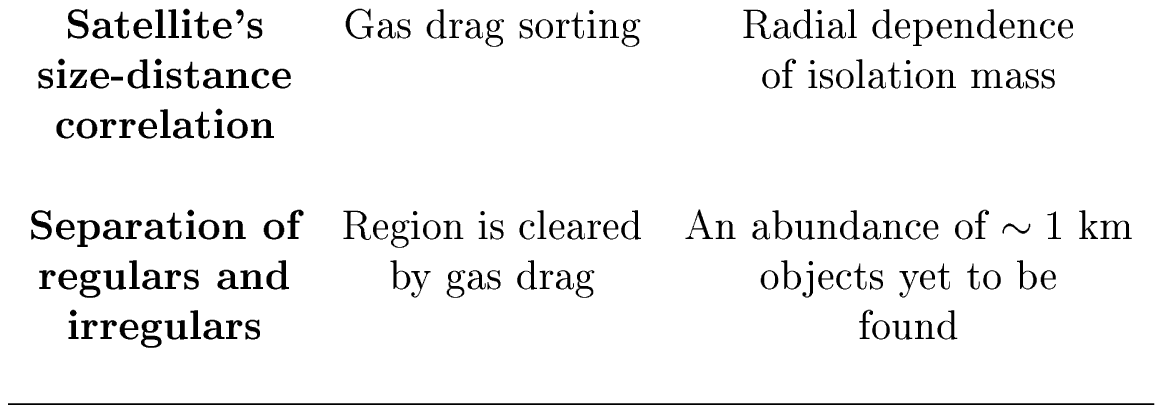}}

\newpage
\centerline{\bf References}

\refbook{Alfven, H., 1971}{Motion of small particles in the solar system.
In {\it Physical Studies of Minor Planets}, proceedings of IAU, colloq. 12
(ed. T. Gehrels), NASA SP 267, p. 315}
\refbook{Alibert, Y., Mousis, O., Benz, W., 2005}{Modeling the Jovian 
subnebula: I - Thermodynamical conditions and migration of proto-satellites.
arXiv:astro-ph/0505367.}
\science{Anderson, J. D., Schubert, G., Jacobson, R. A., Lau, E. L., Moore, 
W. B., Sjogren, W. L, 1998}{Distribution of rock, metals, and ices in 
Callisto}{280}{1573-1576}
\icarus{Anderson, J. D., Jacobson, R. A., McElrath, T. P., Moore, W. B.,
Schubert, G., Thomas, P. C., 2001}{Shape, mean radius, gravity field,
and interior structure of Callisto}{153}{157-161}
\science{Anderson, J. D., Johnson, T. V., Schubert, G., Asmar, S., Jacobson, 
R. A., Johnston, D., Lau, E. L., Lewis, G., Moore, W. B., Taylor, A., Thomas, 
P. C., Weinwurm, G., 2005}{Amalthea's density is less than that of water}
{308}{1291-1293}
\pss{Atreya, S. K., Wong, M. H., Owen, T. C., Mahaffy, P. R., Niemann, H. B., 
de Pater, I., Drossart, P., Encrenaz, T., 1999}{A comparison of the atmospheres
of Jupiter and Saturn: deep atmospheric composition, cloud structure, vertical 
mixing, and origin}{47}{1243-1262}
\rmp{Balbus, S. A., Hawley, J. F. 1998}{Instability, turbulence, and enhanced 
transport in accretion disks}{70}{1-53}
\mnras{Bate, M. R., Lubow, S. H., Ogilvie, G. I., Miller, K. A., 2003}
{Three-dimensional calculations of high- and low-mass planets embedded in 
protoplanetary discs}{341}{213-229}
\icarus{Benz, W., Slattery, W. L., Cameron, A. G. W., 1988}{Collisional 
stripping of Mercury's mantle}{74}{516-528}
\apj{Bryden, G., Chen, X., Lin, D. N. C., Nelson, R. P., Papaloizou, C. B. 
1999}{Tidally induced gap formation in protostellar disks: gap clearing and
supression of protoplanetary growth}{514}{344-367}
\apj{Bryden, G., Rozyczka, M., Lin, D. N. C., Bodenheimer, P., 2000} 
{On the interaction between protoplanets and protostellar disks}{540}
{1091-1101}
\aj{Canup, R. M., Ward, W. R., 2002}{Formation of the Galilean satellites:
conditions of accretion}{124}{3404-3423}
\icarus{Charnoz, S., Morbidelli, A., 2003}{Coupling dynamical and collisional 
evolution of small bodies: an application to the early ejection of 
planetesimals from the Jupiter-Saturn region}{166}{141-156}
\refbook{De, B. R., Alfven, H., Arrhenius, G., 1977}{The critical velocity 
phenomenon and the origin of the regular satellites. In {\it Planetary 
satellites}. Tucson, University of Arizona Press, 1977, p. 472-491.}
\icarus{Dones, L., Tremaine, S., 1993}{On the origin of planetary spins}
{103}{67-92}
\icarus{Dubrulle, B., 1993}{Differential rotation as a source of
angular momentum transport in the solar nebula}{106}{59-76}
\science{Flasar, F. M., Achterberg, R. K., Conrath, B. J., Pearl, J. C., 
Bjoraker, G. L., Jennings, D. E., Romani, P. N., Simon-Miller, A. A., 
Kunde, V. G., Nixon, C. A., Bezard, B., Orton, G. S., Spilker, L. J., 
Spencer, J. R., Irwin, P. G. J., Teanby, N. A., Owen, T. C., Brasunas, J., 
Segura, M. E., Carlson, R. C., Mamoutkine, A., Gierasch, P. J., Schinder, 
P. J., Showalter, M. R., Ferrari, C., Barucci, A., Courtin, R., Coustenis, 
A., Fouchet, T., Gautier, D., Lellouch, E., Marten, A., Prange, R., 
Strobel, D. F., Calcutt, S. B., Read, P. L., Taylor, F. W., Bowles, N., 
Samuelson, R. E., Abbas, M. M., Raulin, F., Ade, P., Edgington, S., 
Pilorz, S., Wallis, B., Wishnow, E. H., 2005a}
{Temperatures, winds, and composition in the Saturnian system}{307}{1247-1251}
\science{Flasar, F. M., Achterberg, R. K., Conrath, B. J., Gierasch, P. J., 
Kunde, V. G., Nixon, C. A., Bjoraker, G. L., Jennings, D. E., Romani, P. N., 
Simon-Miller, A. A., Bezard, B., Coustenis, A., Irwin, P. G. J., Teanby, N. A.,
Brasunas, J., Pearl, J. C., Segura, M. E., Carlson, R. C., Mamoutkine, A., 
Schinder, P. J., Barucci, A., Courtin, R., Fouchet, T., Gautier, D., Lellouch, 
E., Marten, A., Prange, R., Vinatier, S., Strobel, D. F., Calcutt, S. B., 
Read, P. L., Taylor, F. W., Bowles, N., Samuelson, R. E., Orton, G. S., 
Spilker, L. J., Owen, T. C., Spencer, J. R., Showalter, M. R., Ferrari, C., 
Abbas, M. M., Raulin, F., Edgington, S., Ade, P., Wishnow, E. H., 2005b}
{Titan's atmospheric temperatures, winds, and composition}{308}{975-978}
\apj{Gautier, D., Hersant, F., Mousis, O., Lunine, J. I., 2001a}
{Enrichments in volatiles in Jupiter: A new interpretation of the Galileo 
measurements}{550}{L227-L230}
\apj{Gautier, D., Hersant, F., Mousis, O., Lunine, J. I., 2001b}
{Erratum: Enrichments in volatiles in Jupiter: A new interpretation of the 
Galileo measurements}{559}{L183-L183}
\apj{Hartmann, L., Calvet, N., Gullbring, E., D'Alessio, P. 1998}
{Accretion and the evolution of T Tauri disks}
{495}{385}
\apj{Hawley, J. F., Balbus, S. A., Winters, W. F., 1999}
{Local hydrodynamic stability of accretion disks}
{518}{394-404}
\ptps{Hayashi, C. 1981}{Structure of the solar nebula, growth and decay of
magnetic fields, and effects of magnetic and turbulent viscosites on the
nebula}{70}{35-53}
\refbook{Hersant, F., Gautier, D., Lunine, J. I., Tobie, G., 2005}{Volatiles 
in Saturn and Titan: The new lights from Cassini-Huygens. 37th DPS meeting,
Cambridge, England, no. 29.09}
\pss{Hersant, F., Gautier, D., Lunine, J. I., 2004}{Enrichment in volatiles in i
the giant planets of the Solar System}{52}{623-641}
\refbook{Jacobson, R. A., Antreasian, P. G., Bordi, J. J., Criddle, K. E., 
Ionasescu, R., Jones, J. B., Mackenzie, R. A., Meek, M. C., Pelletier, F. J.,
Roth, D. C., Roundhill, I. M., Stauch, J. R., 2005}{The orbits of the major 
Saturnian satellites and the gravity field of the Saturnian system. 36th 
DDA meeting, Santa Barbara, CA}
\nature{Johnson, T. V., Lunine, J. I., 2005}{Density-derived constraints on 
the origin of Saturn's moon Phoebe}{435}{67-71}
\jgr{Johnson, T. V., Brown, R. H., Pollack, J. B., 1987}{Uranus satellites -
densities and composition}{92}{14884-14894}
\aj{Kenyon, S. J., Bromley, B. C., 2004}{The size distribution of Kuiper belt 
objects}{128}{1916-1926}
\apj{Klahr, H. H., Bodenheimer, P., 2003}{Turbulence in accretion disks:
Vorticity generation and angular momentum transport via the global baroclinic
instability}{582}{869}
\jgr{Kuramoto, K., Matsui, T., 1994}{Formation of a hot proto-atmosphere on 
the accreting giant icy satellite: Implications for the origin and evolution 
of Titan, Ganymede, and Callisto}{99}{21,183-21,200}
\apj{Li, H., J. M. Finn, R. V. E. Lovelace, Colgate, S. A., 2000}{Rossby
wave instability of thin accretion disks. II. Detailed linear theory}
{533}{1023-1034}
\icarus{Lissauer, J. J., Kary, D. M., 1991}{The origin of the systematic
component of planetary rotation. I - Planet on a circular orbit}
{94}{126-159}
\apj{Lubow, S. H., Seibert, M., Artymowicz, P., 1999}{Disk accretion onto
high-mass planets}{526}{1001}
\icarus{Lunine. J. I., Stevenson, D. J., 1982}{Formation of the Galilean 
satellites in a gaseous nebula}{52}{14-39}
\refbook{Lunine, J. I., Tittemore, W. C., 1993}{Origins of outer-planet
satellites. In {\it Protostars and Planets III}, University of Arizona
Press, Tucson, p. 1149-1176}
\refbook{Lunine, J. I., Owen, T. C., Brown, R. H., 2000}{The outer solar 
system: Chemical constraints at low temperatures on planet formation. In
{\it Protostars and Planets IV}, University of Arizona Press, Tucson, p. 1055}
\jgr{Mahaffy, P. R., Niemann, H. B., Alpert, A., Atreya, S. K., Demick, J., 
Donahue, T. M., Harpold, D. N., Owen, T. C., 2000}{Noble gas abundance and 
isotope ratios in the atmosphere of Jupiter from the Galileo Probe Mass 
Spectrometer}{105}{15061-15072}
\ssr{Makalkin, A. B., Dorofeeva, V. A., Ruskol, E. L., 1999}{Modeling the
protosatellite circum-jovian accretion disk: An estimate of basic parameters}
{33}{456}
\jgres{McKinnon, W. B., Mueller, S., 1989}{The density of Triton - A 
prediction}{16}{591-594}
\refbook{McKinnon, W. B., Simonelli, D. P., Schubert, G., 1997}{Composition, 
internal structure, and thermal evolution of Pluto and Charon. In {\it Pluto
and Charon}, University of Arizona Press, Tucson, p. 295}
\refbook{Mosqueira, I., Estrada, P. R., Chambers, J. E., 2000}
{Satellitesimal feeding and the formation of the regular satellites.
American Astronomical Society, 32nd DPS meeting}
\refbook{Mosqueira, I., Estrada, P. R., Cuzzi, J. N., Squyres, S. W., 2001}
{Circumjovian disk clearing after gap-opening and the formation of a partially 
differentiated Callisto. 32nd LPSC meeting, March 12-16, 2001, Houston, Texas, 
no. 1989}
\refbook{Mosqueira, I., Kassinos, S., Shariff, K., Cuzzi, J. N., 2003}
{Hydrodynamical shear instability in accretion disks? American Astronomical
Society, 35th DPS meeting 35, 25.05}
\icarus{Mosqueira, I., Estrada, P. R. 2003a}{Formation of the regular 
satellites of giant planets in an extended gaseous nebula I: subnebula model
and accretion of satellites}{163}{198-231}
\icarus{Mosqueira, I., Estrada, P. R. 2003b}{Formation of the regular 
satellites of giant planets in an extended gaseous nebula II: satellite
migration and survival}{163}{232-255}
\refbook{Mosqueira, I., Estrada, P. R., 2005}{On the origin of the Saturnian 
satellite system: Did Iapetus form in-situ? 36th LPSC meeting, Houston, TX,
no. 1951}
\pss{Mousis, O., Gautier, D., 2004}{Constraints on the presence of volatiles 
in Ganymede and Callisto from an evolutionary turbulent model of the Jovian 
subnebula}{52}{361-370}
\icarus{Mousis, O., Gautier, D., Bockelée-Morvan, D., 2002}{An evolutionary 
turbulent model of Saturn's subnebula: Implications for the origin of the 
atmosphere of Titan}{156}{162-175}
\aj{Nesvorny, D., Alvarellos, J. L. A., Dones, L., Levison, H. F., 2003}
{Orbital and collisional evolution of the irregular satellites}{126}{398-429}
\refbook{Niemann, H. B., Atreya, S. K., Bauer, S. J., Carignan, G. R.,
Demick-Montelara, J. E., Frost, R. L., Gautier, D., Haberman, J. A.,
Harpold, D. N., Hunten, D. M., Israel, G., Lunine, J. I., Kasprzak, W. T.,
Owen, T. C., Paulkovich, M., Raulin, F. C., Raaen, E., Way, S. H., 2005}
{Results from the gas chromatograph mass spectrometer (GCMS) experiment
in the Cassini-Huygens probe. Proceedings of the Fall AGU, San Francisco, CA,
December 2005, no. P43B-03}
\refbook{Notesco, G., Bar-Nun, A, 2005}{A $\sim 25$ K temperature of formation 
for the submicron ice grains which formed comets. {\it Science}, in press} 
\icarus{Ohtsuki, K., Ida, S., 1998}{Planetary rotation by accretion of 
planetesimals with nonuniform spatial distribution formed by the planet's 
gravitational perturbation}{131}{393-420}
\pss{Owen, T. S., 2000}{On the origin of Titan's atmosphere}{48}{747-752}
\aj{Pan, M., Sari, R., 2004}{A generalization of the Lagrangian points: 
Studies of resonance for highly eccentric orbits}{128}{1418-1429}
\apj{Papaloizou, J. C. B., Terquem, C., 1999}{Critical protoplanetary Core 
masses in protoplanetary disks and the formation of short-period giant planets}
{521}{823-838}
\refbook{Podolak, M., Hubbard, W. B., Pollack, J. B., 1993}{Gaseous accretion 
and the formation of giant planets. In {\it Protostars and Planets III},
University of Arizona Press, Tucson, p. 1109-1147.}
\icarus{Pollack, J. B., Grossman, A. S., Moore, R., Graboske Jr., H. C., 1976}
{The formation of Saturn's satellites and rings as influenced by Saturn's
contraction history}{29}{35-48}
\refbook{Pollack, J. B., Lunine, J. I., Tittemore, W. C., 1991}{Origin of the 
Uranian satellites. In {\it Uranus}. Tucson, University of Arizona Press, 1991,
p. 469-512.}
\science{Porco, C. C., Baker, E., Barbara, J., Beurle, K., Brahic, A., 
Burns, J. A., Charnoz, S., Cooper, N., Dawson, D. D., Del Genio, A. D., 
Denk, T., Dones, L., Dyudina, U., Evans, M. W., Giese, B., Grazier, K., 
Helfenstein, P., Ingersoll, A. P., Jacobson, R. A., Johnson, T. V., McEwen, A.,
Murray, C. D., Neukum, G., Owen, W. M., Perry, J., Roatsch, T., Spitale, J., 
Squyres, S., Thomas, P. C., Tiscareno, M., Turtle, E., Vasavada, A. R., 
Veverka, J., Wagner, R., West, R., 2005}{Cassini imaging science: Initial 
results on Phoebe and Iapetus}{307}{1237-1242}
\apj{Rafikov, R. R., 2002}{Planet migration and gap formation by tidally 
induced shocks}{572}{566}
\aanda{Richard, D., Zahn, J. P., 1999}{Turbulence in differentially rotating
flows. What can be learned from the Couette-Taylor experiment} 
{347}{734-738}
\aanda{Rudiger, G., Arlt, R., Shalybkov, D., 2002}{Hydrodynamic stability in 
accretion disks under the combined influence of shear and density 
stratification}{391}{781-787}
\refbook{Ruskol, E. L., 1975}{Origin of the Moon. NASA Transl. into english of 
the book ``Proiskhozhdeniye Luny'' Moscow, Nauka Press, 1975 p. 1-188}
\refbook{Ruskol, E. L., 1981}{Formation of planets. In ESA {\it The Solar
System and its Exploration}, p. 107-113 (SEE N82-26087 16-88)}
\refbook{Ruskol, E. L., 1982}{Origin of planetary satellites. {\it Izves. 
Earth Phys.} {\bf 18}, 425-433}
\refbook{Safronov, V. S., Ruskol, E. L., 1977}{The accumulation of satellites.
In {\it Planetary Satellites}. Tucson, University of Arizona Press, 1977, p. 
501-512}
\refbook{Safronov, V. S., Pechernikova, G. V., Ruskol, E. L., Vitiazev, A. V.,
1986}{Protosatellite Swarms. In {\it Satellites} Tucson, AZ, University of 
Arizona Press, 1986, p. 89-116}
\refbook{Schmidt, O. Yu, 1957}{Four lectures on the theory of the origin of the
Earth. 3rd edition, Moscow, NA SSSR, press, 1957}
\zamm{Schultz-Grunow, F., 1959}{On the stability of Couette flow}{39}{101}
\aanda{Shakura, N. I., Sunyaev, R. A., 1973}{Black holes in binary systems.
Observational appearance}{24}{337-355}
\refbook{Shoemaker, E. M. 1984}{Kuiper Prize Lecture, 16th DPS Meeting,
Kona, HI}
\nature{Stern, S. A., Weissman, P. R., 2001}
{Rapid collisional evolution of comets during the formation of the Oort cloud}
{409}{589-591}
\refbook{Stevenson, D. J., A. W. Harris, Lunine, J. I., 1986}{Origins of
satellites. In {\it Satellites} (J. A. Burns and M. S. Matthews, Eds.)
Univ. of Arizona Press, Tucson}
\icarus{Tanaka, H., Ida, S., 1997}{Distribution of planetesimals around a
protoplanet in the nebula gas II. Numerical simulations}{125}{302-316}
\icarus{Tonks, W. B., Melosh, H. J., 1992}{Core formation by giant impacts}
{100}{326-346}
\mnras{Weidenschilling, S. J., 1977}{Aerodynamics of solid bodies in the solar
nebula}{180}{57-70}
\icarus{Weidenschilling, S. J., 1997}{The origin of comets in the solar 
nebula: A unified model}{127}{290-306}
\refbook{Zahn, J. P., 1991}{On the nature of disk viscosity. Proceedings of
the IAU Colloq. 129, Structure and Emission Properties of Accretion
Disks (C. Bertout, S. Collin-Souffrin, and J. P. Lasota, Eds.),
Gif-sur-Yvette: Eds. Frontieres, 1991, p.87}

\newpage
\noindent
{\bf Figure 1}: Plot of the satellitesimal cutoff size $r_2$
(solids line) and solids surface density
$\sigma_2$ (dotted line)
as a function of $r_1$ for a case in which the solar nebula
is ten times minimum mass of solids, a planetesimal feeding
time $\tau_{acc} = 10^6$ years, and $\theta_1 = 4$.
The extremes of $N_c$ are indicated.

\vspace{0.25in}

\noindent
{\bf Figure 2}: Plot of the optical depth $\tau_{z2}$ (solid line) and 
collisional time $\tau_{coll}$ (dotted line)
as a function of $r_1$ for a case in which the solar nebula is
assumed to have ten times the minimum mass of solids, a planetesimal feeding
time of $\tau_{acc} = 10^6$ years, and $\theta_1 = 4$.
The extremes of $N_c$ are indicated.

\vspace{0.25in}

\noindent
{\bf Figure 3}: Same as fig. 1, except that the solids surface
density in the solar nebula is two times minimum mass and
$\theta_1 = 75$ (characteristic of a colder planetesimal
population). In this case, the range of values for $r_1$ are
somewhat broader, but significantly narrower for $r_2$.

\vspace{0.25in}

\noindent
{\bf Figure 4}: Same as fig. 2, except that the solids surface density
in the solar nebula is two times minimum mass and
$\theta_1 = 75$.
About $N_c \sim 5$ collisional cycles are needed to complete
the accretion of the satellites in the case of
$r_1 = 10$ km and $r_2 = 1$ km.

\vspace{0.25in}

\noindent
{\bf Figure 5}: Plot of the mass accreted over $10^6$ years as a function of 
the planetesimal Safronov parameter $\theta_1$. For these curves
$\tau_{z2} \sim 5\times 10^{-6}$, $r_1 = 10$ km, and $r_2 = 1$ km.

\vspace{0.25in}

\noindent
{\bf Figure 6}: The steady-state disk mass $M_{sw}$ as a function of collisional
cycles $N_c$ for $\tau_{acc} = 10^6$ years. The mass $M_{sw}$ is removed
to the inner disk each $\tau_{coll}$ so this mass must be replenished each
cycle.

\vspace{0.25in}

\noindent
{\bf Figure 7}: Mass (dotted lines) and
angular momentum (solid lines) in units of the mass and angular momentum
of the Galilean satellites ($M_{sats}\sim 4\times 10^{26}$ g,
$L_{z,sats}\sim 4\times 10^{43}$
g cm$^2$ s$^{-1}$)
accreted by the circumplanetary disk
in $\tau_{acc} = 10^6$ years
as a function
of the gap size $R_{gap} = \sqrt{\ell_0/\Omega_1}$.
For this case $N_c = 10$, implying a steady-state disk mass is
$M_{sw} = 0.1 M_{sats}$.
The proper budget of mass and
angular momentum to account for the Galilean satellites is achieved
where the solid and
the dotted lines first
intersect at $L_z/L_{z,sats} = M/M_{sats} = 1$. For example,
for the case of $\sigma_1 = 4.5\sigma_{\rm{MM}}$, $R_{gap} \sim 1 R_H$.

\vspace{0.25in}

\noindent
{\bf Figure 8}: Specific angular momentum delivered in units of the specific
angular momentum of the Galilean satellites $\left<\ell_z\right>_{sats}
\sim 1\times 10^{17}$ cm$^2$
s$^{-1}$ as a function of gap size $R_{gap} = \sqrt{\ell_0/\Omega_1}$.
These curves correspond to cases such that
$\left<\ell_z\right>/\left<\ell_z\right>_{sats} = 1$ matches the
mass and angular momentum budget of the
Galilean satellite system for $R_{gap} \sim 0.5 - 1.5 R_H$.

\vspace{0.25in}

\noindent
{\bf Figure 9}: Specific angular momentum delivered in units of the specific
angular momentum of the Galilean satellites $\left<\ell_z\right>_{sats}
\sim 1\times 10^{17}$ cm$^2$
s$^{-1}$ as a function of circumplanetary disk size $R_D$ for three different
values of the
gap size $R_{gap} = \sqrt{\ell_0/\Omega_1}$, $R_{gap}
= R_H$ (solid line), $R_{gap} = 0.5 R_H$ (dotted line), and
$R_{gap} = 0$ (dashed line).

\vspace{0.25in}

\noindent
{\bf Figure 10}: Mass (dotted lines) and
angular momentum (solid lines) in units of the mass and angular momentum
of the Galilean satellites ($M_{sats}\sim 4\times 10^{26}$ g,
$L_{z,sats}\sim 4\times 10^{43}$
g cm$^2$ s$^{-1}$)
accreted by the circumplanetary disk
in $\tau_{acc} = 10^6$ years
as a function
of the gap size $R_{gap} = 5 R_0$.
For this case $N_c = 10$, implying a steady-state disk mass is
$M_{sw} = 0.1 M_{sats}$.
The proper budget of mass and
angular momentum to account for the Galilean satellites is achieved
for the cases of $\epsilon = 0.5$ and $\epsilon = 0.8$
where the solid and
the dotted lines first intersect at $L_z/L_{z,sats} = M/M_{sats} = 1$.
No solution was found for the case $\epsilon = 0.2$.

\vspace{0.25in}

\noindent
{\bf Figure 11}: Specific angular momentum delivered in units of the specific
angular momentum of the Galilean satellites $\left<\ell_z\right>_{sats}
\sim 1\times 10^{17}$ cm$^2$
s$^{-1}$ as a function of gap size $R_{gap} = 5 R_0$.
The curves for $\epsilon = 0.5$ and $\epsilon = 0.8$ correspond to cases
such that
$\left<\ell_z\right>/\left<\ell_z\right>_{sats} = 1$ matches the
mass and angular momentum budget of the
Galilean satellite system for $R_{gap} \sim 0.5 - 1.5 R_H$. 
No solution was found for the case $\epsilon = 0.2$.

\newpage
\begin{figure}
\centerline{\psfig{figure=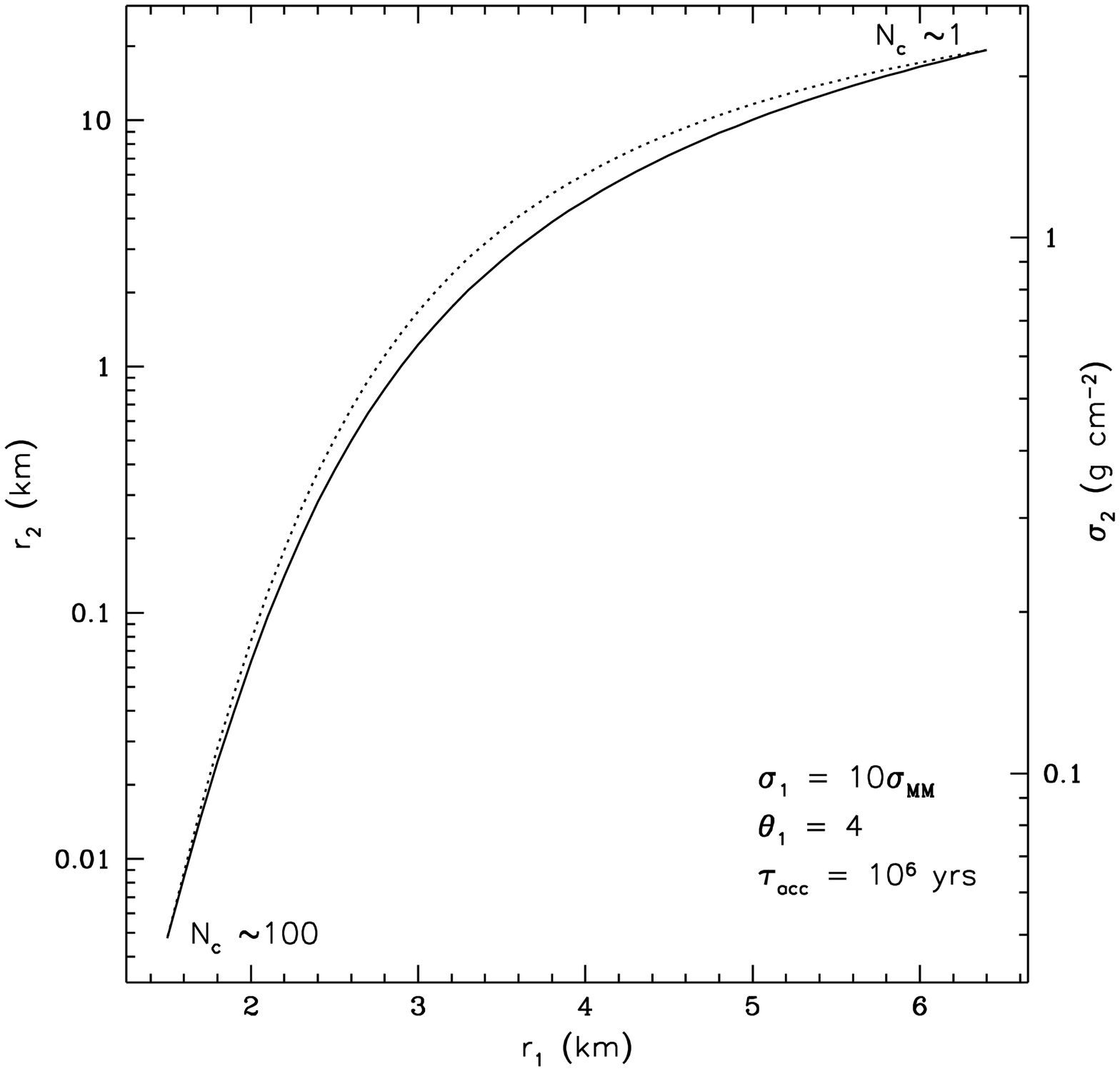,width=7.5in,height=7.5in}}
\begin{center}
Figure 1: Estrada and Mosqueira
\end{center}
\end{figure}

\newpage
\begin{figure}
\centerline{\psfig{figure=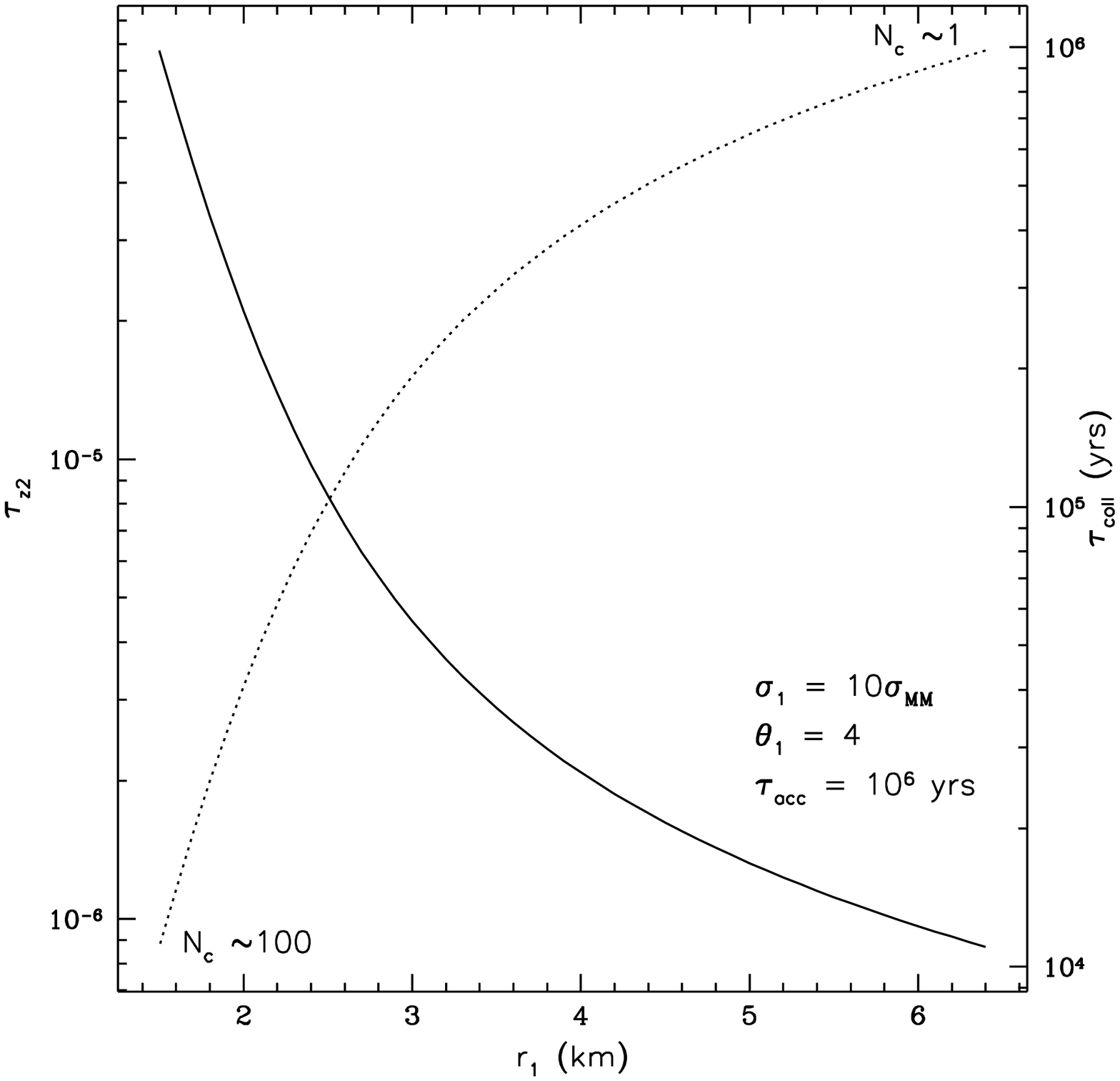,width=7.5in,height=7.5in}}
\begin{center}
Figure 2: Estrada and Mosqueira
\end{center}
\end{figure}

\newpage
\begin{figure}
\centerline{\psfig{figure=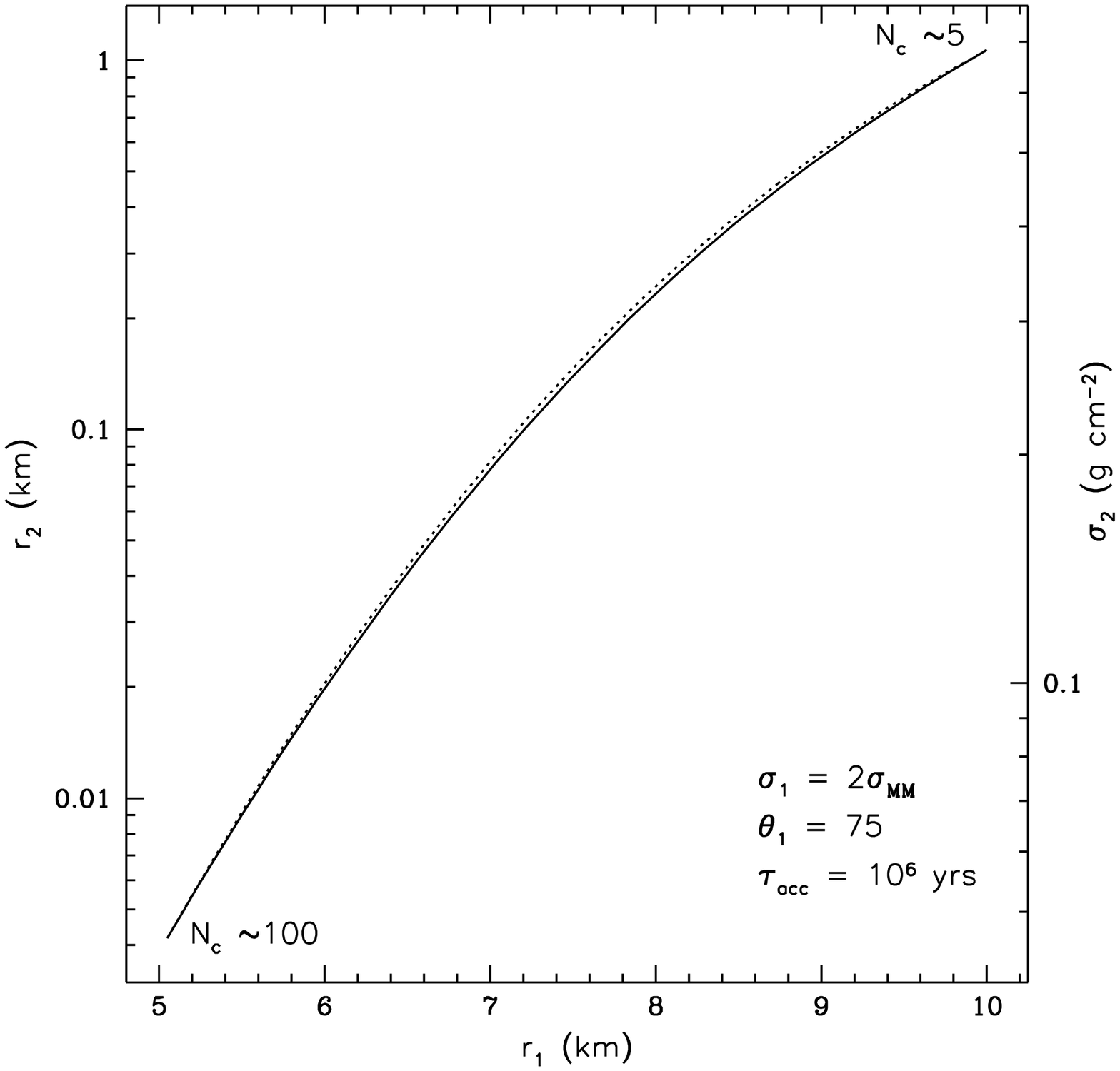,width=7.5in,height=7.5in}}
\begin{center}
Figure 3: Estrada and Mosqueira
\end{center}
\end{figure}

\newpage
\begin{figure}
\centerline{\psfig{figure=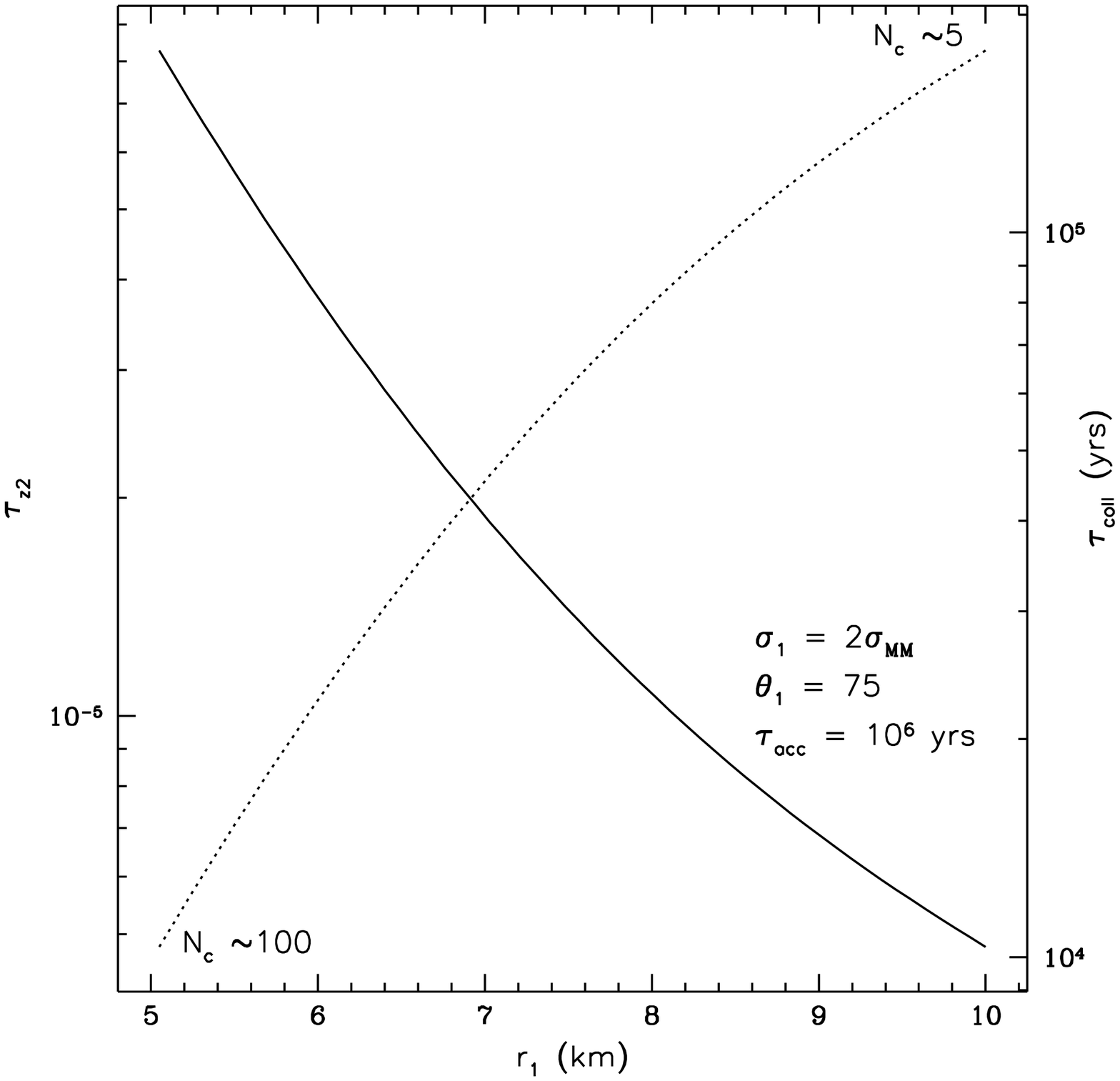,width=7.5in,height=7.5in}}
\begin{center}
Figure 4: Estrada and Mosqueira
\end{center}
\end{figure}

\newpage
\begin{figure}
\centerline{\psfig{figure=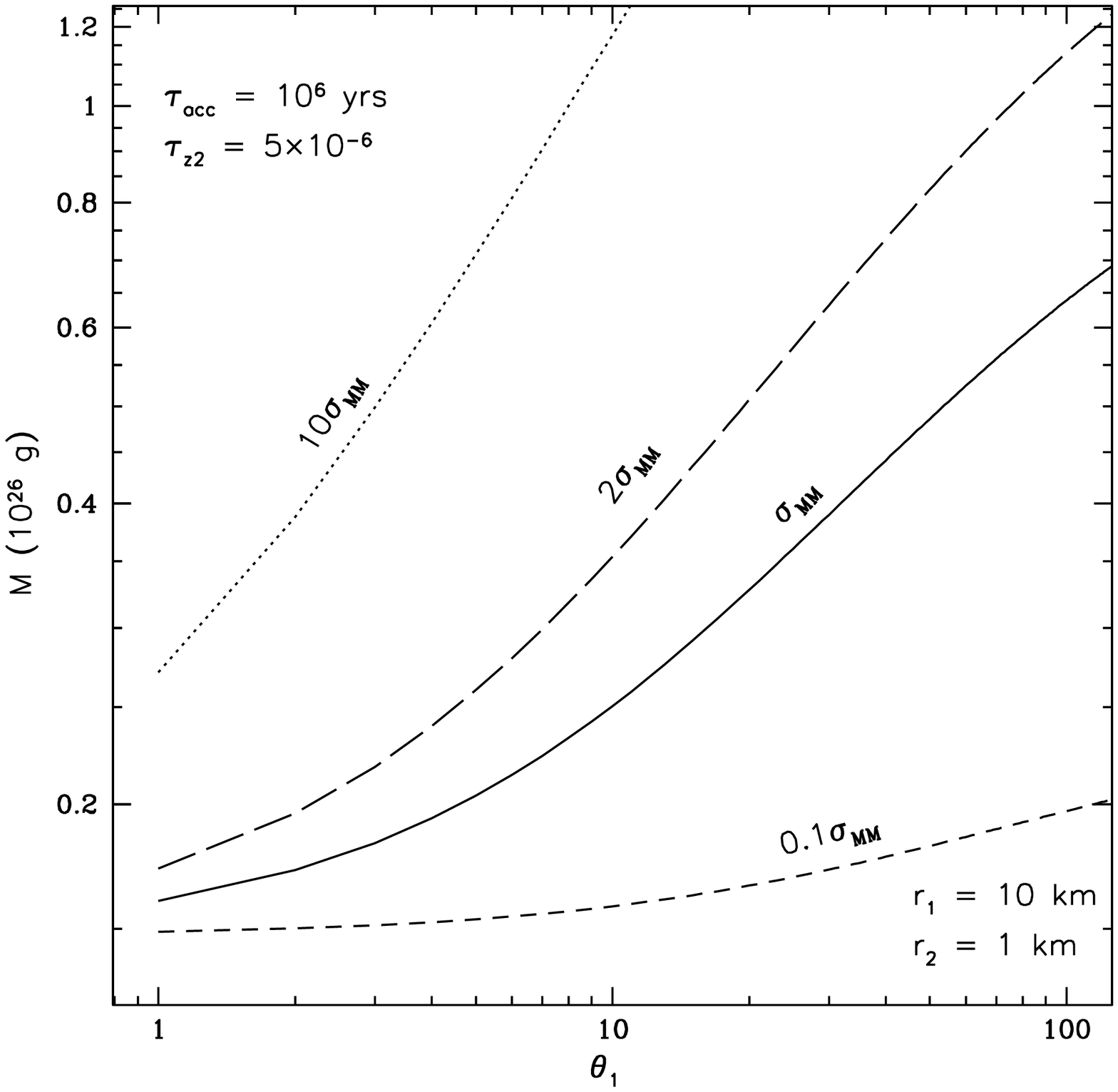,width=7.5in,height=7.5in}}
\begin{center}
Figure 5: Estrada and Mosqueira
\end{center}
\end{figure}

\newpage
\begin{figure}
\centerline{\psfig{figure=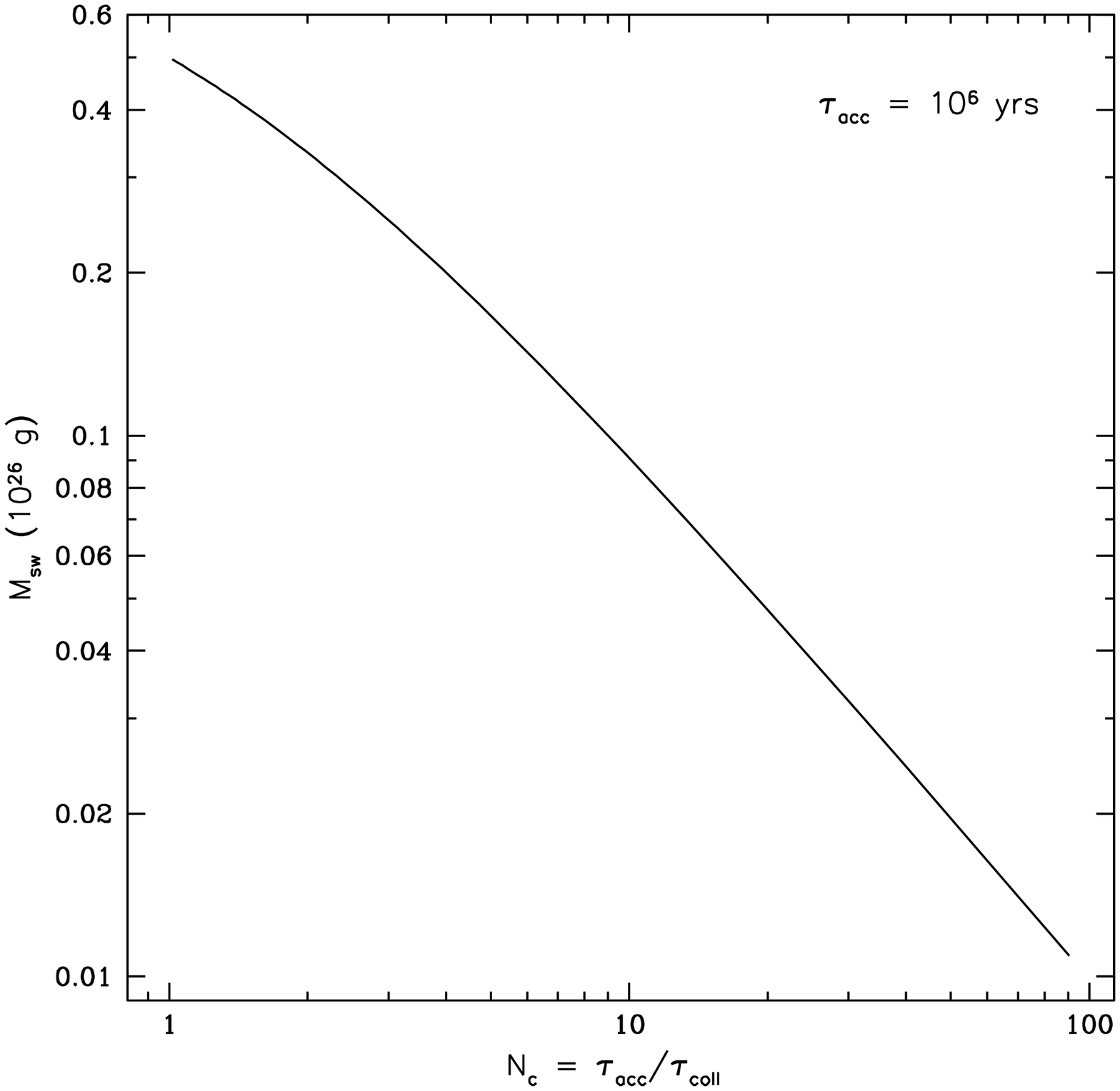,width=7.5in,height=7.5in}}
\begin{center}
Figure 6: Estrada and Mosqueira
\end{center}
\end{figure}

\newpage
\begin{figure}
\centerline{\psfig{figure=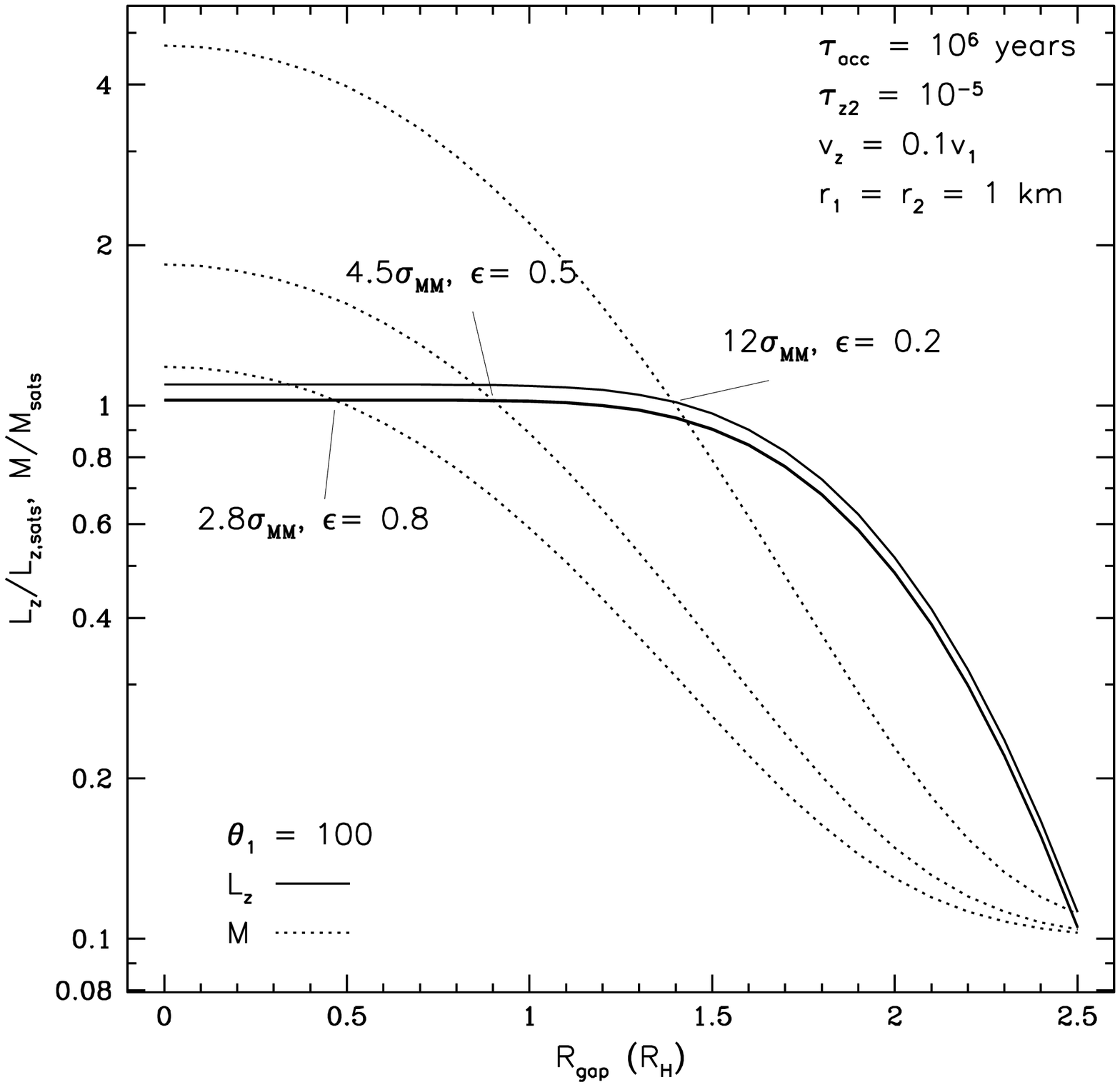,width=7.5in,height=7.5in}}
\begin{center}
Figure 7: Estrada and Mosqueira
\end{center}
\end{figure}

\newpage
\begin{figure}
\centerline{\psfig{figure=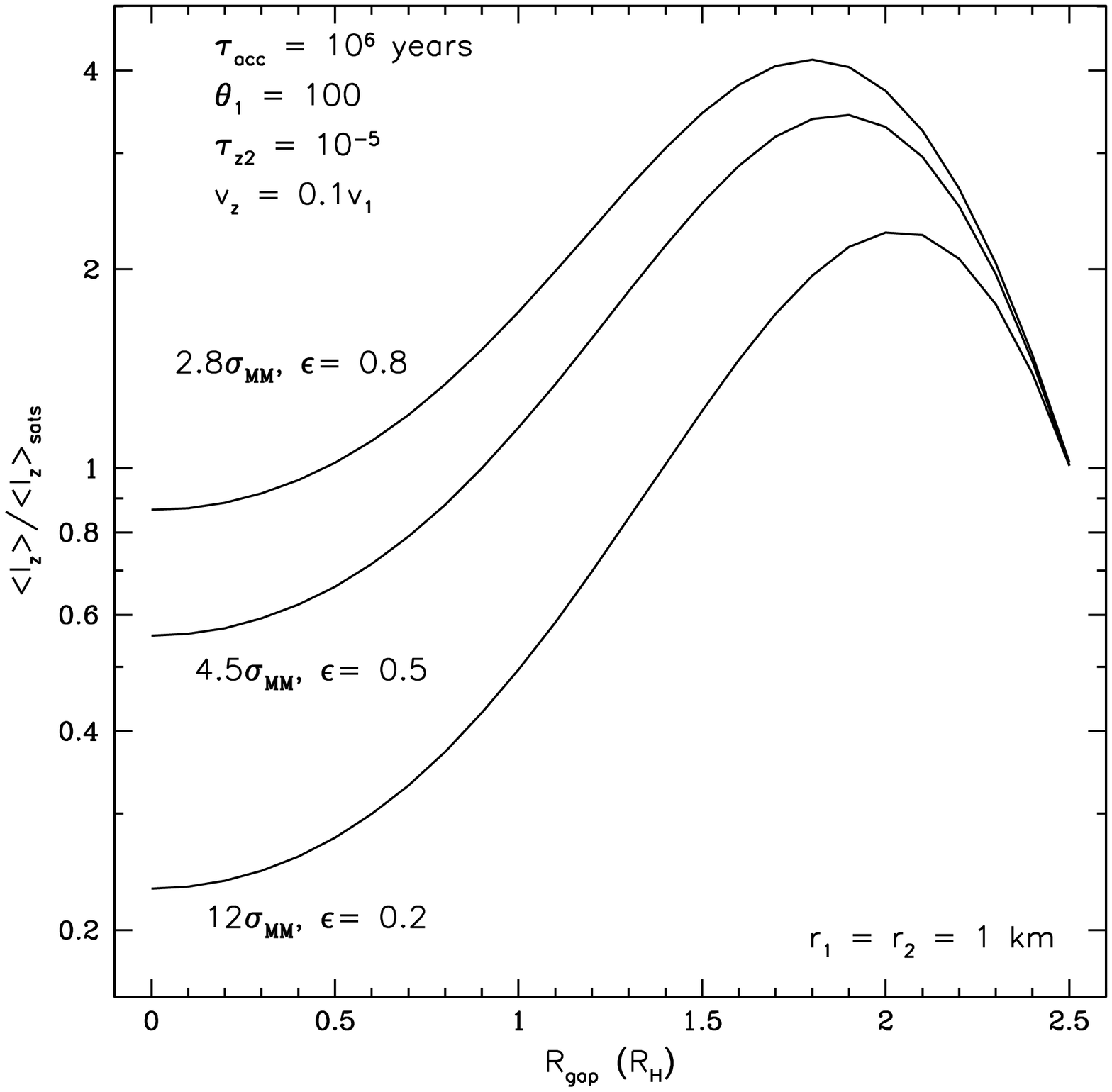,width=7.5in,height=7.5in}}
\begin{center}
Figure 8: Estrada and Mosqueira
\end{center}
\end{figure}

\newpage
\begin{figure}
\centerline{\psfig{figure=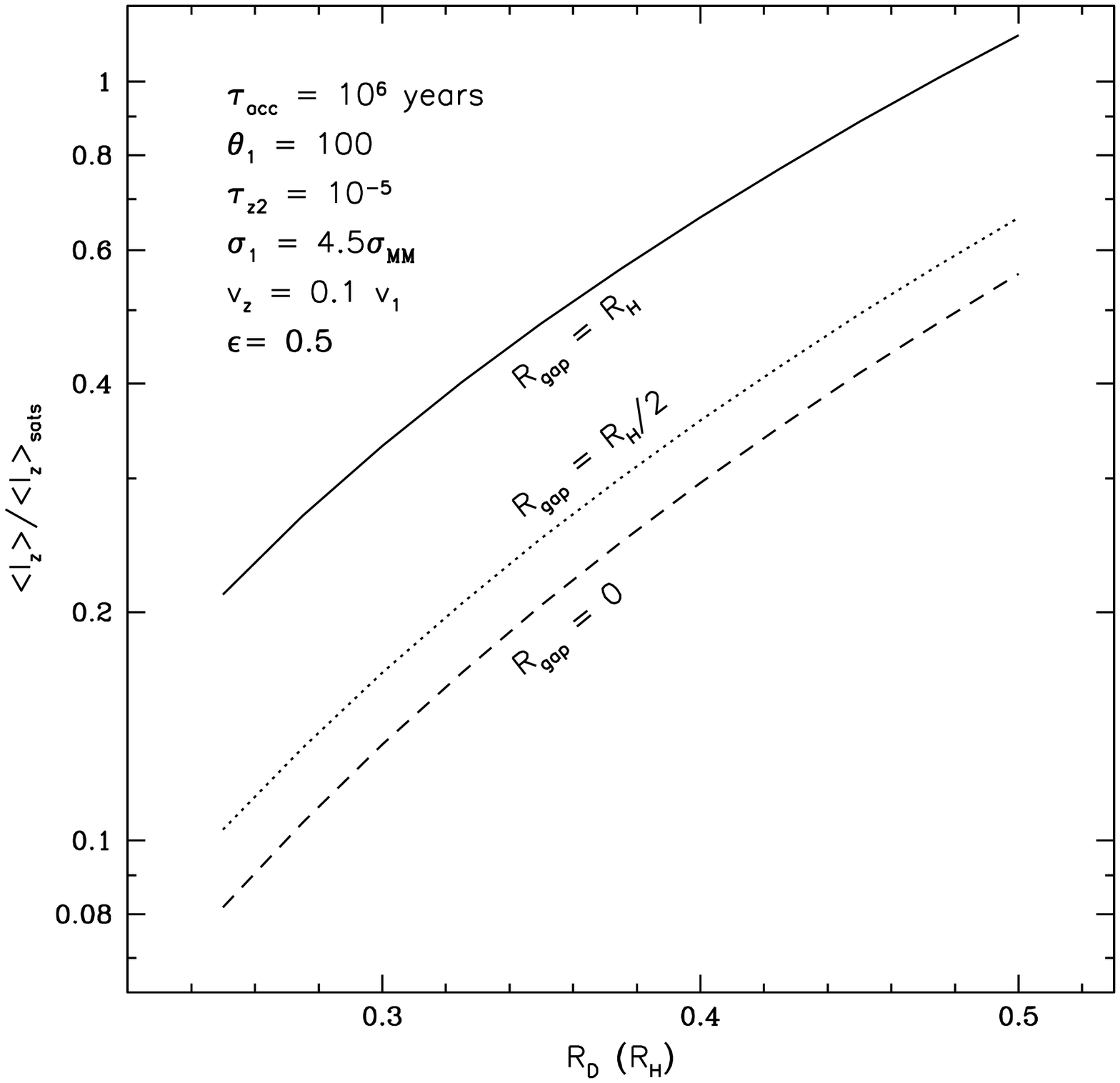,width=7.5in,height=7.5in}}
\begin{center}
Figure 9: Estrada and Mosqueira
\end{center}
\end{figure}

\newpage
\begin{figure}
\centerline{\psfig{figure=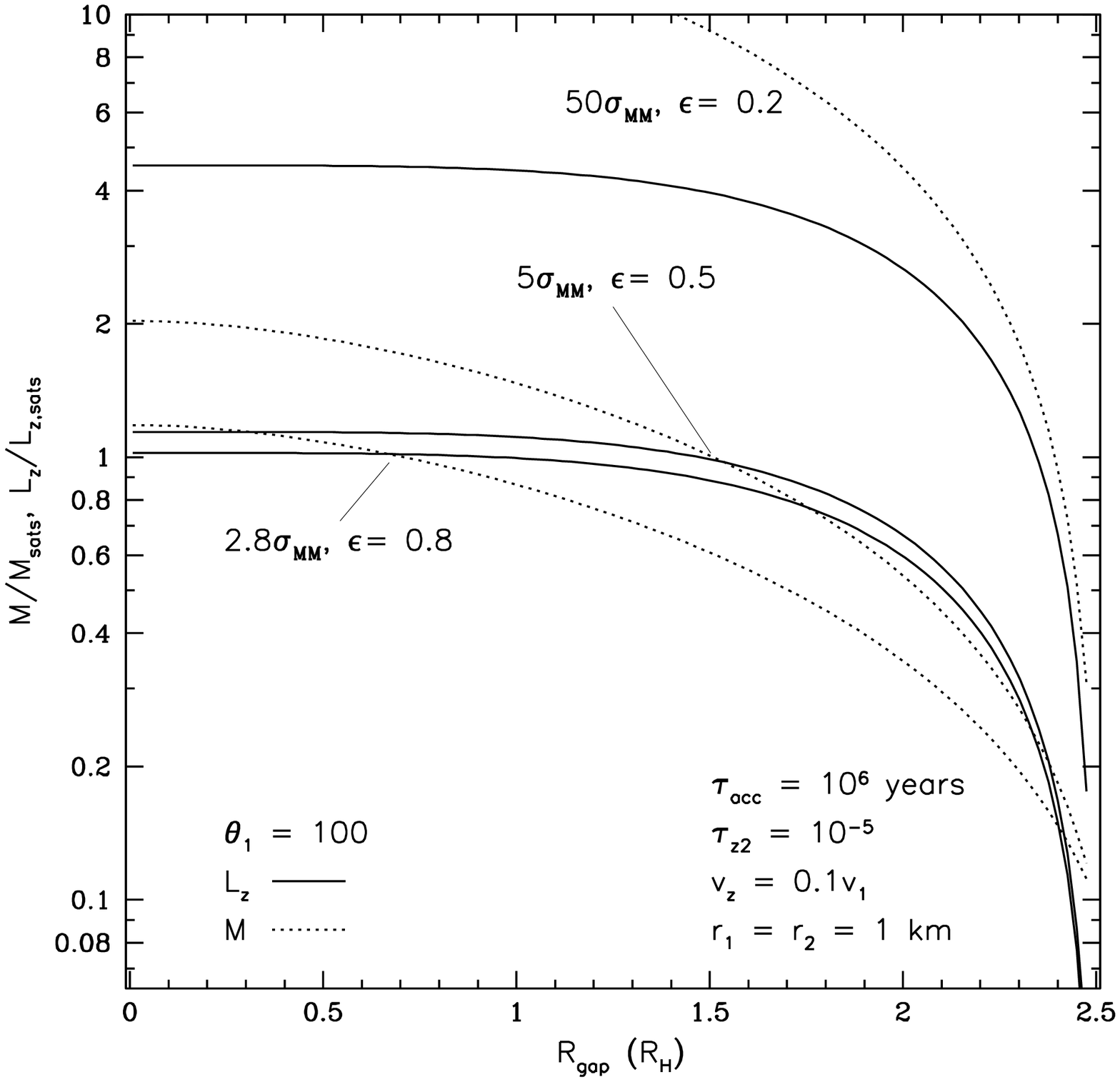,width=7.5in,height=7.5in}}
\begin{center}
Figure 10: Estrada and Mosqueira
\end{center}
\end{figure}

\newpage
\begin{figure}
\centerline{\psfig{figure=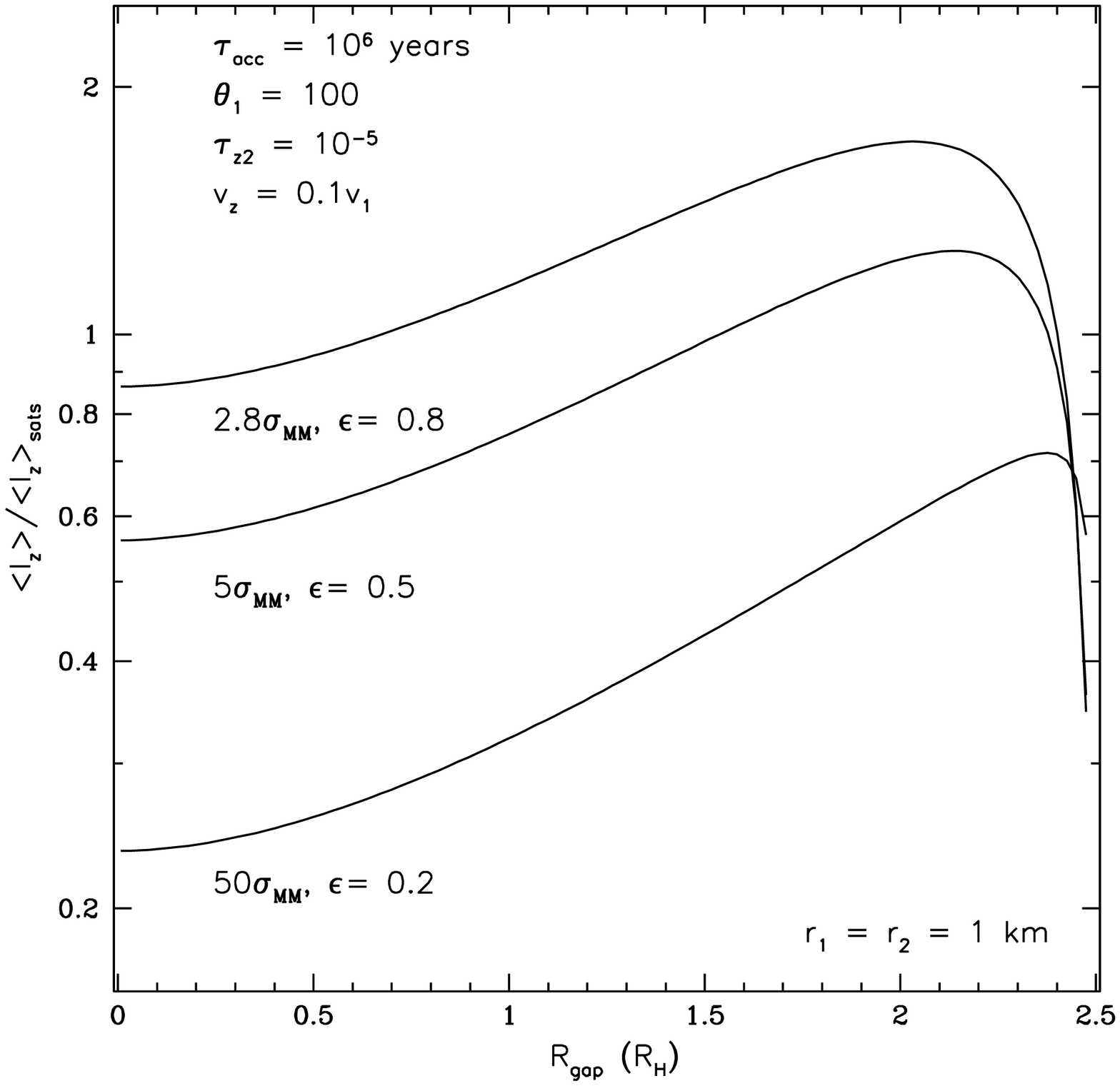,width=7.5in,height=7.5in}}
\begin{center}
Figure 11: Estrada and Mosqueira
\end{center}
\end{figure}

\end{document}